\documentclass[twocolumn]{aastex63}

\usepackage{afterpage}

\usepackage{booktabs} 
\let\tablenum\relax 
\usepackage{siunitx} 
\usepackage{amsmath,amstext,color}
\usepackage{gensymb}
\usepackage{natbib}
\usepackage{hyperref}
\usepackage{longtable}
\usepackage{threeparttablex}
\usepackage{booktabs}
\usepackage{float} 

\usepackage{color}
\usepackage{graphicx}
\usepackage{amsmath}
\usepackage{amssymb}

\usepackage[]{appendix}

\begin{document}
\title{EP\,250108a/SN\,2025kg: Observations of the most nearby Broad-Line Type Ic Supernova Following an Einstein Probe Fast X-ray Transient}
\correspondingauthor{Jillian Rastinejad}
\email{jillianrastinejad2024@u.northwestern.edu}

\shorttitle{SN\,2025kg}
\shortauthors{Rastinejad et al.}
\newcommand{\NU}{\affiliation{Center for Interdisciplinary Exploration and Research in Astrophysics (CIERA) and Department of Physics and Astronomy, Northwestern University, Evanston, IL 60208, USA}}

\newcommand{\Radboud}{\affiliation{Department of Astrophysics/IMAPP, Radboud University, 6525 AJ Nijmegen, The Netherlands}}

\newcommand{\Leicester}{\affiliation{School of Physics and Astronomy, University of Leicester, University Road, Leicester, LE1 7RH, UK}}

\newcommand{\LANL}{\affiliation{Center for Nonlinear Studies, Los Alamos National Laboratory, Los Alamos, NM 87545 USA}}

\newcommand{\ESA}{\affiliation{European Space Agency (ESA), European Space Astronomy Centre (ESAC), Camino Bajo del Castillo s/n, 28692 Villanueva de la Cañada, Madrid, Spain}}

\newcommand{\INAFbologna}{\affiliation{INAF–Osservatorio di Astroﬁsica e Scienza dello Spazio, via Piero Gobetti 93/3, I-40129 Bologna, Italy}}

\newcommand{\INAFRoma}{\affiliation{INAF–Osservatorio Astronomico di Roma, via Frascati 33, I-00040 Monte Porzio Catone, Italy}}

\newcommand{\Caltech}{\affiliation{Department of Astronomy and Astrophysics, California Institute of Technology, Pasadena, CA 91125, USA}}

\newcommand{\Birmingham}{\affiliation{School of Physics and Astronomy, University of Birmingham, Birmingham B15 2TT, UK}}

\newcommand{\BirmIGWA}{\affiliation{Institute for Gravitational Wave Astronomy, University of Birmingham, Birmingham B15 2TT}}

\newcommand{\IfAUH}{\affiliation{Institute for Astronomy, University of Hawaii, 2680 Woodlawn Drive, Honolulu, HI 96822, USA}}

\newcommand{\Oxford}{\affiliation{Astrophysics sub-Department, Department of Physics, University of Oxford, Keble Road, Oxford, OX1 3RH, UK}}

\newcommand{\Belfast}{\affiliation{Astrophysics Research Centre, School of Mathematics and Physics, Queen’s University Belfast, BT7 1NN, UK}}
\author[0000-0002-9267-6213]{Jillian C.~Rastinejad}
\NU

\author[0000-0001-7821-9369]{Andrew J. Levan}
\Radboud
\affil{Department of Physics, University of Warwick, Coventry, CV4 7AL, UK}

\author[0000-0001-5679-0695]{Peter G. Jonker}
\Radboud

\author[0000-0002-5740-7747]{Charles~D.~Kilpatrick}
\NU

\author[0000-0003-2624-0056]{Christopher L.~Fryer}
\LANL

\author[0000-0003-2700-1030]{Nikhil Sarin}
\affiliation{The Oskar Klein Centre, Department of Physics, Stockholm University, AlbaNova, Stockholm, SE-106 91, Stockholm, Sweden}
\affiliation{Nordita, Stockholm University and KTH Royal Institute of Technology, Hannes Alfvéns väg 12, Stockholm, SE-106 91, Stockholm, Sweden}

\author[0000-0002-5826-0548]{Benjamin P. Gompertz}
\Birmingham
\BirmIGWA

\author[0000-0002-7866-4531]{Chang Liu}
\NU

\author[0000-0002-8775-2365]{Rob A. J. Eyles-Ferris}
\affiliation{School of Physics and Astronomy, University of Leicester, University Road, Leicester, LE1 7RH, UK}

\author[0000-0002-7374-935X]{Wen-fai Fong}
\NU

\author[0000-0002-2942-3379]{Eric Burns}
\affil{Department of Physics and Astronomy, Louisiana State University, Baton Rouge, Louisiana 70803, USA}

\author[0000-0002-8094-6108]{James H. Gillanders}
\Oxford

\author[0000-0002-6134-8946]{Ilya Mandel}
\affiliation{School of Physics and Astronomy, Monash University, Clayton VIC 3800, Australia}
\affiliation{ARC Centre of Excellence for Gravitational-wave Discovery (OzGrav), Melbourne, Australia}

\author[0000-0002-7517-326X]{Daniele Bj\o{}rn Malesani}
\affiliation{Cosmic Dawn Center (DAWN), Denmark}
\affiliation{Niels Bohr Institute, University of Copenhagen, Jagtvej 128, Copenhagen, 2200, Denmark}

\author[0000-0002-5128-1899]{Paul T. O'Brien}
\affiliation{School of Physics and Astronomy, University of Leicester, University Road, Leicester, LE1 7RH, UK}

\author[0000-0003-3274-6336]{Nial R. Tanvir}
\affiliation{School of Physics and Astronomy, University of Leicester, University Road, Leicester, LE1 7RH, UK}

\author[0000-0002-8648-0767]{Kendall Ackley}
\affil{Department of Physics, University of Warwick, Coventry, CV4 7AL, UK}

\author[0000-0002-9928-0369]{Amar Aryan}
\affiliation{Graduate Institute of Astronomy, National Central University, 300 Jhongda Road, 32001 Jhongli, Taiwan}

\author[0000-0002-8686-8737]{Franz E. Bauer}
\affil{Instituto de Alta Investigaci{\'{o}}n, Universidad de Tarapac{\'{a}}, Casilla 7D, Arica, Chile}

\author{Steven Bloemen}
\Radboud

\author[0000-0001-5486-2747]{Thomas de Boer}
\IfAUH

\author[0000-0003-4383-2969]{Cl\'ecio R. Bom}
\affiliation{Centro Brasileiro de Pesquisas F\'isicas, Rua Dr. Xavier Sigaud 150, 22290-180 Rio de Janeiro, RJ, Brazil}

\author[0009-0000-6374-3221]{Jennifer A. Chac\'on}
\affil{Instituto de Astrof\'isica, Facultad de F\'isica, Pontificia Universidad Cat\'olica de Chile, Campus San Joaquín, Av. Vicu\~{n}a Mackenna 4860, Macul Santiago, Chile, 7820436}
\affil{Millennium Institute of Astrophysics, Nuncio Monse\~{n}or S\'otero Sanz 100, Of 104, Providencia, Santiago, Chile}

\author[0000-0001-6965-7789]{Ken Chambers}
\IfAUH

\author[0000-0002-1066-6098]{Ting-Wan Chen}
\affil{Graduate Institute of Astronomy, National Central University, 300 Jhongda Road, 32001 Jhongli, Taiwan}

\author[0000-0001-9842-6808]{Ashley A. Chrimes}
\Radboud\ESA

\author[0009-0007-6927-7496]{Joyce N. D. van Dalen}
\Radboud

\author[0000-0003-3703-4418]{Valerio D'Elia}
\affil{Space Science Data Center (SSDC) - Agenzia Spaziale Italiana (ASI), I-00133 Roma, Italy}

\author[0000-0002-4036-7419]{Massimiliano De Pasquale}
\affil{Department of Mathematics and Computer Sciences, Physical Sciences and Earth Sciences, University of Messina, Via F. S. D'Alcontres 31, 98166, Messina, Italy}

\author[0000-0003-1916-0664]{Michael D. Fulton}
\Belfast

\author[0000-0002-4488-726X]{Paul J. Groot}
\Radboud
\affil{Department of Astronomy, University of Cape Town, Private Bag X3, Rondebosch 7701, South Africa}
\affil{South African Astronomical Observatory, P.O. Box 9, Observatory 7935, South Africa}

\author[0000-0003-4905-7801]{Rahul Gupta}
\altaffiliation{NASA Postdoctoral Program Fellow}
\affil{Astrophysics Science Division, NASA Goddard Space Flight Center, Mail Code 661, Greenbelt, MD 20771, USA}

\author[0000-0002-8028-0991]{Dieter H. Hartmann}
\affil{Clemson University, Department of Physics \& Astronomy, Clemson, SC 29631 USA}

\author[0009-0005-5404-2745]{Agnes P.C. van Hoof}
\Radboud

\author[0000-0003-1059-9603]{Mark E.~Huber}
\IfAUH

\author[0000-0002-7965-2815]{Luca Izzo}
\affil{INAF-Osservatorio Astronomico di Capodimonte, Salita Moiariello 16, I-80131, Napoli, Italy}
\affil{DARK, Niels Bohr Institute, University of Copenhagen, Jagtvej 155, 2200 Copenhagen N, Denmark}

\author[0000-0002-3934-2644]{Wynn Jacobson-Galan}
\altaffiliation{NASA Hubble Fellow}
\Caltech

\author[0000-0002-9404-5650]{P\'all Jakobsson}
\affil{Center for Astrophysics and Cosmology, Science Institute, University of Iceland, Dunhagi 5, 107 Reykjavik, Iceland}

\author[00000-0002-5105-344X]{Albert Kong}
\affiliation{Institute of Astronomy, National Tsing Hua University, Hsinchu 30013, Taiwan}

\author[0000-0003-1792-2338]{Tanmoy Laskar}
\affiliation{Department of Physics \& Astronomy, University of Utah, Salt Lake City, UT 84112, USA}
\Radboud

\author{Thomas B. Lowe}
\affil{University of Hawaii Institute for Astronomy, 34 Ohia Ku Street, Makawao, HI 96768, USA}

\author[0000-0002-7965-2815]{Eugene A. Magnier}
\affil{Institute for Astronomy, University of Hawai‘i at Manoa, 2680 Woodlawn Drive, Honolulu, HI 96822, USA}

\author[0000-0003-2593-4355]{Elisabetta Maiorano}
\INAFbologna

\author[0000-0001-5108-0627]{Antonio Martin-Carrillo}
\affiliation{School of Physics and Centre for Space Research, University College Dublin, Belfield, Dublin 4, Ireland}

\author{Lluis Mas-Ribas}
\affil{Department of Astronomy and Astrophysics, University of California, 1156 High Street, Santa Cruz, CA 95064, USA}
\affil{University of California Observatories, 1156 High Street, Santa Cruz, CA 95064, USA}

\author[0000-0003-0245-9424]{Daniel Mata S\'anchez}
\affil{Instituto de Astrof\'isica de Canarias, E-38205 La Laguna, Tenerife, Spain}
\affil{Departamento de Astrofísica, Univ. de La Laguna, E-38206 La Laguna, Tenerife, Spain}

\author[0000-0002-2555-3192]{Matt Nicholl}
\Belfast

\author[0000-0002-2137-4146]{Christopher J. Nixon}
\affiliation{School of Physics and Astronomy, Sir William Henry Bragg Building, Woodhouse Ln., University of Leeds, Leeds LS2 9JT, UK}

\author[0000-0001-9309-7873]{Samantha R. Oates}
\affil{Department of Physics, Lancaster University, Lancaster, LA1 4YB, UK}

\author[0000-0002-6639-6533]{Gregory Paek}
\IfAUH

\author[0000-0002-9408-1563]{Jesse Palmerio}
\affil{Universit\'e Paris-Saclay, Universit\'e Paris Cit\'e, CEA, CNRS, AIM, 91191 Gif-sur-Yvette, France}

\author[0000-0002-7409-8114]{Diego Paris}
\INAFRoma

\author[0000-0003-3114-2733]{Dani\"elle L. A. Pieterse}
\Radboud

\author[0000-0003-3457-9375]{Giovanna Pugliese}
\affil{Astronomical Institute Anton Pannekoek, University of Amsterdam, 1090 GE Amsterdam, The Netherlands}

\author[0000-0001-8602-4641]{Jonathan A. Quirola Vasquez}
\Radboud

\author{Jan van Roestel}
\affil{Anton Pannekoek Institute for Astronomy, University of Amsterdam, 1090 GE Amsterdam, The Netherlands}

\author[0000-0002-8860-6538]{Andrea Rossi}
\INAFbologna

\author[0000-0003-3937-0618]{Alicia Rouco Escorial}
\ESA

\author[0000-0002-9393-8078]{Ruben Salvaterra}
\affil{INAF—Istituto di Astrofisica Spaziale e Fisica Cosmica di Milano, Via A. Corti 12, 20133 Milano, Italy}

\author[0000-0003-4876-7756]{Benjamin Schneider}
\affiliation{Aix Marseille University, CNRS, CNES, LAM, Marseille, France}

\author[0000-0002-8229-1731]{Stephen J. Smartt}
\Oxford
\Belfast

\author[0000-0001-9535-3199]{Ken Smith}
\Oxford
\Belfast

\author[0000-0001-8605-5608]{Ian A. Smith}
\IfAUH

\author[0000-0003-4524-6883]{Shubham Srivastav}
\Oxford

\author[0000-0003-0245-9424]{Manuel A. P. Torres}
\affil{Instituto de Astrof\'isica de Canarias, E-38205 La Laguna, Tenerife, Spain}
\affil{Departamento de Astrofísica, Univ. de La Laguna, E-38206 La Laguna, Tenerife, Spain}

\author[0009-0009-9751-9215]{Chiara Ventura}
\INAFRoma

\author{Paul Vreeswijk}
\Radboud

\author[0000-0002-1341-0952]{Richard Wainscoat}
\IfAUH

\author[0000-0001-9108-573X]{Yi-Jung Yang}
\affil{Graduate Institute of Astronomy, National Central University, 300 Jhongda Road, 32001 Jhongli, Taiwan}

\author[0000-0002-2898-6532]{Sheng Yang}
\affil{Institute for Gravitational Wave Astronomy, Henan Academy of Sciences, Zhengzhou 450046, Henan, China}

\begin{abstract}
With a small sample of fast X-ray transients (FXTs) with multi-wavelength counterparts discovered to date, their progenitors and connections to $\gamma$-ray bursts (GRBs) and supernovae (SNe) remain ambiguous. Here, we present photometric and spectroscopic observations of SN\,2025kg, the SN counterpart to the FXT EP\,250108a. At $z=0.17641$, this is the closest known SN discovered following an \textit{Einstein Probe} ({\it EP}) FXT. 
We show that SN\,2025kg's optical spectra reveal the hallmark features of a broad-lined Type Ic SN. Its light curve evolution and expansion velocities are comparable to those of GRB-SNe, including SN\,1998bw, and two past FXT SNe. We present JWST/NIRSpec spectroscopy taken around SN\,2025kg's maximum light, and find weak absorption due to He I $\lambda 1.0830, \lambda 2.0581$ $\mu$m and a broad, unidentified emission feature at $\sim$ 4--4.5 $\mu$m. 
Further, we observe broadened H$\alpha$ in optical data at 42.5~days that is not detected at other epochs, indicating interaction with H-rich material. From its light curve, we derive a $^{56}$Ni mass of 0.2 - 0.6 $M_{\odot}$. 
Together with our companion paper \citep{EylesFerris25}, our broadband data are consistent with a trapped or low energy ($\lesssim 10^{51}$ ergs) jet-driven explosion from a collapsar with a zero-age main sequence mass of 15--30 $M_{\odot}$. Finally, we show that the sample of {\it EP} FXT SNe support past estimates that low-luminosity jets seen through FXTs are more common than successful (GRB) jets, and that similar FXT-like signatures are likely present in at least a few percent of the brightest Ic-BL SNe.
\end{abstract}


\section{Introduction}

Though first discovered in the era of sounding rockets \citep[e.g.][]{cooke76}, and highlighted by discoveries with {\em Chandra} and {\em XMM-Newton} \citep{Jonker2013,AlpLarsson20,QuirolaVasquez+23}, the nature of fast X-ray transients (``FXTs''; non-repeating transients detected in the X-ray) and their connection to the deaths of massive stars and luminous, jetted $\gamma$-ray bursts (GRBs) remains uncertain. 
By definition, FXTs represent a broad category of transients. In some cases, FXTs could represent the tail of the emission from classical GRBs whose spectra peak at tens to hundreds of keV. FXTs also encompass the population of so-called X-ray Flashes (``XRFs''; \citealt{Heise+01,Sakamoto+05}), a historical ``sub-type'' of GRBs defined by a greater ratio of fluence in the X-ray compared to the $\gamma$-ray bands. Most notably, this XRF/FXT group includes the nearby, serendipitous discoveries of Type Ib/c supernovae (SNe), SN\,2006aj \citep{Campana+06,Pian+06,Soderberg+06} identified in the long GRB/XRF 060218 and the X-ray-only discovery of the much less luminous SN\,2008D \citep{Soderberg+08}. These discoveries showed that at least some FXTs are produced by the core-collapse of massive, stripped-envelope stars. Though sharing similarities with the broad-line Type Ic (Ic-BL) SNe accompanying GRBs (GRB-SNe), the SNe associated with XRFs demonstrate considerable diversity compared to GRB SNe. For example, two SNe (SN\,2008D and SN\,2010bh; \citealt{Soderberg+08}) following these XRFs were $\approx 5-100$ times less luminous compared to the prototypical GRB-SN, GRB\,980425/SN\,1998bw. Further, one (SN\,2008D) showed absorption due to He \citep{Soderberg+06,Modjaz+09}, potentially representing a stage in the continuum of massive star explosions (e.g., \citealt{Galama+98,Clocchiatti+11,Soderberg+08,AlpLarsson20}). However, with a limited sample of such events to date, the mapping of GRBs to FXTs, XRFs or core-collapse SNe without high-energy emission continues to be an open question.  

Previously, the lack of a dedicated discovery mission meant that the FXT sample was limited to rare, serendipitous detections (e.g., \citealt{Soderberg+06,Soderberg+08}) or archival searches in narrow-field instruments such as {\em Chandra} or {\em XMM-Newton} \citep{Jonker2013,Glennie2015,Bauer2017,AlpLarsson20,Novara+20,2022ApJ...927..211L,QuirolaVazquez+22,QuirolaVasquez+23}. The small ($\lesssim$40) population size of FXTs and paucity of multi-wavelength counterparts resulted in large uncertainties on their rates, local environments, and counterpart properties. Uncertainties on these properties inhibit a full understanding of how FXTs fit within the landscape of transients. In particular, many events discovered in the narrow-field searches were of long duration (hundreds to thousands of seconds) compared to GRBs. At present, it is unclear if these long durations represent a selection effect or fundamentally different physical processes \citep[e.g.][]{Levan+24_ep}. At the same time, the realization of substantial diversity in the progenitors of long GRBs, including events likely produced by compact object mergers \citep[e.g.][]{Rastinejad+22,Troja+22,Yang+22,Levan+24}, erases the assumption that FXTs arise exclusively via core-collapse. 

The 2024 launch of a wide-field soft X-ray monitor onboard the {\it Einstein Probe} ({\it EP}), opened a new route to discover FXTs and characterize their multi-wavelength counterparts \citep{Yuan+15,Yuan+22}. Critically, {\it EP} is able to detect FXTs with its Wide Field X-ray Telescope (WXT; 3600 sq. degree field of view), localize them via WXT to $\sim 3$ arcminute precision, and subsequently re-point a narrow-field follow-up telescope to obtain $\sim 10$ arcsecond position. As such, the latency of FXT announcements has reduced from days (and, often, year timescales) to minutes (e.g., \citealt{240315a_disc,GCN_EP250108a_disc}). This speed increase offers the unprecedented ability to locate multi-wavelength counterparts and unveil their origins.  

In its first year alone, {\it EP} discovered dozens of extragalactic FXTs, revealing a diversity of counterparts and environments. While several FXTs have been associated with GRBs (e.g., \citealt{Yin+24,GCN.37071,Liu+25}), others have no observed prompt GRB counterpart despite constraints from $\gamma$-ray facilities, prompting speculation that their progenitors are distinct from GRBs produced following core-collapse (e.g., \citealt{Sun+24,Bright+24_EP240414,vanDalen+24,Busmann+25,O'Connor+25}). At least four {\it EP} events are known to originate at redshifts $z>3.5$ \citep{EP240804_z,GCN38593,Levan+24_ep,Liu+25}, representing an exciting new path to exploring the distant Universe but limiting detailed studies of multi-wavelength counterparts. Notably, the counterpart of EP\,240414a was localized to a galaxy at $z=0.401$ and revealed the spectroscopic signatures of a Ic-BL SN. Critically, this event established that at least some {\it EP} FXTs originate from the explosions of massive stars \citep{vanDalen+24,Sun+24}. However, at this redshift, late-time spectroscopic and photometric follow-up of the SN was limited \citep{Sun+24,vanDalen+24,SCG+25}. More recently, the event EP\,250207b was potentially associated with a galaxy at $z=0.082$ \citep{EP250207b_z}. However, at present, no evidence of an SN counterpart has been announced and published observational limits appear to rule out a SN \citep{ep250207b_gtc}. 

On 8 January 2025 at 12:30:28.34 UT, {\it EP} discovered a new FXT, EP\,250108a, with a duration of $\sim 960^{+3092}_{-208}$~s \citep{GCN_EP250108a_disc,Li+25_ep250108a}. Later analysis of \textit{Fermi}-GBM observations place a conservative, sky-averaged upper limit on associated emission in the 10-1000 keV band of $f_{\gamma} < 2.6 \times 10^{-8}$ erg cm$^{-2}$ s$^{-1}$ at 415~s following the initial trigger (\textit{Fermi}'s view of the localizaton was occulted by Earth prior to this time; \citealt{EP250108a_fermi}). Prompt follow-up observations were not taken, but imaging obtained $\sim 1.5$ days after the outburst revealed a blue and rapidly-fading counterpart with a redshift of $z = 0.17641$ \citep{EylesFerris25}. Beginning at $\approx 6$~days following the initial trigger, the cooling, optical counterpart began to rise in brightness. Spectra taken by the Nordic Optical Telescope (NOT) and Gemini-North at $\delta t \sim 10$~days (where $\delta t$ refers to the observer-frame time since the {\it EP} trigger) showed the signatures of a Ic-BL SN, dubbed SN\,2025kg (see Section~\ref{sec:spec} for further analysis; \citealt{ep250108a_gmos_sn,ep250108a_not_sn,Srinivasaragavan+25}). SN\,2025kg offers a unique first opportunity for a detailed comparison between the multi-band light curves and spectroscopic properties of FXT SNe, GRB-SNe and stripped-envelope SNe observed without high-energy counterparts, shedding light on FXT progenitors.

Here, we present optical and infrared observations and analysis of SN\,2025kg, including {\it James Webb Space Telescope} (JWST) spectroscopy. This work represents a companion paper to \citet{EylesFerris25} which contains an in-depth analysis of the early ($\lesssim 6$~days post-trigger) observations as well as analysis of the radio and high-energy data. In Section~\ref{sec:obs} we detail our extensive photometric and spectroscopic campaign to characterize SN\,2025kg. In Section~\ref{sec:spec_analysis} we analyze our observations, highlight key features in the JWST spectrum, and compare the properties of SN\,2025kg to those of past stripped envelope SNe associated with FXTs, GRBs and discovered without high-energy triggers.  In Section~\ref{sec:mod} we describe our photometric modeling of SN\,2025kg. In Section~\ref{sec:discussion}, we infer properties of the progenitor from our observations, including its zero-age main sequence (ZAMS) mass, constraints on a He shell due to common envelope mass ejection, and contextualize FXT SNe in the wider population of Ic-BL SNe. Throughout this work, we assume a Planck cosmology, \citep{Planck20} report all magnitudes in the AB system, and assume an event redshift of $z=0.17641 \pm 0.0003$ \citep{EylesFerris25}.

\section{Observations}
\label{sec:obs}

\subsection{Photometry}

\begin{figure}
\centering
\includegraphics[angle=0,width = .49\textwidth]{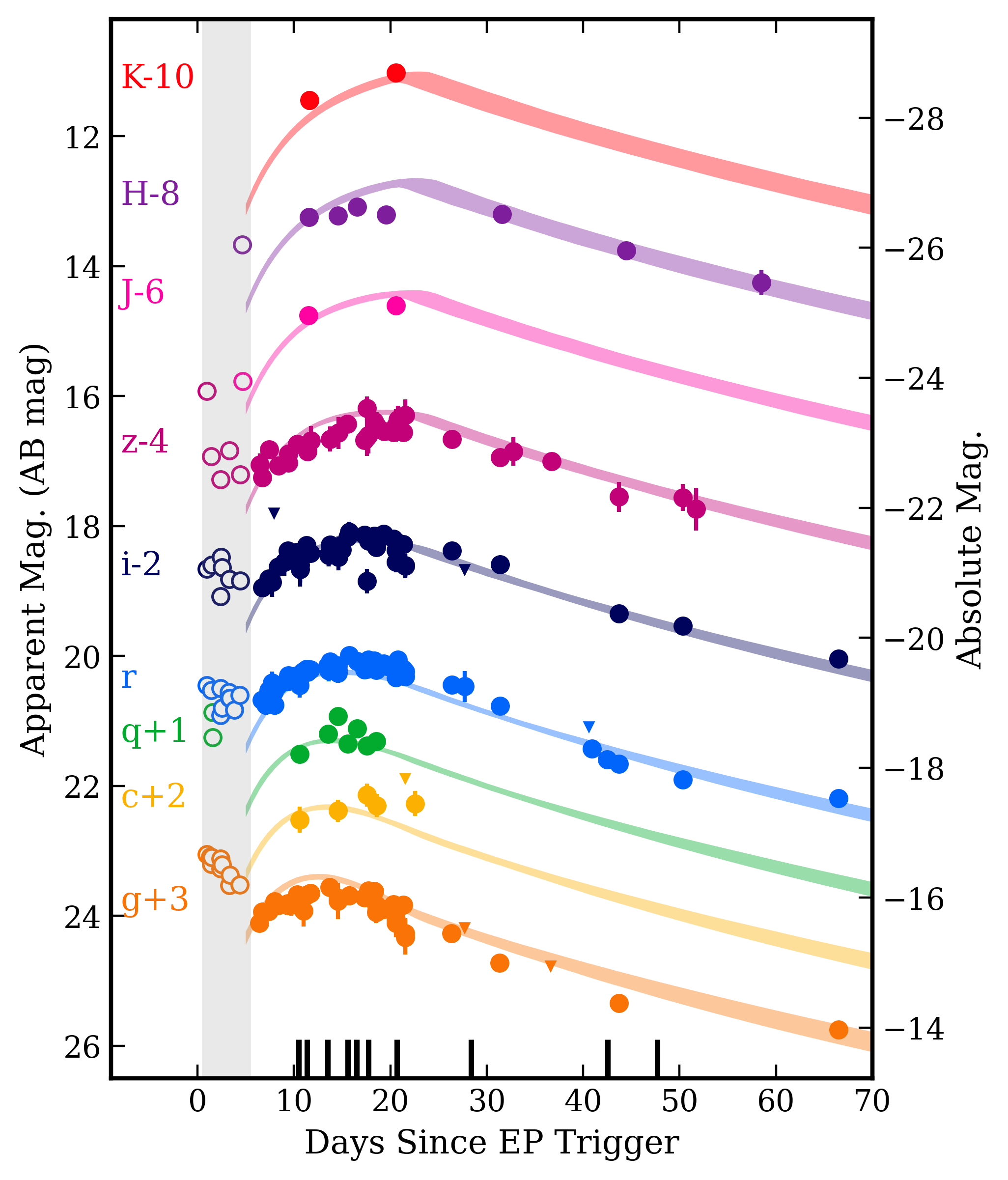}
\caption{Optical-near-IR detections (circles) and upper limits (triangles) of the counterpart to EP\,250108a. The data at $\delta t > 6$~days is well-described by a one-zone radioactive decay model (lines; \citealt{Arnett82,Sarin+24}) with an ${}^{56}$Ni  mass of $0.6 \pm 0.1 M_{\odot}$  (Section~\ref{sec:mod}). Observation during the fast-cooling phase prior to the emergence of SN\,2025kg ($\delta t < 6$~days; open symbols and grey vertical band) are discussed in detail in  \citet{EylesFerris25}. We also show the times of the spectroscopic observations with black vertical lines at the bottom of the figure.}
\label{fig:lc}
\end{figure}

Here, we present optical and near-infrared (IR) photometry of SN\,2025kg covering $6 < \delta t < 66.5$~days (observed frame; observations at $\delta t \lesssim 6$~days are presented in \citealt{EylesFerris25}). We obtained optical observations with the Alhambra Faint Object Spectrograph and Camera (ALFOSC) on the NOT, the IO:O on the Liverpool Telescope (LT), the Gemini Multi-Object Spectrographs (GMOS) on the Gemini-North and South Telescopes, Pan-STARRS, BlackGEM (for $q$-band description see \citealt{BlackGEM24}), the Multi-Object Double-beam Spectrograph (MODS) on the Large Binocular Telescope (LBT), the Goodman spectrograph on the Southern Astrophysical Research (SOAR) Telescope, the Lulin Observatory, MegaCam on the Canada-France-Hawaii Telescope (CFHT), and T80S-Cam on T80S. All optical images were processed using standard techniques, including Gemini DRAGONS \citep{DRAGONS19}, a custom python pipeline, POTPyRI\footnote{https://github.com/CIERA-Transients/POTPyRI}, {\tt photpipe} \citep[see][for details]{Rest+05,Santos+24}, a dedicated LBT data reduction pipeline \citep{Fontana2014a}, a modified version of the \texttt{photometry-sans-frustration} package \texttt{psf} \citep{Nicholl23}, the \texttt{Elixir} pipeline \citep{elixir+04}, a custom built pipeline\footnote{\href{https://hdl.handle.net/11296/98q6x4}{https://hdl.handle.net/11296/98q6x4}}, as well as a combination of standard \texttt{PyRAF} tasks \citep{Pyraf2012} and the \texttt{LACosmic} task \citep{van_Dokkum2001}. 

We obtain photometry directly on each image using a common set of standard stars drawn from Pan-STARRS \citep{Flewelling+20}. We do not anticipate significant ($\gtrsim 0.6$~mag) contamination from the underlying host galaxy given its faintness relative to the transient as measured in archival Legacy Survey imaging ($g=23.40 \pm 0.04$ mag, $r=23.13 \pm 0.02$~mag, $i=22.99 \pm 0.06$~mag, $z=22.89 \pm 0.08$~mag; \citealt{Dey+19}). In addition, we obtained near-IR imaging with FLAMINGOS2 on Gemini-South and the MMT and Magellan Infrared Spectrograph (MMIRS) on the MMT. We process all near-IR images using DRAGONS and POTPyRI and obtain photometry using a common set of standard stars from the Two Micron All Sky Survey (2MASS; \citealt{2MASS}). 

We provide details for all photometric programs in Appendix Table~\ref{tab:photprog}.  We further incorporate detections from the GCNs \citep{ep250108a_dfot,ep250108a_gmg,ep250108a_mephisto} and ATLAS forced photometry ($co$-bands; \citealt{Tonry+18}). We present all photometry in AB magnitudes in Appendix Table~\ref{tab:phot} and show our observations in Figure~\ref{fig:lc}.

\subsection{Spectroscopy}
\label{sec:spec}

\begin{figure}
\centering
\includegraphics[angle=0,width=.5\textwidth]{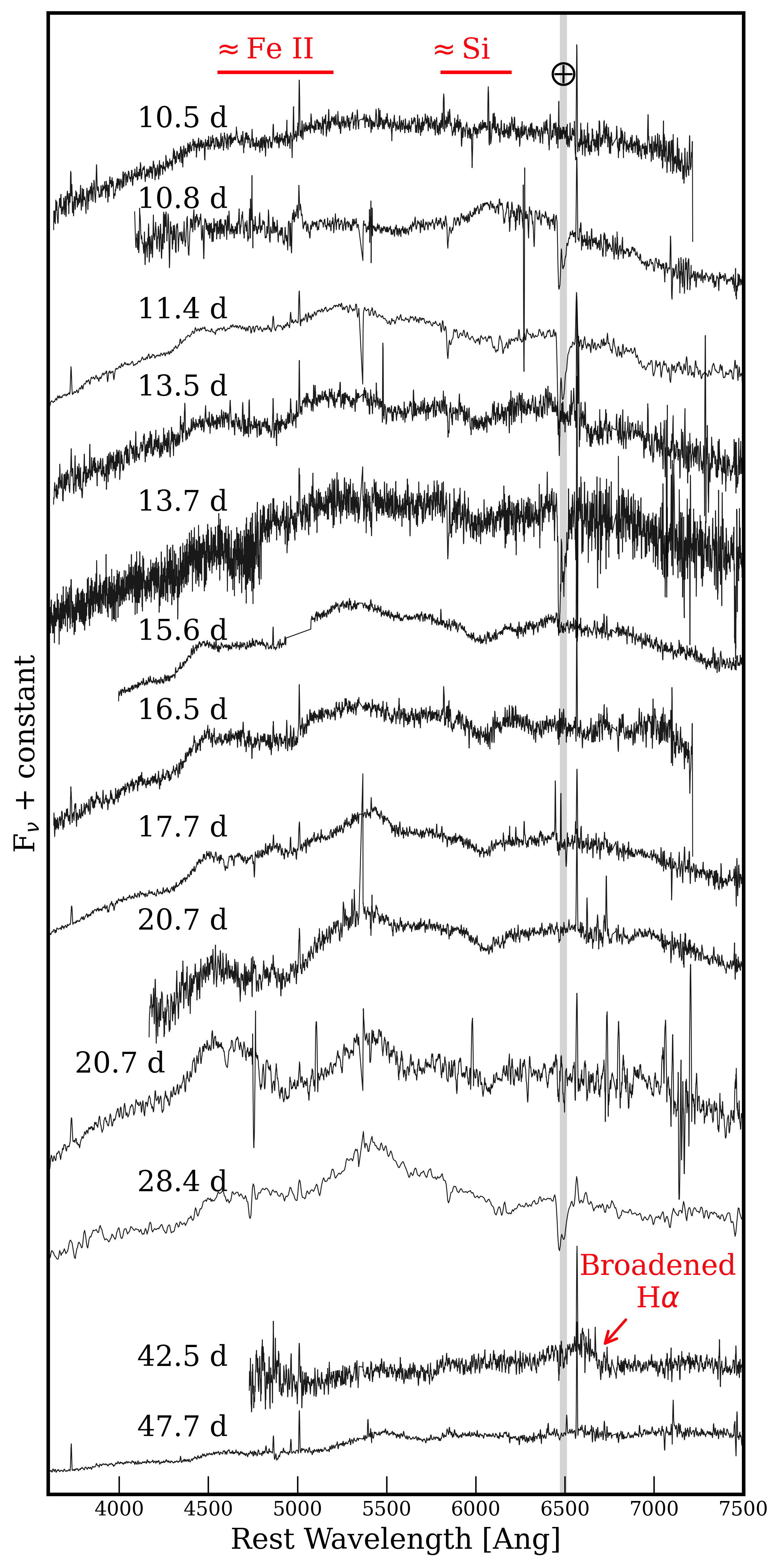}
\caption{Our spectroscopic sequence of SN\,2025kg. We observe broad absorption features around $\sim$5000 \AA\ and $\sim$6100 \AA\ that we ascribe to Fe II $\lambda 5169$ and Si $\lambda 6355$. In the spectrum at $\delta t = 42.5$~days we observe broadened H$\alpha$ emission. We mark the location of a telluric feature with a $\bigoplus$.}
\label{fig:spectra_series}
\end{figure}

\begin{figure*}
\centering
\includegraphics[angle=0,width=.65\textwidth]{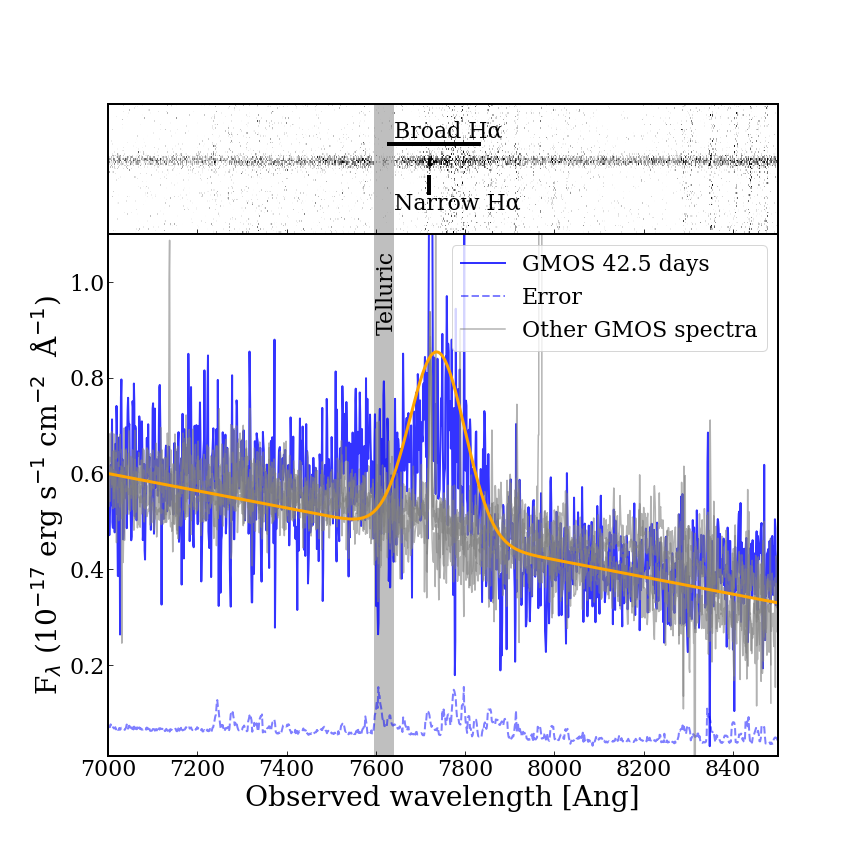}
\vspace{-.7cm}
\caption{Images of the 2D (top) and 1D (bottom) GMOS spectrum of SN\,2025kg taken at $\delta t = 42.5$~days, centered around H$\alpha$ $\lambda 6562.8$ \AA. For comparison, we show our other GMOS spectra in grey, which show only the narrow H$\alpha$ component. In addition to the narrow component, which is present in other epochs (Figure~\ref{fig:spectra_series}) and likely due to the underlying host galaxy, we observe broad, underlying emission, indicating an H-rich source is interacting with SN\,2025kg at this epoch. We show an orange Gaussian superimposed on the continuum for visual purposes.}
\label{fig:1d2dspectrum_halpha}
\end{figure*}

We present thirteen optical spectra of SN\,2025kg taken over $\delta t = 10.5 - 47.7$~days. Spectroscopy prior to the emergence of SN features ($2.6 < \delta t < 4.6$~days) is presented and analyzed in our companion paper \citep{EylesFerris25}. Our optical spectroscopy was obtained with GMOS on the Gemini-North and South Telescopes (Program IDs GN-2024B-Q-107, GN-2024B-Q-131, GS-2024B-Q-105, GS-2025A-Q-107, P.I.s Rastinejad, Huber), OSIRIS+ on the Gran Telescopio Canarias (Program ID GTC1-24ITP; P.I.~Jonker), the Multi Unit Spectroscopic Explorer (MUSE) on the Very Large Telescope (VLT; Program ID 111.259Q.001, P.I.~Jonker), and the Low Resolution Imaging Spectrometer (LRIS) on Keck Observatory (P.I.s Harrison, Prochaska). Details of each spectroscopic set up are logged in Appendix Table~\ref{tab:spec}.

We reduce Keck and Gemini spectroscopy using \texttt{PypeIt} \citep{PypeIt}. To account for instrumental flexure, sub-pixel shifts in the position of the [O I] 6300 \AA\ sky emission line were measured and applied as corrections. We reduce GTC spectra using \texttt{Molly} to correct for the Earth's motion relative to the target and observations of spectrophotometric standard stars taken the same night for flux calibration. We reduce the VLT spectrum using ESO-reflex \citep{esoreflex}. 
We process the LBT spectrum using the Spectroscopic Interactive Pipeline and Graphical Interface (SIPGI) tool \citep{sipgi2022} and perform wavelength calibration using arc lamp frames. To obtain flux-calibrated LBT spectrum, we applied the sensitivity function derived from the spectro-photometric standard star Feige 34. 
We correct all spectra for Galactic extinction in the direction of the burst ($A_V = 0.049$~mag; \citealt{SchlaflyFinkbeiner11}) and show the optical spectroscopic series in Figure~\ref{fig:spectra_series}. 

In each spectrum, we identify the distinct signatures of Type Ic-BL SNe. For instance, we observe at least two broad absorption features around the expected locations of Fe II $\lambda 5169$ (or a blend of the Fe triplet) and Si $\lambda 6355$ (Figure~\ref{fig:spectra_series}) and no prominent absorption around the optical features of H or He. In several of the spectra we observe narrow emission lines, including those at the rest-frame locations of H$\alpha$, H$\beta$, and [OIII]$\lambda 4958, 5007$, which we attribute to the underlying host galaxy. To compare SN\,2025kg to other SNe, we utilize the spectral template matching tool, \texttt{gelato} \citep{Harutyunyan+08}, which matches an input spectrum to a large bank of observed SNe spectra at a range of phases. For the SN\,2025kg spectrum observed at $\delta t = 17.7$~days \texttt{gelato} finds a strong match to the Type Ic-BL SN\,2006aj (associated with GRB/XRF 060218) at 0.8~days past peak provides the strongest fit. For the SN\,2025kg spectrum observed at $\delta t = 28.4$~days, \texttt{gelato} finds strong matches to several Type Ic-BL SNe, including SNe 1997ef, 1998bw, 2002ap and 2006aj. This analysis definitively confirms our classification of SN\,2025kg as a Type Ic-BL SN.  

In the GMOS spectrum at $\delta t  = 42.5$~days (Figure~\ref{fig:1d2dspectrum_halpha}) we observe a broadened H$\alpha$ emission feature, distinct from the narrow H$\alpha$ emission observed in previous epochs (Figure~\ref{fig:spectra_series}). We inspected the individual four 2D spectral frames of the $\delta t = 42.5$~days spectrum and observe evidence for excess emission associated with this broadened feature in each. We do not see significant broadened H$\alpha$ emission in the previous ($\delta t = 28.4$~days) or subsequent ($\delta t = 47.7$~day) spectra. We discuss this feature in greater detail in Section~\ref{subsec:spec_comp}.

\begin{figure*}
\centering
\includegraphics[angle=0,width=\textwidth]{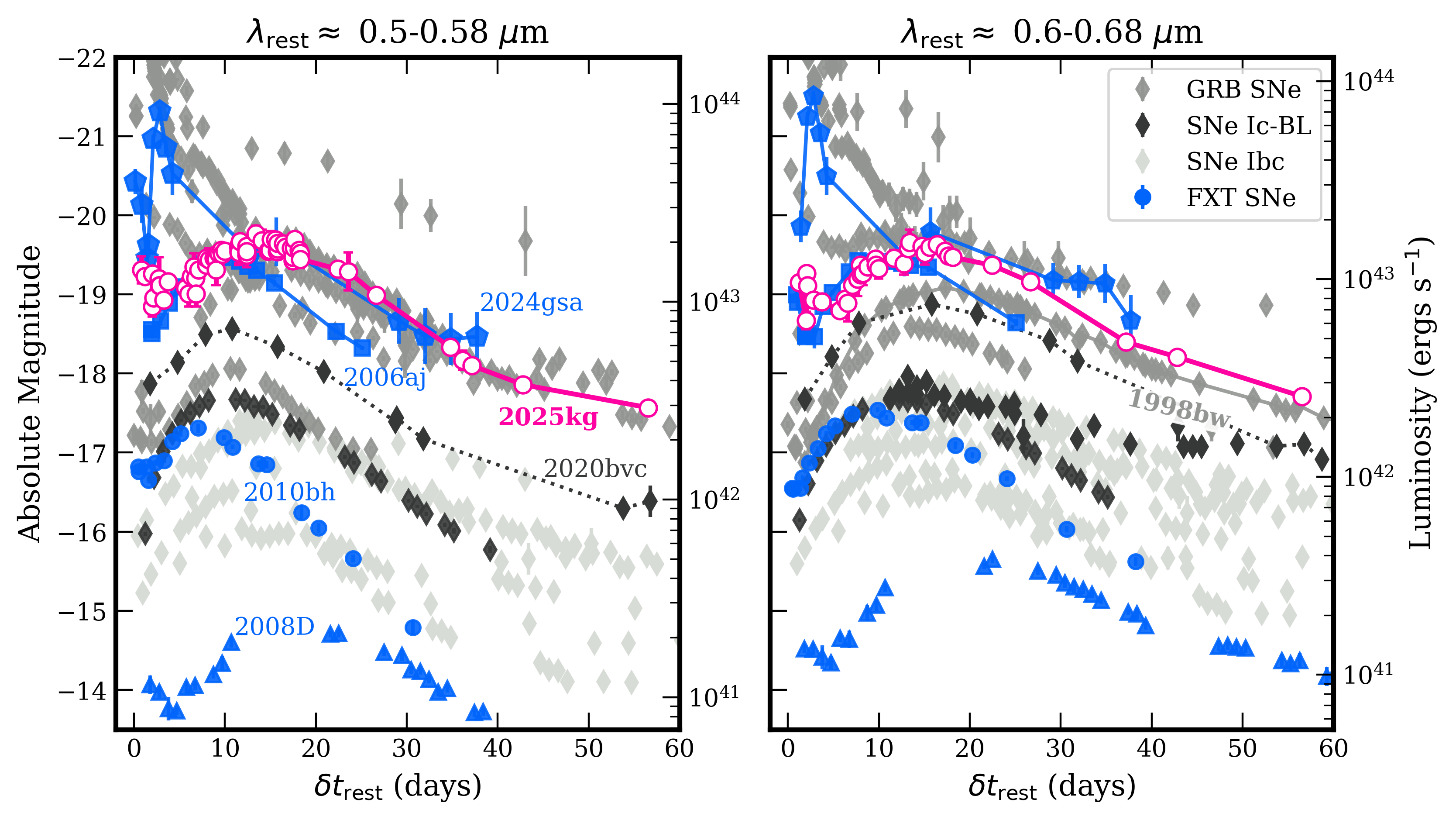}
\caption{Light curve comparison of SN\,2025kg (pink circles) and SNe associated with FXTs (blue), GRB-SNe (grey diamonds), SNe Ic-BL observed without GRB counterparts (dark grey diamonds), and SNe Ib/c (light grey diamonds). Observations are corrected for Milky Way extinction \citep{SchlaflyFinkbeiner11} and presented in the nearest rest-frame band using the event's known redshift. The light curve of SN\,2025kg strongly resembles those of SN\,1998bw, SN\,2006aj and SN\,2024gsa in its peak luminosity and temporal evolution.}
\label{fig:lc_comp}
\end{figure*}

In addition, we obtained two near-IR spectra of SN\,2025kg.  At $\delta t = 12.7$~days post-burst we obtained a cross-dispersed spectrum covering $0.82-2.5~\mu$m with the Gemini Near-InfraRed Spectrograph (GNIRS) on Gemini-North (Program ID GN-2024B-Q-107, P.I.~Rastinejad). These data were processed in {\tt Pypeit} with additional post-processing performed following methods described in \citet{Tinyanont+24}.  Due to low signal-to-noise over the majority of the bandpass, we consider only the portion $\lesssim 1.25~\mu$m usable. Further, at $\delta t = 17.2$ days, we obtained fixed slit spectroscopy with NIRSpec on JWST (Program 6133; P.I.~Gompertz) using the prism (covering 0.5-5 microns) at low resolution, and with an exposure time of 6302 seconds. At the time of the JWST spectroscopy the source is point-like and several magnitudes brighter than the anticipated magnitude of its compact underlying host galaxy. Therefore, we utilize the default pipeline reduction and 1D extraction of the NIRSpec observations. 

\section{Analysis \& Comparison to Past Events}
\label{sec:spec_analysis}

\subsection{SN Light Curves}
\label{sec:lcs}

To place SN\,2025kg in context of past GRB-SNe, FXT-SNe, and other stripped-envelope SNe, we compare its light curve to those of relevant past events. We include eleven events with X-ray or GRB counterparts: SN\,1998bw (GRB\,980425; \citealt{Clocchiatti+11}), SN\,2003dh (GRB\,030329; \citealt{Matheson+03}), SN\,2006aj (GRB/XRF\,060218; \citealt{Mirabal+06,Soderberg+06}), SN\,2008D (XRF\,080109; \citealt{Soderberg+08}) SN\,2010bh (GRB/XRF\,100316D; \citealt{Cano+11,Olivares+12}), SN\,2011kl (GRB\,111209A; \citealt{Levan+14_ULGRBs,Greiner+15}) SN\,2013cq (GRB\,130427A; \citealt{Perley+14}), SN\,2016jca (GRB 161219B; \citealt{Cano+17}), SN\,2017iuk (GRB\,171205A; \citealt{Izzo+19}), SN\,2019oyw (GRB\,190829A; \citealt{Hu+21,Rastinejad+24}), and SN\,2024gsa (EP\,240414a; \citealt{SCG+25}). For events without high-energy counterparts, we show a subset of Type Ib/c and Ic-BL SNe from the literature \citep{Foley+02,Drout+11,Izzo+20,Ho+20}, which we normalize in time to the peak of SN\,2025kg in each respective filter. We show all light curves in the nearest rest-frame filter using their known redshifts (Figure~\ref{fig:lc_comp}).

This comparison (Figure~\ref{fig:lc_comp}) highlights that SN\,2025kg's luminosity is comparable to most GRB-SNe, particularly SN\,1998bw, and distinct from SNe observed without high-energy counterparts. Amongst the SNe associated with FXTs, the closest analogs of SN\,2025kg are SN\,2006aj and SN\,2024gsa, the latter of which is also associated with an {\it EP} FXT. The optical light curve of SN\,2024gsa is marked with a large flare between $\delta t = 1.5 - 3$~days (e.g., \citealt{SCG+25,vanDalen+24}), which we do not observe in early observations of SN\,2025kg \citep{EylesFerris25}.

\subsection{SN\,2025kg Spectra \& Broadened H$\alpha$ at $\delta t = 42.5$~days}
\label{subsec:spec_comp}

\begin{figure}
\centering
\includegraphics[width=.47\textwidth]{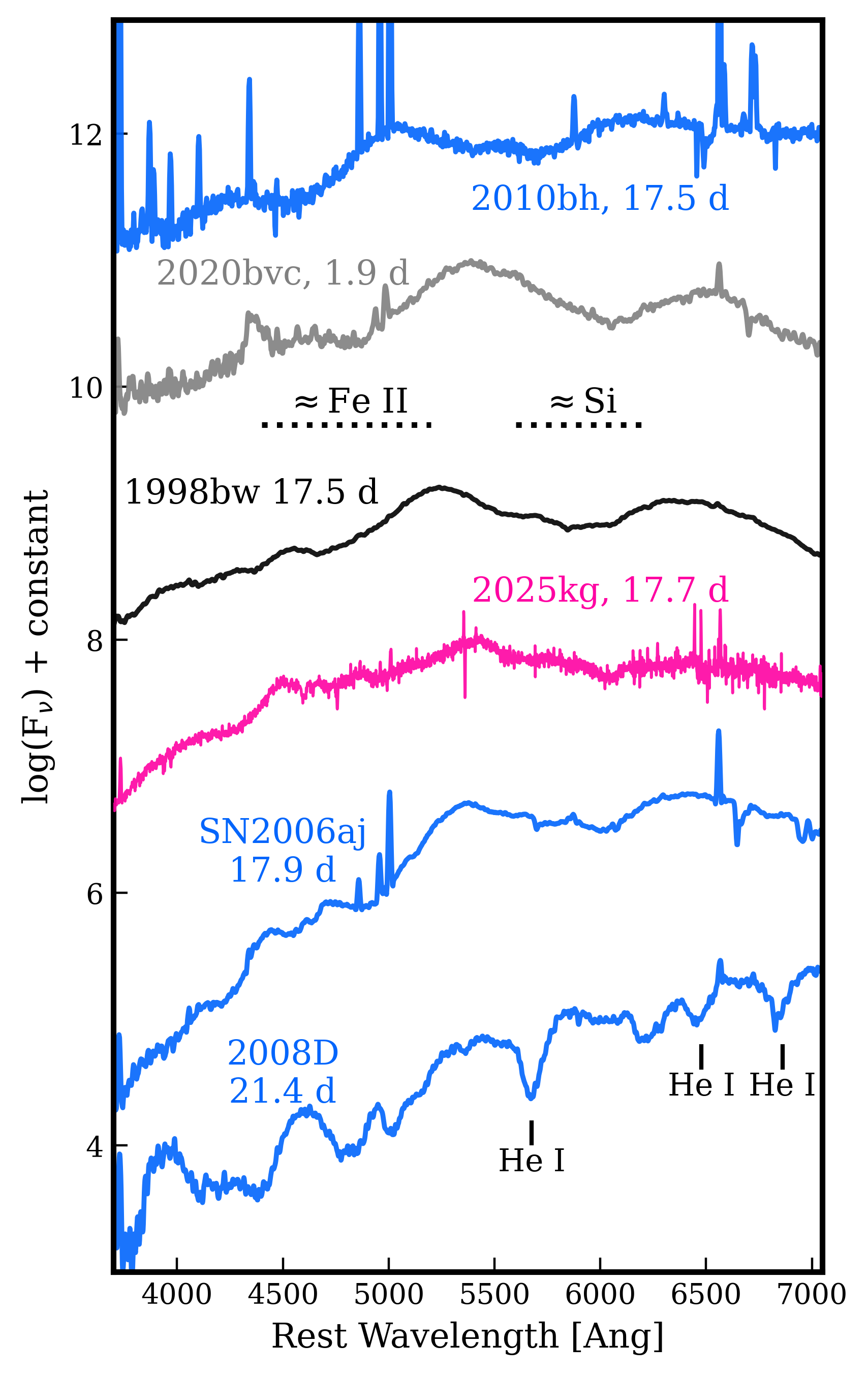}
\caption{Comparison of SN\,2025kg spectra (pink) to SN\,1998bw (black; \citealt{Patat+01}), and three SNe associated with FXTs (blue), SN\,2006aj \citep{Pian+06}, SN\,2008D \citep{Modjaz+09,Modjaz+14} and SN\,2010bh \citep{Bufano+12} around their SNe peak. We also show a spectrum of the Ic-BL SN\,2020bvc \citep{2020bvc_spec} taken $\sim$1.9 days post-explosion. We label the approximate locations of Fe II $\lambda 5169$ \AA and Si $\lambda 6355$ \AA. Due to its broad-lined features and lack of He lines (labeled in SN\,2008D) SN\,2025kg is comparable to SN\,1998bw and SN\,2006aj.}
\label{fig:spec_comp}
\end{figure}

We next attempt to compare our spectra of SN\,2025kg to past GRB-SNe, FXT SNe, Ic-BL SNe without a high-energy trigger, and AT\,2018cow, the latter of which is a benchmark event showing broadened H$\alpha$ emission \citep{Prentice+18,Margutti+19,Perley+19}. We download spectra from WISeREP \citep{Yaron+12} of the prototypical GRB-SN, SN\,1998bw \citep{Patat+01}, three SNe associated with FXTs, SN\,2006aj \citep{Pian+06}, SN\,2008D \citep{Modjaz+06,Modjaz+14} and SN\,2010bh \citep{Bufano+12} and SN\,2020bvc \citep{2020bvc_spec}. Further, we incorporate spectra of the luminous fast blue optical transient, AT\,2018cow, in which broadened H$\alpha$ was prominently observed starting at $\approx$15 days following the event's detection (e.g., \citealt{Margutti+19,Perley+19,Xiang+21}).

In Figure~\ref{fig:spec_comp} we show the spectra for GRB-SNe and FXT SNe within a week of the SN peak ($\delta t = 17.5 - 21.4$~days) and the spectrum of SN\,2020bvc taken 1.9~days after the explosion (no later spectra are available on WISEReP). Spectra of the Type Ib SN\,2008D show prominent He I $\lambda$ 4471, 5876, 6678, 7061 \AA~absorption lines \citep{Modjaz+09}, which we do not observe in SN\,2025kg. Considering the remaining spectra, at both epochs, SN\,2010bh is less visually comparable to SN\,2025kg than SN\,2006aj and SN\,1998bw, which we attribute in part to its larger ejecta velocities (see Section~\ref{subsec:velocities}; \citealt{Chornock+10}). Similar to their light curve properties (Section~\ref{sec:lcs}) SN\,1998bw and SN\,2006aj are spectroscopically comparable objects to SN\,2025kg (Section~\ref{sec:spec}). SN\,2020bvc also shows a comparable shape to SN\,2025kg. We note that SN\,2020bvc was also comparable to SN\,2025kg in terms of photometric color, blackbody radius, and X-ray properties during the early ($\delta t \lesssim 6$~days) fast-cooling phase, as shown in our companion paper \citep{EylesFerris25}.

\begin{figure}
\centering
\includegraphics[angle=0,width=.45\textwidth]{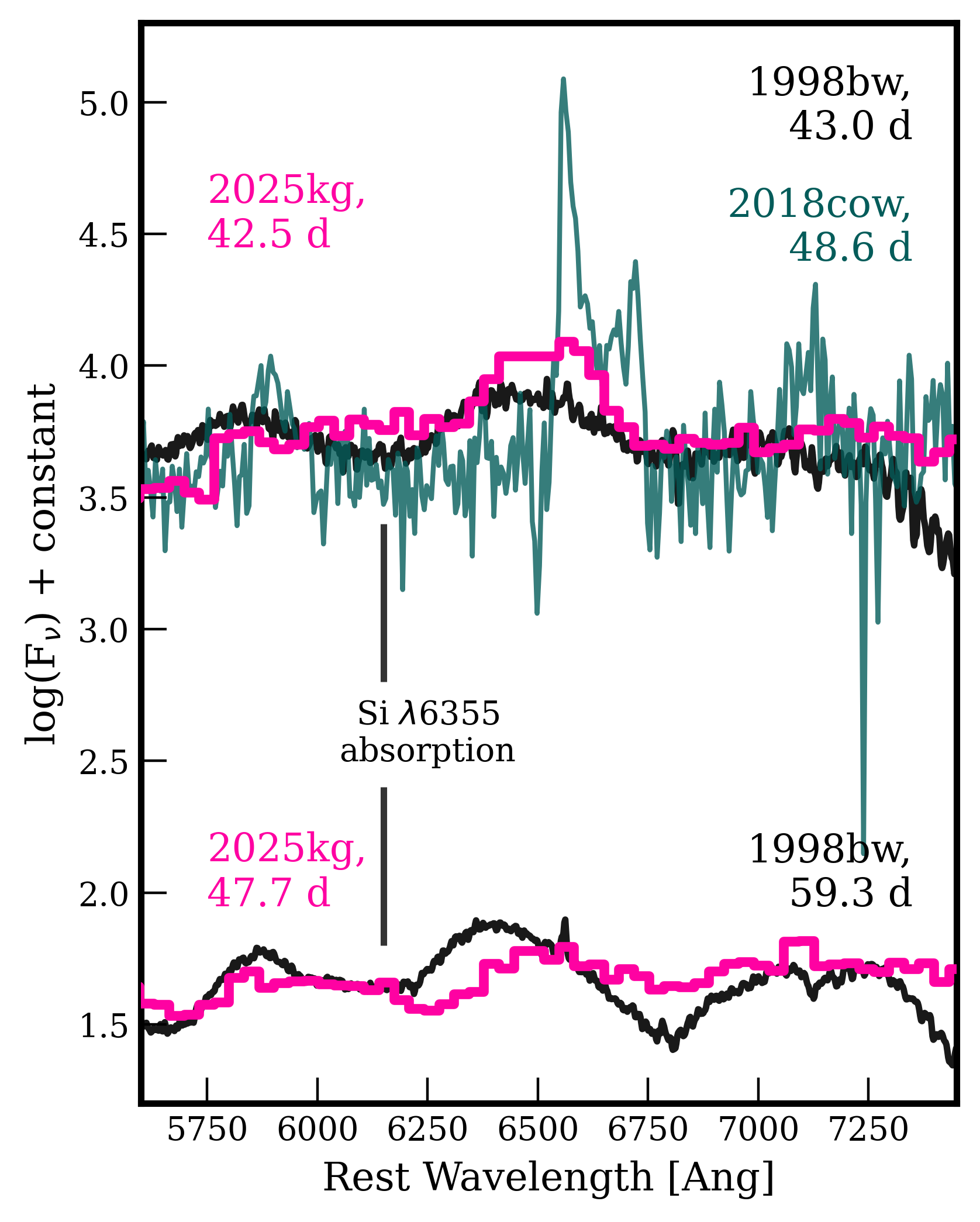}
\caption{Spectra of SN\,2025kg centered around H$\alpha$ at $\delta t = 42.5$ and 47.7~days binned at 40 \AA~binned spectra (pink) in which regions affected by the telluric and narrow H$\alpha$ have been removed. We observe a broadened H$\alpha$ emission feature in the SN\,2025kg Gemini-South/GMOS spectrum at $\delta t = 42.5$~days. We observe broadened H$\alpha$ in the ESO/NTT spectrum of AT\,2018cow (green) but not in the Danish/DFOSC spectrum of SN\,1998bw (black) at similar epochs.
}
\label{fig:halpha}
\end{figure}

Turning to the broadened H$\alpha$ emission, we fit a simple Gaussian profile to the feature (omitting the narrow component using \texttt{curvefit}, finding a full width at half maximum value of $\sim$50 \AA. We next compare our SN\,2025kg spectrum at $\delta t = 42.5$~days to those of SN\,1998bw, SN\,2010bh and AT\,2018cow at $\delta t = 42.5 - 48.6$~days in the region of H$\alpha$ (Figure~\ref{fig:halpha}). Spectra of SN\,2006aj beyond 20~days of the initial high-energy trigger are not available on WISeREP. Compared to the spectrum of SN\,2025kg binned at a resolution of 40 \AA, the broadened emission feature is stronger and more narrow in AT\,2018cow and absent or very weak in SN\,2010bh. Though a comparable bump in the spectrum of SN\,1998bw is observed, we note that this likely due to two absorption features on either side of H$\alpha$. Further, in SN\,1998bw the spectral shape in this region is is broadly consistent across several weeks of observations. In contrast, in SN\,2025kg the continuum is fairly flat over $\sim 5700-6000$~\AA at most epochs, but exhibits broad H$\alpha$ at one single epoch (Figure~\ref{fig:1d2dspectrum_halpha}). Thus, we conclude that absorption features cannot explain the broadened emission feature in SN\,2025kg, and that its most likely interpretation is that it arises from H interaction. 

Whether this feature is common in FXT SNe but not GRB-SNe, or if SN\,2025kg is an outlier amongst other FXT SNe remains to be seen. Indeed, since the H appears only in one spectrum it may be a transient feature that could occur more frequently in other Ic-BL SNe but is often missed because it appears only at later times, and is only visible for a short period. One potential explanation for this feature in SN\,2025kg is interaction with an extended H shell of circumstellar material (CSM), as was considered for AT\,2018cow \citep{Margutti+19,Perley+19}. However, in AT\,2018cow the broadened H$\alpha$ signature was stronger compared to the continuum and persisted over several weeks, and was generally narrower, whereas in SN\,2025kg the feature is weaker and absent in the spectrum at $\delta t =47.7$~days (Figure~\ref{fig:spectra_series}). Alternatively, this could be a signature of a stellar companion, as is observed in a small (1-5\%) fraction of Type Ia SNe \citep{Maguire+16,Tucker+20}. A larger sample of spectra at $\delta t \gtrsim 30$~days is critical to exploring the presence and temporal evolution of broadened H$\alpha$ in future FXT SNe. We further discuss implications of this feature in Section~\ref{sec:discuss_halpha}.

\subsection{Broad Line Velocity Measurements}
\label{subsec:velocities}

\begin{figure}
\centering
\includegraphics[angle=0,width=.49\textwidth]{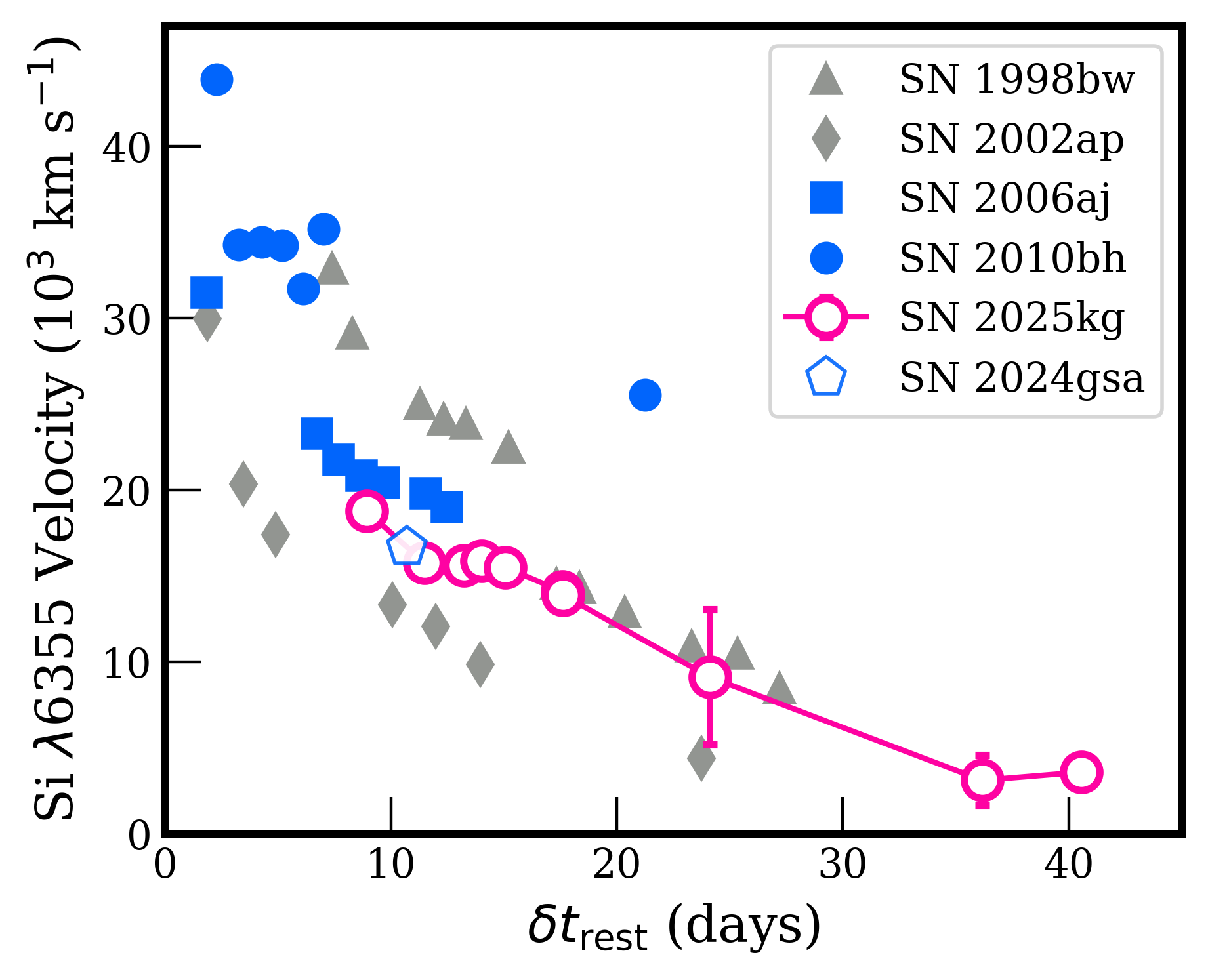}
\caption{
Si $\lambda 6355$ velocities of SN\,2025kg (pink circles) along with literature values for GRB-SNe, FXT SNe and SNe Ic-BL 2002ap \citep{Chornock+10}. SN\,2025kg comparable in absorption velocity to SNe\,1998bw and 2006aj.}
\label{fig:velocities}
\end{figure}

We compute the ejecta expansion velocity using the Doppler shift of the absorption minima ascribed to Fe II $\lambda 5169$ \AA and Si $\lambda 6355$ \AA, which are well-characterized in GRB-SNe (e.g., \citealt{Chornock+10,Bufano+12}). For each spectrum, we fit the wavelength region around the absorption feature with a linear continuum and Gaussian profile using \texttt{scipy.curvefit}, and determine the velocity using the absorption minima's shift relative to the rest-frame wavelength. We show the results from our fitting over time and compare to values for several GRB-SNe and an SN Ic-BL \citep{Chornock+10} in Figure~\ref{fig:velocities}. 

We derive the velocity evolution for both Fe II $\lambda 5169$ and Si $\lambda 6355$. Though slight discrepancies between the velocities are present at several epochs, we note that measurements of expansion velocity using different lines are known to produce inconsistent values, in part due the locations of each elements within different layers of the ejecta \citep{Modjaz+16,Finneran+24}. We next compare to the Si $\lambda 6355$ velocities of SNe 1998bw, 2002ap, 2006aj and 2010bh \citep{Chornock+10} and the only other confirmed {\it EP} SN, SN\,2024gsa (EP\,240124A; \citealt{Sun+24,vanDalen+24}), for which one measurement at $\delta t \approx 15$~days is available (Figure~\ref{fig:velocities}, bottom panel; \citealt{Sun+24}). At this epoch, the expansion velocities of SN\,2025kg and SN\,2024gsa are comparable. Compared to the wider sample of events, SN\,2025kg appears most closely related to SN\,2006aj, although the temporal coverage is not equivalent. SN\,2025kg has larger measured velocities compared to the Ic-BL SN\,2002ap and, generally, lower velocities compared to SN\,1998bw and SN\,2010bh.

\subsection{Near-IR Evidence for Helium}
\label{subsec:helium}

Motivated by questions of the unknown stellar progenitors of FXTs, we next investigate to what degree He is present in the spectra of SN\,2025kg. In our optical spectra, we do not observe prominent He I $\lambda$ 4471, 5876, 6678, 7061 \AA~absorption lines, as were observed in SN\,2008D (Figure~\ref{fig:spec_comp}; \citealt{Soderberg+08}). However, as He produces stronger signatures in the near-IR compared to the optical, we focus our search on the He I lines $\lambda 1.0830, \lambda 2.0581$ $\mu$m in the near-IR spectra. He I $\lambda 1.0830$ is a prominent feature that dominates the 1 $\mu$m region in Type Ib SNe, and may be  blended with C I $\lambda 1.0693$ $\mu$m and Mg II $\lambda 1.0927$ $\mu$m lines in Type Ic/Ic-BL SNe (e.g., \citealt{Shahbandeh+22,Tinyanont+24}).  Though it is well-established that the He I $\lambda 1.0830$ $\mu$m line is a more dominant line compared to He I $\lambda 2.0581$ $\mu$m, the redder He I line has a greater offset from other potential absorption features and, thus, may be easier to distinguish as He I \citep{Shahbandeh+22,Tinyanont+24}. In the GNIRS spectrum, we do not observe any significant absorption features in the 1 $\mu$m regime, though we note that the majority of the spectrum has a low signal-to-noise. 

\begin{figure*}
\centering
\includegraphics[angle=0,width=.75\textwidth,trim={0 0 0 8.6cm},clip]{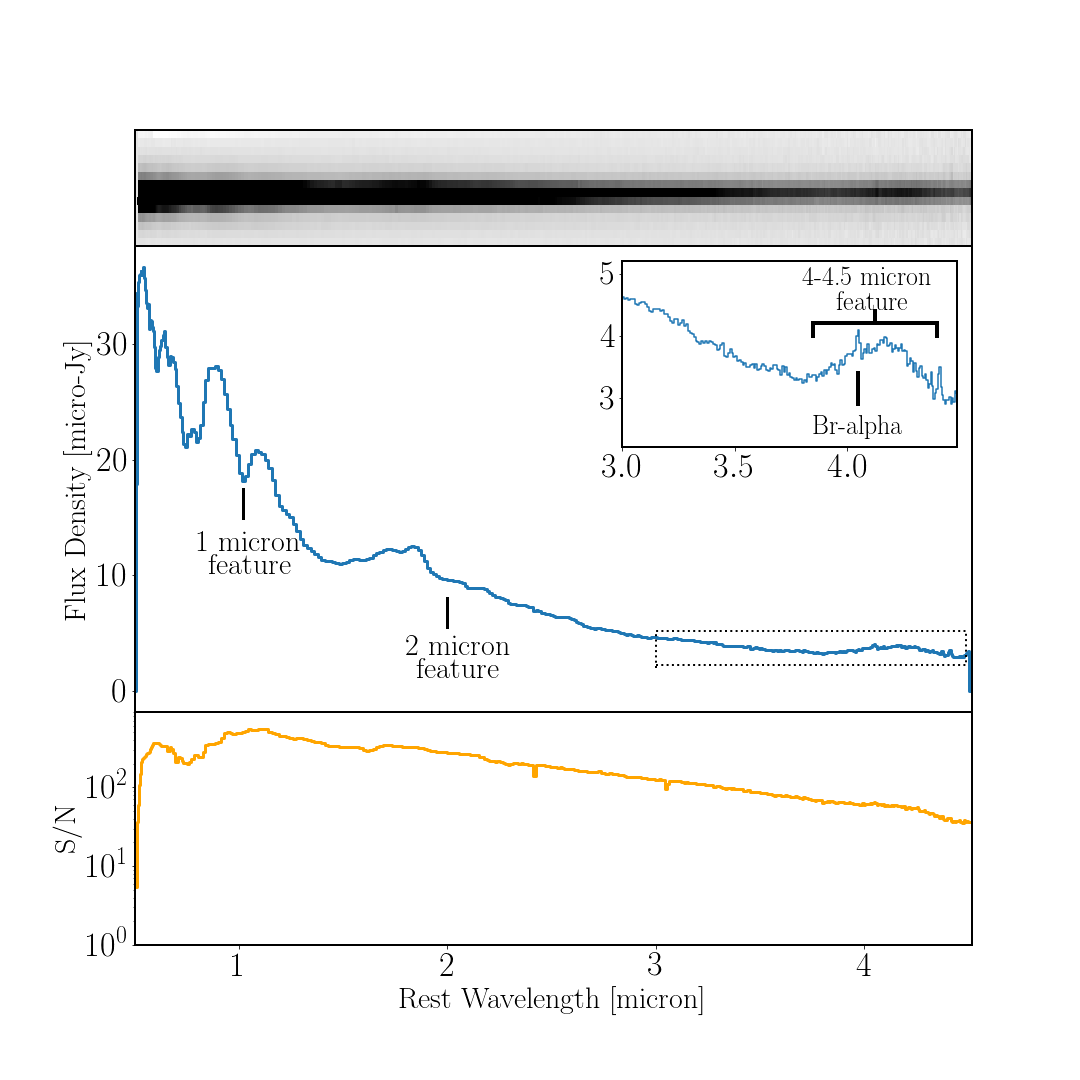}
\caption{The JWST spectrum of SN\,2025kg near maximum light. In the top and bottom panels we show the 1D spectrum and the signal-to-noise ratio (S/N), respectively. In the top panel, we highlight the three features discussed in detail in Section~\ref{subsec:helium} (the 1 and 2~$\mu$m features) and Section~\ref{subsec:jwst45} (the 4-4.5 $\mu$m feature, shown in the inset panel).}
\label{fig:jwst}
\end{figure*}

Turning to the JWST spectrum, we observe a prominent, broadened absorption feature $\sim$ 1 $\mu$m and, potentially, a shallower absorption feature $\sim$ 2 $\mu$m (Figure~\ref{fig:jwst}). To investigate the source and significance of these features, we utilize a Bayesian toolkit to jointly model the 1 and 2 $\mu$m features as blended absorption features from multiple species with Gaussian components \citep{Liu+23}\footnote{\url{https://github.com/slowdivePTG/BayeSpecFit}}. We give the priors for our fits in Appendix Section~\ref{sec:priors_appendix}. We note that the errors on velocity found through this method do not encompass uncertainties in the fit to the continuum. Thus, the true uncertainties are likely larger than those reported here. 

First, we test the significance of the possible feature at $\sim$ 2 $\mu$m as, if real, the feature would provide a second observational constraint on the presence of He I. We apply two models and compare their $\chi^2$ and their Bayesian Information Criterion (BIC). The first model fits the 2 $\mu$m region with a linear function, a reasonable approximation for the continuum in the limited wavelength regime, while the second applies a Gaussian to represent an absorption feature. We find that the Gaussian model results in an improvement in $\chi^2$ and BIC $\sim 3500$ compared to the linear fit, suggesting that the 2 $\mu$m feature is real.

\begin{figure}
\centering
\includegraphics[angle=0,width=.45\textwidth]{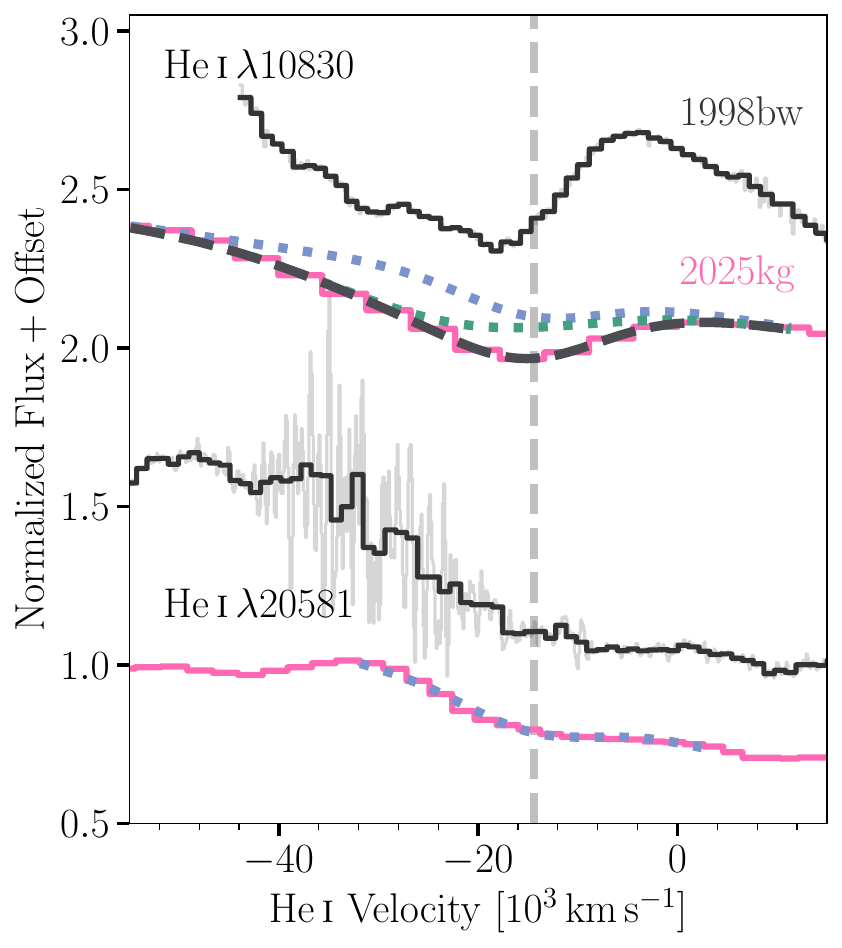}
\caption{The JWST/NIRSpec spectrum of SN\,2025kg (pink) and the near-IR spectrum of SN\,1998bw (black; binned to mitigate telluric noise in the spectrum) in velocity space $\sim$ the 1 $\mu$m (upper) and 2 $\mu$m (lower) features. We show our combined fit (grey dashed) and individual contributions to the 1 $\mu$m feature due to C I ($\lambda 1.0693, \lambda 2.1259$ $\mu$m; dotted turquoise) and He I ($\lambda 10830, \lambda 20587$; dotted blue) absorption. We also mark the velocity derived from a single Gaussian fit to the 1 $\mu$m feature with a vertical grey line. In the top spectrum, the superposition of the C I and He I absorption provides a strong fit to the observed feature. Our analysis (Section~\ref{subsec:helium}) shows that a small ($\lesssim 0.5 M_{\odot}$) amount of He is likely present in SN\,2025kg.
}
\label{fig:hei}
\end{figure}

We next evaluate fits to the 1 and 2 $\mu$ features for He I, C I and/or Mg II. The larger width of the 1 $\mu$m feature relative to the 2 $\mu$m feature (Figure~\ref{fig:hei}) suggests a larger dispersion velocity. In turn, this implies that the 1 $\mu$m feature is not due to one element alone and is instead likely a blend of He I, C I and/or Mg II. Thus, we first fit both features for He I $\lambda 1.0830, \lambda 2.0581$ $\mu$m and C I $\lambda 1.0693, \lambda 2.1259$ $\mu$m (Figure~\ref{fig:hei}). This fit finds He dominates the 2 $\mu$m feature and is blended with C I in the 1 $\mu$m feature. Both the C I (17,600 $\pm$ 200 km s$^{-1}$) and He I (14,500 $\pm$ 100 km s$^{-1}$) expansion velocities are similar to the Si velocities at a similar epoch (Figure~\ref{fig:velocities}). A second fit replacing C I with Mg II results in a similar He I velocity and a higher Mg II velocity (21,700 $\pm$ 100 km s$^{-1}$). Finally, we explore if the 1 and 2 $\mu$m features can be explained with only C I and Mg II or a combination of the two (i.e., no He I). We find expansion velocities in the range of 24,000 - 41,000 km s$^{-1}$, $\approx$10,000 km s$^{-1}$ faster than the Si expansion velocities at a similar epoch (Figure~\ref{fig:velocities}). This finding implies that the 1 and 2 $\mu$m features are most reasonably fit with the presence of He I.

The most consistent expansion velocities with optical lines (Section~\ref{subsec:velocities}) are derived from fits to these regions that include some He I. We also note that the best-fit pseudo-equivalent width ratio EW(He I $\lambda 1.0830\,\mu$m)/EW(He I $\lambda 2.0581\,\mu$m) from our first fit is $\approx 1$, in line with predictions for these components \citep{Lucy91}, also consistent with the presence of He. Blending of He I and another feature is supported by a higher-resolution near-IR spectrum of SN\,1998bw spectrum showing two ``sub-features'' within the 1 $\mu$m feature (Figure~\ref{fig:hei}). 

We conclude that there is likely a small amount of He in the ejecta of SN\,2025kg, though a precise estimate is outside the scope of the current work. 
Based on the absence of He I lines in the optical (Figures~\ref{fig:spectra_series} and \ref{fig:spec_comp}) we place a conservative upper limit on He in the ejecta of $M_{\rm He} \lesssim 0.5 M_{\odot}$ \citep[see, e.g., models for SN\,Ib/c optical spectra in][which are independent of engine type]{Dessart+20}. The extent of He mixing in the ejecta may significantly impact the observed signature (e.g., \citealt{Dessart+20,Dessart+24}).

In Figure~\ref{fig:hei} we show the SN\,2025kg JWST spectrum and a near-IR SN\,1998bw spectrum (8~days post-peak; \citealt{Patat+01}) in velocity space relative to the He I and Mg II features. Both spectra show evidence for absorption around the 1 and 2 $\mu$m features, with the 1 $\mu$m feature being most prominent. Near-IR spectra of SN\,1998bw were taken at 8, 33 and 51~days post-peak, and each shows a prominent feature near 1 $\mu$m ascribed to He I \citep{Patat+01} which was not observed in a near-IR spectrum of SN\,2010bh \citep{Chornock+10}. Further, in the SN\,1998bw spectrum at +51 days post-peak \citet{Patat+01} also claim the detection of He I $\lambda 2.0581$ $\mu$m and state this line is weakly detected in the spectra at +8 and +33~days post-peak \citep{Patat+01}. We observe a similar shape $\sim$ 2 $\mu$m in the spectra of SN\,2025kg and SN\,1998bw around peak (Figure~\ref{fig:hei}). Altogether, this comparison demonstrates that small amounts of He are likely common in GRB-SNe and FXT SNe, but are not typically observed due to a paucity of near-IR spectroscopic coverage.

\subsection{4-4.5 $\mu$m Feature}
\label{subsec:jwst45}

We next turn to the broad emission feature centered at a rest-frame wavelength of $\sim 4.2\mu$m in JWST/NIRSpec spectrum taken around SN\,2025kg's peak (Figure~\ref{fig:jwst}). Here, we discuss several potential explanations for this feature, but ultimately determine that a larger sample of observed events is critical to establishing its origin.

First, we investigate the possibility that this feature is a signature of $r$-process nucleosynthesis. Infrared excesses have been suggested as possible evidence of $r$-process, which may occur in the accretion disk after the collapse of massive, rapidly-rotating stars, that are thought to be the progenitors of GRBs and some FXTs (e.g., \citealt{SiegelBarnesMetzger2019,Zenati+20}). To test this possibility, we explore if the $4.2$\,$\mu$m feature is consistent with a thermal signature. We find that this condition is not satisfied, as the feature is too narrow to be well fit with a Planck function. Further, the peak wavelength implies a temperature of $\sim700$\,K, from which we infer a radius of $\sim2 \times 10^{16}$\,cm to match the luminosity. This would require an average expansion velocity of $0.9$c over the first two rest-frame weeks, which is incompatible with both observations (Section~\ref{subsec:velocities}) and theoretical expectations.

Second, we consider the possibility that this feature is due to dust.
To investigate this, we consider comparable spectroscopic observations of SNe at mid-IR wavelengths to investigate their timescales for dust emergence, which are thus far limited in the literature. JWST NIRSpec and MIRI spectroscopic observations of the Type IIP SN\,2022acko show no evidence for dust at 50~days following the explosion \citep{Shahbandeh+24}. In SN\,1987A, near-IR spectra cover the range $1.05$\,$\mu$m -- $4.1$\,$\mu$m \citep{Meikle89}. The 4\,$\mu$m region is dominated by a narrow emission feature that is attributed to Brackett $\alpha$ ($\lambda_{\rm peak} = 4.053$\,$\mu$m). H features are not expected in a Ic-BL SN like SN 2025kg, although we note a narrow excess at this wavelength in our spectrum (Figure~\ref{fig:jwst}), which may be emission from the underlying host galaxy. In SN 1987A, at around 260 days post-peak a comparable broad underlying feature emerged and was tentatively suggested to be due to an overtone of silicon monoxide (SiO) \citep{Meikle89}. Both of these epochs are significantly later than our JWST spectrum of SN\,2025kg, which was taken at $\approx 15$ rest-frame days after the FXT trigger. Additionally, at this point, the ejecta temperature is likely too high for dust molecules to form, supporting that this feature is not SiO. Further, spectral synthesis modeling indicates that the SiO overtone emission emerges at $\gtrsim100$~days following core-collapse in Type Ic-BL SNe \citep{Liljegren23}.

Finally, the feature may be the result of nebular Fe lines. This interpretation would require regions of the ejecta to become optically thin, which may be possible with high expansion velocities. Expansion velocities of $\sim0.1$c were inferred at early time in \citet{EylesFerris25}, which may be sufficient to produce this signature. Looking forward, the assembly of a sample of mid-IR spectra of stripped envelope SNe or other FXTs will be important to understand how ubiquitous this feature is, and shed light on its physical origin.

\subsection{Comparison of SN and High-Energy Properties}
\label{sec:sn_xraycomp}

\begin{figure}
\centering
\includegraphics[angle=0,width=.48\textwidth]{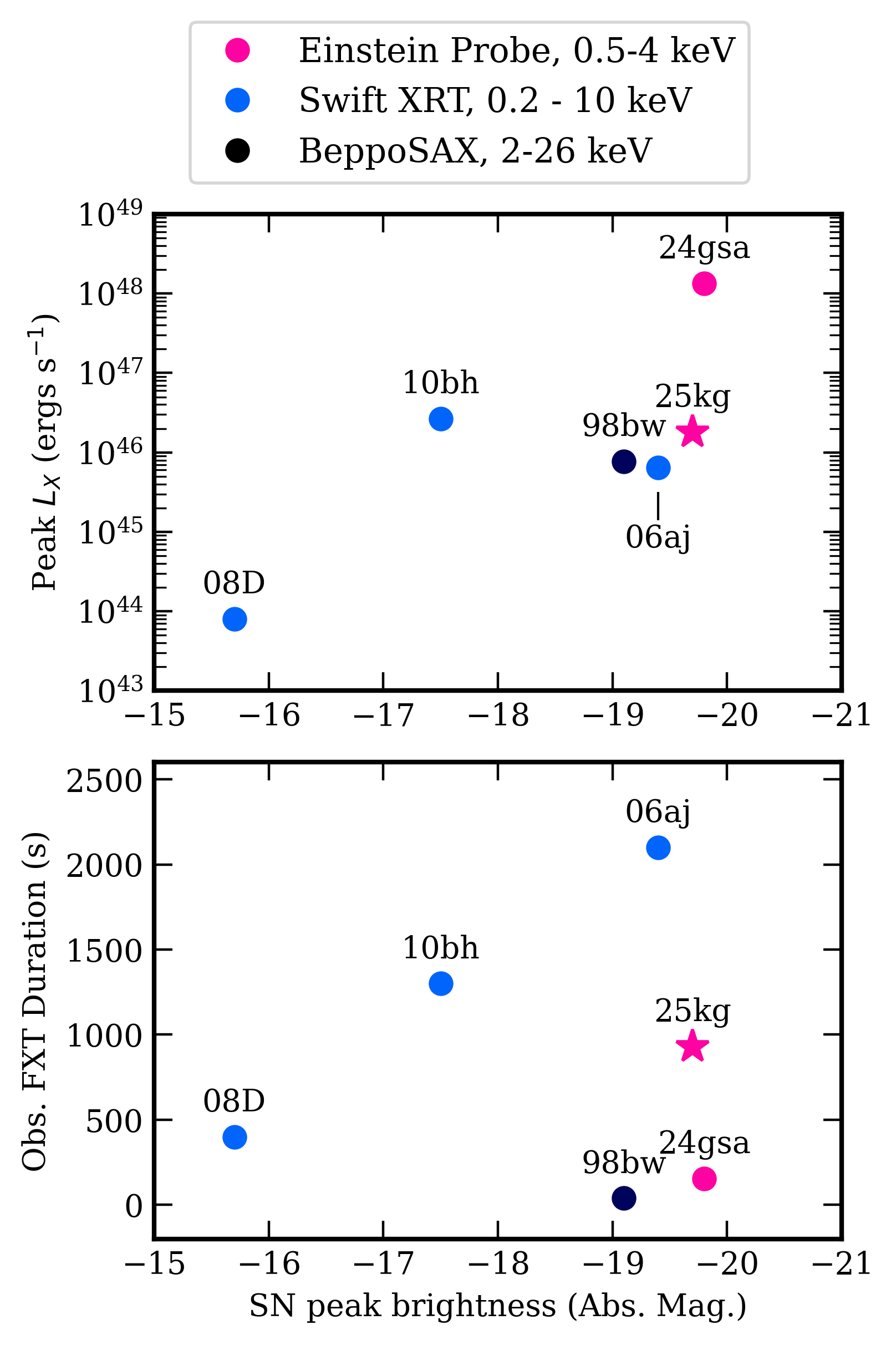}
\caption{Comparison of the SN peak time and FXT properties of several FXT events with SN counterparts (SNe 1998bw, 2006aj, 2008D, 2010bh, 2024gsa, 2025kg). EP\,250108a/SN\,2025kg is marked with a pink star in the first two rows. We caution that the different instruments' bandpasses and slewing strategies impact both the X-ray peak luminosities and durations, making direct comparisons inadvisable. However, the span of several orders of magnitude in some parameters mitigate some of these biases.}
\label{fig:xray_sn_space}
\end{figure}

In our companion paper, we show that EP\,250108a/SN\,2025kg's prompt FXT, early ($\delta t \lesssim 6$~days) optical light curve, and radio upper limits are consistent with a trapped or weak jet from a collapsar \citep{EylesFerris25}. Motivated to explore if this event is an typical FXT SN, we consider any potential correlations between the high-energy and SN properties of FXTs (Figure~\ref{fig:xray_sn_space}).  

In addition to EP\,250108a/SN\,2025kg, we include the events SNe\,1998bw, 2006aj, 2008D, 2010bh, and 2024gsa in this comparison. We utilize the prompt X-ray (2-26 keV) emission of SN\,1998bw detected by BeppoSAX \citep{Galama+98,Pian+00}, X-ray detections by \textit{Swift}-XRT of SN\,2006aj, SN\,2008D, and SN\,2010bh pulled from UKSSDC \citep{Evans+07,Evans+09}, and the prompt X-ray emission of SN\,2024gsa and SN\,2025kg observed by {\it EP} \citep{Sun+24,Li+25_ep250108a}. We caution that these properties are drawn from a variety of instruments, each with different energy ranges and slewing strategies. However, we note that these comparisons span several orders of magnitude, mitigating some biases. We mark the different instruments in Figure~\ref{fig:xray_sn_space}. For SN properties, we incorporate SN peak magnitudes derived from the light curves presented in Sections~\ref{sec:lcs} and Si $\lambda 6355$ velocities presented in Section~\ref{subsec:velocities}.

Figure~\ref{fig:xray_sn_space} highlights that SN\,2025kg, SN\,1998bw and SN\,2006aj occupy a similar region in the parameter space of peak SN and peak X-ray luminosity (see also Figure 13 of \citet{EylesFerris25}). We observe a potential trend in these properties for the objects outside of this cluster, though a significantly larger sample is needed to test this trend further. In terms of duration, the \textit{Swift} events show a potential trend between SN peak brightness and observed FXT duration. We do not observe any correlations between the remaining events.

\section{Radioactive Decay SN Modeling}
\label{sec:mod}

With this extensive data set, we are motivated to constrain the physical properties of SN\,2025kg, which enable comparisons to previous SN and possible properties of the progenitor. Thus, we fit the light curve for SN\,2025kg using {\sc Redback}~\citep{Sarin+24}. We fit the light curve using two models: first, a one-zone ``Arnett'' model~\citep{Arnett82} and second, a model that accounts for mixing of ${}^{56}$Ni into the outer ejecta. 

For both models, we fit the entire light curve at $\delta t \gtrsim 6.5$~days  assuming a Gaussian likelihood. For each photometric point, we include a systematic error of 0.15 mag added in quadrature to the statistical errors to capture any discrepancies caused by differences in photometric reduction, filter transmission curves, or in the model itself. We explore the model parameter space using Bayesian Inference via the {\sc Pymultinest} nested sampler~\citep{pymultinest} through the {\sc Bilby} library~\citep{ashton19}.

\subsection{One-zone Model}
\label{subsec:arnett}

We begin with a one-zone ``Arnett'' model that assumes a one-zone, homologously-expanding ejecta powered by a centrally-located heating term (i.e., the ${}^{56}$Ni) with a gray ejecta opacity that is constant in time. This model is commonly invoked to describe the physical properties of SNe. We consider $7$ model parameters for the one-zone model here: ejecta mass, 
$M_{\rm ej}$; fraction of ${}^{56}$Ni in the ejecta, $f_{\rm Ni}$; ejecta velocity, $v_{\rm ej}$; gray opacity, $\kappa$; $\gamma$-ray opacity, $\kappa_{\gamma}$; temperature at which the photosphere begins to recede, $T_{\rm floor}$; and host extinction in magnitudes, $A_{\rm v}$. For the final parameter, we assume a standard Fitzpatrick extinction law with an $R_{v}=3.1$~\citep{Fitzpatrick1999}. We use broad uninformative priors on each parameter. 

Our model provides a strong fit to the data. Our fits with this model are shown in Figure~\ref{fig:lc} highlighting broad agreement with the data. We infer $1.4 \pm 0.15 M_{\odot}$ of ejecta with a ${}^{56}$Ni mass of $0.6 \pm 0.1 M_{\odot}$ and an initial ejecta velocity of $19,000 \pm 700$~km s$^{-1}$. All uncertainties at $68\%$ credible interval for the one-dimensional marginalized posterior. Our estimate of ejecta mass and energy is consistent with median values inferred from Type Ic-BL SNe~(e.g., \citealt{Taddia2019,Rodriguez+23,Srinivasaragavan+24}) using a similar model. However, our estimate of the ${}^{56}$Ni is marginally inconsistent with the Type Ic-BL SNe medial values of \citet{Taddia2019} (at $1\sigma$), which could be due to differences in assumptions about $\gamma$-ray opacities or hinting towards a limitation with the one-zone model. We note that our analysis here is on the photometry, while past works have directly fit a reconstructed bolometric luminosity which could also explain the discrepancy.

\subsection{Caveats to One-zone Model}
\label{subsec:onezonecaveats}

There are clear limitations with one-zone model in its assumption of a centrally-located ${}^{56}$Ni and constant opacity.  In reality, the opacity is almost certainly not gray or constant with time~\citep[e.g.,][]{Niblett2025} and ${}^{56}$Ni is likely mixed out to large radii due to Rayleigh-Taylor or jet-driven instabilities~\citep[e.g.,][]{Khatami2019, Reichert2023}. The latter effect significantly impacts the overall bolometric light curve. In particular, the classical ``Arnett-like'' one-zone model often over-predicting the quantity of ${}^{56}$Ni compared to models with mixing~\citep[e.g.,][]{Bersten2014} for two reasons. First, the central heating term is forced to overcome adiabatic losses compared to heating from ejecta at larger radii. Second, energy through heating has to diffuse through more ejecta. Therefore, our ${}^{56}$Ni mass measurement above is likely an overestimate. Additional luminosity in the form of a central engine could also provide an alternative path to powering the light curve which would naturally reduce the quantity of ${}^{56}$Ni.

In Appendix Section~\ref{subsec:mixing} we present an alternate model in which ${}^{56}$Ni is mixed with the outer SN layers, resulting in a lower best-fit value for the ${}^{56}$Ni mass (0.2 $M_{\odot}$). Though this model offers several advantages compared to the one described in Section~\ref{sec:mod}, there are still uncertainties, such as the assumed mass profile, therefore, we conservatively use the ${}^{56}$Ni estimate from both models. These masses and mixing provide some strong hints into the progenitor, as we discuss in Section~\ref{sec:discussion}.

Finally, we have assumed that the peak emission is powered entirely by ${}^{56}$Ni decay.  Shock heating can contribute to the peak emission, reducing the requirements on the ${}^{56}$Ni mass~\citep{Niblett2025}.

\section{Discussion}
\label{sec:discussion}

Our analysis of the light curves and spectra of SN\,2025kg place strong constraints on the nature of the progenitor. In our companion paper \citep{EylesFerris25} we argue that the engine behind this transient is jet-driven, supported by shock break-out model fits to the prompt X-ray emission, upper limits on radio emission and the blue and rapidly-fading light curve prior to the onset of the SN. In this section, we combine the inferences of both papers to discuss constraints on the progenitor and implications for the population of FXTs from collapsars.

\subsection{Constraints on the ZAMS Mass of the Progenitor}

Here, we determine the ZAMS mass of the progenitor by combining the inferred masses on the ejecta yield, the $^{56}$Ni yield (which dictates the disk mass) and the compact remnant mass. We note that the $^{56}$Ni mass may be overestimated due to luminosity contribution from additional sources, but do not anticipate this significantly impacts our order-of-magnitude estimates. To do this, we set the mass of the star at explosion to the C/O core mass as SN\,2025kg shows evidence for stripping of its H and most of its He layers; Section~\ref{sec:spec}. We assume that the C/O core mass will be equivalent to the summation of the ejecta mass, the disk mass and the compact object remnant mass. For the first component, we use the results from our light curve fits, which indicate a total ejecta mass of $0.8-1.4 \, {\rm M_\odot}$ with a $^{56}$Ni mass of $0.2-0.6 \, {\rm M_\odot}$ in SN\,2025kg (Section~\ref{sec:mod}). We explore two scenarios to determine the latter two components: a black hole accretion disk jet model and a model invoking a neutron star accretion disk.  If the remnant is a black hole, we assume the accretion disk forms around a $3-5\,M_\odot$ black hole.  If it is a neutron star, we assume an initial remnant mass of $1.4 \,M_\odot$.

The large amount of $^{56}$Ni ejecta mass required to explain the SN light curve peak (Section~\ref{sec:mod}) indicates that a considerable mass must be processed through an accretion disk. In a jet-driven explosion, the $^{56}$Ni is primarily produced in the high-temperature disk and ejected through a disk wind~\citep{2006ApJ...643.1057S}.  Typically models predict that $\sim 20$\% of the disk is lost through a wind~\citep[e.g][]{2017MNRAS.471.4806A, 2017PhRvL.119w1102S}.  This means that the mass that flows into the disk must be greater than $5$ times that of the $^{56}$Ni ejecta mass.  To eject $0.2-0.6 \, {\rm M_\odot}$ of $^{56}$Ni, roughly $1-3 \, {\rm M_\odot}$ would have to be processed through a disk.  Combined with the non-nickel ejecta mass and the compact remnant mass, this corresponds to a C/O core of $\sim 3.0-5.2 \, {\rm M_\odot}$ for a neutron star accretion disk model or $\sim 4.6-8.8 \, {\rm M_\odot}$ for a black hole accretion disk model.  Using the C/O star masses from low-metallicity stars simulated by \citep{2002RvMP...74.1015W}, the C/O core masses correspond to a ZAMS progenitor masses of $15-21 \, {\rm M_\odot}$ if the compact remnant is a neutron star and $19-30 \, {\rm M_\odot}$ if the compact remnant is a black hole.  These mass ranges are consistent with our current expectations for the progenitors of neutron stars and black holes respectively~\citep{1999ApJ...522..413F}. The uncertainty in the $^{56}$Ni mass dominates the wide range of our predicted progenitor masses. A late-time observation of the light curve will help constrain the $^{56}$Ni mass and place stronger constraints on the progenitor's ZAMS mass. Finally, we note that our above estimates are based on single star models with a fixed set of parameters for stellar evolution (e.g. mixing; \citealt{2002RvMP...74.1015W}). Models that include the effects of binary mass transfer and ejection are likely to alter or broaden the predicted range of progenitors. However, these results provide a first pass of the stellar conditions.

\subsection{Constraints on a He Shell}

In our companion paper \citep{EylesFerris25} we find that the early optical light curve of the counterpart of EP\,250108a ($\delta t \lesssim 6$~days) is marginally consistent with interaction of a $0.2-0.9 \, {\rm M_\odot}$ shell at a distance of $7 \times 10^{14} {\rm \, cm}$. While we note that the preferred model to explain the early light curve is a shocked cocoon from a jet \citep{EylesFerris25}, here we briefly consider the implications of this alternate possibility for a shell around the progenitor of EP\,250108a/SN\,2025kg from our observations. This distance is consistent with being produced by a common envelope mass ejection of a He shell, and just below the lower radii predicted for He shell mass ejection from a common envelope if it occurs at C ignition~\citep{2024arXiv241010378F}. This distance is also consistent with a He merger event~\citep{1998ApJ...502L...9F,2025Ap&SS.370...11G}. 

Using the He shell models of \cite{2002RvMP...74.1015W} models, we predict a large amount of He in this shell ($>1\,M_{\odot}$). This He shell will be swept up into the ejecta and behind the photosphere at the time of our optical spectra. Notably, this predicted shell mass is inconsistent with both the upper limit on the He mass of $<0.5 \, {\rm M_\odot}$ from the optical spectra (\citealt{Dessart+20}; Section~\ref{subsec:helium}) and the mass of the shell derived in our companion paper \citep{EylesFerris25}. This He shell mass could be reduced if significant Wolf-Rayet mass loss occurs prior to the final mass ejection episode that produces the shell at $7 \times 10^{14} {\rm \, cm}$.  Alternatively, enhanced mixing~\citep[e.g.][]{2013ApJ...773L...7F} could reduce the size of the He shell. In conclusion, for our models to fit the data presented in \citet{EylesFerris25} and this work, either considerable wind mass-loss or stellar mixing is required.

\subsection{Implications of Hydrogen Observed at $\delta t = 42.5$~days}
\label{sec:discuss_halpha}

In Section~\ref{sec:spec} we show that broadened H$\alpha$ is observed at $\delta t = 42.5$~days, or when the shock reaches $\sim 10^{16} {\rm \, cm}$. Similarly, the Type I superluminous SN iPTF13ehe showed a similar late-time H$\alpha$ feature initially observed in spectra at +251~days from peak light, which was interpreted as coming from a detached shell of H around 10$^{16}$~cm from the progenitor star \citep{Yan+17}. 

Here, we briefly consider several potential explanations for this feature. We find it unlikely that the H is due to the progenitor star's wind, as models robustly predict that a Wolf-Rayet wind would evacuate the H in a region with radial extent of  0.1-1\,pc~\citep{1977ApJ...218..377W}. 

One potential explanation could be that the H$\alpha$ signature could be produced through interaction with a companion star. However, such a wide separation would argue that this companion star had no impact on the stellar evolution of the collapsing star.  To ensure that the supernova blastwave sweeps up enough H to produce the H$\alpha$ line, this companion would have to be evolved, either in a giant or supergiant phase~\citep{2018ApJ...864..119H,2020MNRAS.499.1154H,2023MNRAS.523.6011H}. Since we are likely to require a close binary to explain the rotation and loss of the He shell, this would then imply that the progenitor of SN\,2025kg is in a triple system. However, we disfavor the triple companion explanation due the giant or supergiant phase requirement and the temporal likelihood of being in this phase. Furthermore, based on hydrodynamical simulations, material stripped from a companion by the supernova ejecta will have velocities of order the escape velocity from the companion, too low to match the observed H broadening.

Alternatively, the H$\alpha$ signature could be due to clumps of H-rich material surrounding the SN produced by an asymmetric common envelope mass ejection phase (e.g., \citealt{Quataert+16,Fryer+20,Fryer+23}). These clumps could be sufficiently dense that they could persist despite the strong Wolf-Rayet wind (e.g., \citealt{Owocki+84,Puls+08}). Post-SN binary interactions that cause ongoing energy injection through feedback from H-rich material accretion or through magnetar winds ablating an H-rich companion (E.g., \citealt{Zhu+24}) could produce a similar dense cloud. We favor this explanation for the H$\alpha$ signature observed at $\delta t = 42.5$~days.

\subsection{A Growing Population of FXT-SNe \& Connections to GRB-SNe and Ic-BL SNe}

Early works on XRF/FXTs with SNe found that the volumetric rates of such events would be at least an order of magnitude higher than those of collapsar GRBs \citep{Soderberg+08}. EP\,250108a/SN\,2025kg joins a population of three FXT-SNe (EP\,240414a/SN\,2024gsa, EP\,250304a; \citealt{vanDalen+24,Sun+24,2025GCN.39851}) discovered in just the first year of {\it EP} operations. Of those with published SNe light curves and spectra (EP\,240414a/SN\,2024gsa and now EP\,250108a/SN\,2025kg) these events are comparable in their peak optical luminosities, optical spectra, and Si $\lambda$6355 \AA\ absorption velocities (Sections~\ref{sec:lcs}-\ref{subsec:velocities}) to GRB-SNe and the XRF/FXT\,060218/SN\,2006aj. Further, in Section~\ref{sec:sn_xraycomp}, we demonstrated that the peak X-ray luminosities of GRB\,980425, XRF/FXT\,060218, and EP\,250108a are also comparable. Taken together, these support a causal link between the progenitors and mechanisms driving GRBs and FXTs with SNe. 

The three recent {\it EP} FXT detections and the rates of GRB-SNe discovered over the last 20 years ($\lesssim$1 per year; e.g., \citealt{Dainotti+22}) support earlier calculations \citep{Soderberg+08} that FXT are significantly more common than GRB-SNe. While a robust calculation of the relative rates of FXT SNe and GRB-SNe is outside the scope of this work, we note that the three FXT SNe discovered in the first year of {\it EP} operations were at $z=0.40, 0.17$ and $0.20$ and detected by an instrument covering $\sim 9\%$ of the sky. From this, we derive an estimate of the intrinsic volumetric rates of at least several tens Gpc$^{-3}$ yr$^{-1}$, 1-2 orders of magnitude larger than the rate of on-axis luminous GRBs \citep{Sun15}. 

In our companion paper, \citet{EylesFerris25} we demonstrate that EP\,250108a/SN\,2025kg is consistent with being driven by a collapsar-powered jet that either fails to break out of the dense circumstellar material or has an energy weaker than $\sim 10^{50}-10^{51}$~ergs. Taken together with the rates, this indicates that trapped or weak jets are more common than the successful jets that produce GRBs. The lack of radio and X-ray detections at later times following EP\,250108a/SN\,2025kg \citep{EylesFerris25} implies that observing any relativistic material may be challenging from these FXT-producing jets, in keeping with late-time radio surveys of Ic-BL SNe observed without high-energy counterparts (e.g., \citealt{Corsi+16}). Zooming out, it is likely that FXT-producing, ``trapped'' jets accompany a substantially larger fraction of Ic-BL SNe than those with successful jets and GRBs, representing at least a few percent of the Ic-BL SNe population.

Finally, we note that the SNe of two past FXT events, SN\,2008D and SN\,2010bh, are $\sim$5-100 times less luminous than FXT SNe 2006aj, 2024gsa, and 2025kg, and GRB-SNe and, in the former's case, show prominent He absorption (Figure~\ref{fig:spec_comp}; \citealt{Soderberg+08,Modjaz+09}). The lower optical luminosities of these SNe are echoed in the lower peak X-ray luminosities of their FXTs (Figure~\ref{fig:xray_sn_space}). Future {\it EP} detections of FXT SNe will reveal both the rates of such events and the full span in stellar progenitors of FXT SNe.

\section{Conclusion}

Along with our companion paper \citep{EylesFerris25}, we have presented the most detailed dataset to date of an SN accompanying an {\it EP} FXT. Our main conclusions are as follows:

\begin{itemize}
    \item Optical spectra of SN\,2025kg are characterized by broad absorption features due to Fe II $\lambda 5169$ and Si $\lambda 6355$ and do not show obvious emission due to He I, leading to a Type Ic-BL classification. 
    \item SN\,2025kg is a close analog of GRB-SNe, particularly SN\,1998bw, and the FXT SNe, SN\,2006aj and SN\,2024gsa, in terms of its peak luminosity, expansion velocities, and spectral evolution. We also find a strong match between SN\,2025kg and the Ic-BL SN\,2020bvc, in keeping with comparisons to the early light curve \citep{EylesFerris25}. 
    \item We observe absorption features $\sim$1 and 2 $\mu$m features in the JWST spectrum taken around maximum light, and find that they are well-fit with a combination of He I and C I. From optical spectra, we conservatively conclude that the mass of He in the ejecta is $\lesssim 0.5 M_{\odot}$. 
    \item We observe a broad feature at 4-4.5 $\mu$m in the JWST spectrum. We investigate several explanations for this feature, including thermal emission due to $r$-process nucleosynthesis, nebular Fe lines, and He emission, but cannot definitively identify its origin with present information. 
    \item We fit SN\,2025kg with a one-zone radioactive decay model and a model that accounts for mixing of $^{56}$Ni into the outer SN ejecta. We derive a range in the  $^{56}$Ni mass of $0.2 - 0.6~M_{\odot}$.
    \item Together, observations presented in our companion paper \citep{EylesFerris25} and this work favor a ZAMS progenitor mass for EP\,250108a/SN\,2025kg of $15-21 \, {\rm M_\odot}$ or $21-30 \, {\rm M_\odot}$ if the compact remnant is a neutron star or black hole, respectively. It is plausible that the progenitor was surrounded by a He shell ejected during a common envelope episode, and either H-rich clumps due to an prior, asymmetric common envelope ejection or a giant/supergiant tertiary stellar companion.
    \item Our analysis supports a causal link between GRB-SNe and FXT SNe, in which GRBs are produced by successful jets and FXT SNe are produced by trapped or weak jets. Thus far, the rate of SN detections following {\it EP} FXTs indicates that trapped or weak jets are substantially more common than successful (GRB) jets.
\end{itemize}

EP\,250108a is only the second {\it EP} transient with an observed SN counterpart to date. Our detailed observational study of SN\,2025kg has revealed both similarities with previous FXT and GRB SNe, and unexpected signals, including H at 42.5~days and the broad 4-4.5$\mu$m feature, that cannot be fully accounted for with current models. Looking forward, the opportunities for detailed observational studies to shed light on the stellar progenitors of FXTs and their catastrophic explosions is only increasing with the continued operations of {\it EP} and coordinated rapid follow-up efforts.

\section*{Acknowledgements}

We dedicate this work to the memory of Alicia M. Soderberg, a pioneer in the study of XRFs and FXTs. 

We are deeply grateful to Tom Marsh for developing the {\sc molly} software, one of his many contributions to advancing the field of compact objects. We thank Jennifer Andrews, the T-80 South technical team, and others for their support during observations.

J.C.R. acknowledges support from the Northwestern Presidential Fellowship. 
P.G.J. J.N.D.D., J.S.S., J.Q.V., and A.P.C.H.~are supported by the European Union (ERC, Starstruck, 101095973). Views and opinions expressed are however those of the author(s) only and do not necessarily reflect those of the European Union or the European Research Council Executive Agency. Neither the European Union nor the granting authority can be held responsible for them. 
N.S. acknowledges support from the Knut and Alice Wallenberg Foundation through the ``Gravity Meets Light" project and by and by the research environment grant ``Gravitational Radiation and Electromagnetic Astrophysical Transients'' (GREAT) funded by the Swedish Research Council (VR) under Dnr 2016-06012. 
B.P.G acknowledges support from STFC grant No.~ST/Y002253/1 and The Leverhulme Trust grant No.~RPG-2024-117
C.D.K. gratefully acknowledges support from the NSF through AST-2432037, the HST Guest Observer Program through HST-SNAP-17070 and HST-GO-17706, and from JWST Archival Research through JWST-AR-6241 and JWST-AR-5441. 
W.F. gratefully acknowledges support by the David and Lucile Packard Foundation, the Alfred P. Sloan Foundation, and the Research Corporation for Science Advancement through Cottrell Scholar Award 28284. 
P.O.'B. acknowledges support from the UKRI grant: ST/W000857/1. 
The work by C.L.F. was supported by the US Department of Energy through the Los Alamos National Laboratory. Los Alamos National Laboratory is operated by Triad National Security, LLC, for the National Nuclear Security Administration of U.S.\ Department of Energy (Contract No.\ 89233218CNA000001). 
W.J-G. is supported by NASA through the NASA Hubble Fellowship grant HSTHF2-51558.001-A awarded by the Space Telescope Science Institute, which is operated by the Association of Universities for Research in Astronomy, Inc., for NASA, under contract NAS5-26555. 
L.M.R acknowledges support from NSF grants AST-1911140, AST-1910471, and AST-2206490. 
S.Y. acknowledges the funding from the National Natural Science Foundation of China under grant No.12303046 and from the Startup Research Fund of Henan Academy of Sciences No.241841217.  
A.A.C. acknowledges support through the European Space Agency (ESA) research fellowship programme.
C.R.B. acknowledges the financial support from CNPq (316072/2021-4) and from FAPERJ (grants 201.456/2022 and 210.330/2022) and the FINEP contract 01.22.0505.00 (ref. 1891/22).  
D.M.S. and M.A.P.T. acknowledge support by the Spanish Ministry of Science via the Plan de Generacion de conocimiento PID2020-120323GB-I00. DMS also acknowledges support via a Ramon y Cajal Fellowship RYC2023-044941.
M.N. is supported by the European Research Council (ERC) under the European Union’s Horizon 2020 research and innovation programme (grant agreement No.~948381)..
R.G. was sponsored by the National Aeronautics and Space Administration (NASA) through a contract with ORAU. The views and conclusions contained in this document are those of the authors and should not be interpreted as representing the official policies, either expressed or implied, of the National Aeronautics and Space Administration (NASA) or the U.S. Government. The U.S. Government is authorized to reproduce and distribute reprints for Government purposes notwithstanding any copyright notation herein. 
S.J.S., J.G., S.S., and K.S. 
acknowledge funding from STFC Grant ST/Y001605/1, a Royal Society Research Professorship and the Hintze Charitable Foundation. 
C.J.N. acknowledges support from the Science and Technology Facilities Council (grant No. ST/Y000544/1) and from the Leverhulme Trust (grant No. RPG-2021-380).
A.A. acknowledges the Yushan Young Fellow Program by the Ministry of Education, Taiwan for the financial support (MOE-111-YSFMS-0008-001-P1).
T.-W.C. acknowledges the Yushan Fellow Program by the Ministry of Education, Taiwan for the financial support (MOE-111-YSFMS-0008-001-P1).
C.L. is supported by DoE award \#DE-SC0025599.
A.R.E. is supported by the European Space Agency (ESA) Research Fellowship.

Observations reported here were obtained at the MMT Observatory, a joint facility of the University of Arizona and the Smithsonian Institution. MMT Observatory access was supported by Northwestern University and the Center for Interdisciplinary Exploration and Research in Astrophysics (CIERA).

Based on observations obtained at the international Gemini Observatory (Program IDs GN-2024B-Q-131, GN-2024B-Q-107, GS-2024B-Q-105, GS-2025A-Q-107), a program of NOIRLab, which is managed by the Association of Universities for Research in Astronomy (AURA) under a cooperative agreement with the National Science Foundation on behalf of the Gemini Observatory partnership: the National Science Foundation (United States), National Research Council (Canada), Agencia Nacional de Investigaci\'{o}n y Desarrollo (Chile), Ministerio de Ciencia, Tecnolog\'{i}a e Innovaci\'{o}n (Argentina), Minist\'{e}rio da Ci\^{e}ncia, Tecnologia, Inova\c{c}\~{o}es e Comunica\c{c}\~{o}es (Brazil), and Korea Astronomy and Space Science Institute (Republic of Korea). Data was processed using the Gemini DRAGONS (Data Reduction for Astronomy from Gemini Observatory North and South) package.

This work is based in part on observations made with the NASA/ESA/CSA James Webb Space Telescope. The data were obtained from the Mikulski Archive for Space Telescopes at the Space Telescope Science Institute, which is operated by the Association of Universities for Research in Astronomy, Inc., under NASA contract NAS 5-03127 for JWST. These observations are associated with program \#6133. The observation analyzed in this work can be accessed via \dataset[DOI: 10.17909/a0xj-1h19]{https://doi.org/10.17909/a0xj-1h19}. 

Data for this paper has in part been obtained under the International Time Programme of the CCI (International Scientific Committee of the Observatorios de Canarias
of the IAC) under programm ID ITP24 PI Jonker with the NOT and GTC operated on the island of La Palma by the Roque de los Muchachos. Observations have been made in part with the ALFOSC instrument, which is provided by the Instituto de Astrofisica de Andalucia (IAA) under a joint agreement with the University of Copenhagen and the Nordic Optical Telescope, owned in collaboration by the University of Turku and Aarhus University, and operated jointly by Aarhus University, the University of Turku and the University of Oslo, representing Denmark, Finland and Norway, the University of Iceland and Stockholm University at the Observatorio del Roque de los Muchachos, La Palma, Spain, of the Instituto de Astrofisica de Canarias.

This publication has made use of data collected at Lulin Observatory, partly supported by MoST grant 109-2112-M-008-001 and TAOVA with NSTC grant 113-2740-M-008-005.

Based on observations obtained with MegaPrime/MegaCam, a joint project of CFHT and CEA/DAPNIA, at the Canada-France-Hawaii Telescope (CFHT) which is operated by the National Research Council (NRC) of Canada, the Institut National des Science de l'Univers of the Centre National de la Recherche Scientifique (CNRS) of France, and the University of Hawai'i. The observations at the Canada-France-Hawaii Telescope were performed with care and respect from the summit of Maunakea which is a significant cultural and historic site.

Based on observations with the BlackGEM telescope array. The BlackGEM telescope array is built and run by a consortium consisting of Radboud University, the Netherlands Research School for Astronomy (NOVA), and KU Leuven with additional support from Armagh Observatory and Planetarium, Durham University, Hamburg Observatory, Hebrew University, Las Cumbres Observatory, Tel Aviv University, Texas Tech University, Technical University of Denmark, University of California Davis, the University of Barcelona, the University of Manchester, University of Potsdam, the University of Valparaiso, the University of Warwick, and Weizmann Institute of science. BlackGEM is hosted and supported by ESO at La Silla.

Pan-STARRS is primarily funded to search for near-Earth asteroids through NASA grants NNX08AR22G and NNX14AM74G. The Pan-STARRS science products for transient follow-up are made possible through the contributions of the University of Hawaii Institute for Astronomy and Queen's University Belfast.

This work has made use of data from the Asteroid Terrestrial-impact Last Alert System (ATLAS) project. The Asteroid Terrestrial-impact Last Alert System (ATLAS) project is primarily funded to search for near earth asteroids through NASA grants NN12AR55G, 80NSSC18K0284, and 80NSSC18K1575; byproducts of the NEO search include images and catalogs from the survey area. This work was partially funded by Kepler/K2 grant J1944/80NSSC19K0112 and HST GO-15889, and STFC grants ST/T000198/1 and ST/S006109/1. The ATLAS science products have been made possible through the contributions of the University of Hawaii Institute for Astronomy, the Queen’s University Belfast, the Space Telescope Science Institute, the South African Astronomical Observatory, and The Millennium Institute of Astrophysics (MAS), Chile.

The Legacy Surveys consist of three individual and complementary projects: the Dark Energy Camera Legacy Survey (DECaLS; Proposal ID \#2014B-0404; PIs: David Schlegel and Arjun Dey), the Beijing-Arizona Sky Survey (BASS; NOAO Prop. ID \#2015A-0801; PIs: Zhou Xu and Xiaohui Fan), and the Mayall $z$-band Legacy Survey (MzLS; Prop. ID \#2016A-0453; PI: Arjun Dey). DECaLS, BASS and MzLS together include data obtained, respectively, at the Blanco telescope, Cerro Tololo Inter-American Observatory, NSF’s NOIRLab; the Bok telescope, Steward Observatory, University of Arizona; and the Mayall telescope, Kitt Peak National Observatory, NOIRLab. Pipeline processing and analyses of the data were supported by NOIRLab and the Lawrence Berkeley National Laboratory (LBNL). The Legacy Surveys project is honored to be permitted to conduct astronomical research on Iolkam Du\'ag (Kitt Peak), a mountain with particular significance to the Tohono O\'odham Nation.

\bibliographystyle{aasjournal}
\bibliography{refs}

\begin{thebibliography}{}
\expandafter\ifx\csname natexlab\endcsname\relax\def\natexlab#1{#1}\fi
\providecommand{\url}[1]{\href{#1}{#1}}
\providecommand{\dodoi}[1]{doi:~\href{http://doi.org/#1}{\nolinkurl{#1}}}
\providecommand{\doeprint}[1]{\href{http://ascl.net/#1}{\nolinkurl{http://ascl.net/#1}}}
\providecommand{\doarXiv}[1]{\href{https://arxiv.org/abs/#1}{\nolinkurl{https://arxiv.org/abs/#1}}}

\bibitem[{{Aktar} {et~al.}(2017){Aktar}, {Das}, {Nandi}, \& {Sreehari}}]{2017MNRAS.471.4806A}
{Aktar}, R., {Das}, S., {Nandi}, A., \& {Sreehari}, H. 2017, \mnras, 471, 4806, \dodoi{10.1093/mnras/stx1893}

\bibitem[{{Alp} \& {Larsson}(2020)}]{AlpLarsson20}
{Alp}, D., \& {Larsson}, J. 2020, \apj, 896, 39, \dodoi{10.3847/1538-4357/ab91ba}

\bibitem[{{Arnett}(1982)}]{Arnett82}
{Arnett}, W.~D. 1982, \apj, 253, 785, \dodoi{10.1086/159681}

\bibitem[{{Ashton} {et~al.}(2019){Ashton}, {H{\"u}bner}, {Lasky}, {Talbot}, {Ackley}, {Biscoveanu}, {Chu}, {Divakarla}, {Easter}, {Goncharov}, {Hernandez Vivanco}, {Harms}, {Lower}, {Meadors}, {Melchor}, {Payne}, {Pitkin}, {Powell}, {Sarin}, {Smith}, \& {Thrane}}]{ashton19}
{Ashton}, G., {H{\"u}bner}, M., {Lasky}, P.~D., {et~al.} 2019, \apjs, 241, 27, \dodoi{10.3847/1538-4365/ab06fc}

\bibitem[{{Bauer} {et~al.}(2017){Bauer}, {Treister}, {Schawinski}, {Schulze}, {Luo}, {Alexander}, {Brandt}, {Comastri}, {Forster}, {Gilli}, {Kann}, {Maeda}, {Nomoto}, {Paolillo}, {Ranalli}, {Schneider}, {Shemmer}, {Tanaka}, {Tolstov}, {Tominaga}, {Tozzi}, {Vignali}, {Wang}, {Xue}, \& {Yang}}]{Bauer2017}
{Bauer}, F.~E., {Treister}, E., {Schawinski}, K., {et~al.} 2017, \mnras, 467, 4841, \dodoi{10.1093/mnras/stx417}

\bibitem[{{Bersten} {et~al.}(2014){Bersten}, {Benvenuto}, \& {Nomoto}}]{Bersten2014}
{Bersten}, M.~C., {Benvenuto}, O., \& {Nomoto}, K. 2014, in IAU Symposium, Vol. 296, Supernova Environmental Impacts, ed. A.~{Ray} \& R.~A. {McCray}, 58--62, \dodoi{10.1017/S174392131300923X}

\bibitem[{{Bochenek} {et~al.}(2024){Bochenek}, {Xu}, {Zhu}, {Izzo}, {Palmerio}, {Saccardi}, {D'Elia}, {De Cia}, {Vergani}, {Tanvir}, {An}, \& {Stargate Collaboration}}]{EP240804_z}
{Bochenek}, A., {Xu}, D., {Zhu}, Z.~P., {et~al.} 2024, GRB Coordinates Network, 37039, 1

\bibitem[{{Bright} {et~al.}(2024){Bright}, {Carotenuto}, {Fender}, {Choza}, {Mummery}, {Jonker}, {Smartt}, {DeBoer}, {Farah}, {Matthews}, {Pollak}, {Rhodes}, \& {Siemion}}]{Bright+24_EP240414}
{Bright}, J.~S., {Carotenuto}, F., {Fender}, R., {et~al.} 2024, arXiv e-prints, arXiv:2409.19055, \dodoi{10.48550/arXiv.2409.19055}

\bibitem[{{Buchner} {et~al.}(2014){Buchner}, {Georgakakis}, {Nandra}, {Hsu}, {Rangel}, {Brightman}, {Merloni}, {Salvato}, {Donley}, \& {Kocevski}}]{pymultinest}
{Buchner}, J., {Georgakakis}, A., {Nandra}, K., {et~al.} 2014, \aap, 564, A125, \dodoi{10.1051/0004-6361/201322971}

\bibitem[{{Bufano} {et~al.}(2012){Bufano}, {Pian}, {Sollerman}, {Benetti}, {Pignata}, {Valenti}, {Covino}, {D'Avanzo}, {Malesani}, {Cappellaro}, {Della Valle}, {Fynbo}, {Hjorth}, {Mazzali}, {Reichart}, {Starling}, {Turatto}, {Vergani}, {Wiersema}, {Amati}, {Bersier}, {Campana}, {Cano}, {Castro-Tirado}, {Chincarini}, {D'Elia}, {de Ugarte Postigo}, {Deng}, {Ferrero}, {Filippenko}, {Goldoni}, {Gorosabel}, {Greiner}, {Hammer}, {Jakobsson}, {Kaper}, {Kawabata}, {Klose}, {Levan}, {Maeda}, {Masetti}, {Milvang-Jensen}, {Mirabel}, {M{\o}ller}, {Nomoto}, {Palazzi}, {Piranomonte}, {Salvaterra}, {Stratta}, {Tagliaferri}, {Tanaka}, {Tanvir}, \& {Wijers}}]{Bufano+12}
{Bufano}, F., {Pian}, E., {Sollerman}, J., {et~al.} 2012, \apj, 753, 67, \dodoi{10.1088/0004-637X/753/1/67}

\bibitem[{{Busmann} {et~al.}(2025){Busmann}, {O'Connor}, {Sommer}, {Gruen}, {Beniamini}, {Gill}, {Moss}, {Palmese}, {Riffeser}, {Yang}, {Troja}, {Dichiara}, {Ricci}, {Klingler}, {G{\"o}ssl}, {Hu}, {Rau}, {Ries}, {Ryan}, {Schmidt}, {Yadav}, \& {Zeimann}}]{Busmann+25}
{Busmann}, M., {O'Connor}, B., {Sommer}, J., {et~al.} 2025, arXiv e-prints, arXiv:2503.14588.
\newblock \doarXiv{2503.14588}

\bibitem[{{Campana} {et~al.}(2006){Campana}, {Mangano}, {Blustin}, {Brown}, {Burrows}, {Chincarini}, {Cummings}, {Cusumano}, {Della Valle}, {Malesani}, {M{\'e}sz{\'a}ros}, {Nousek}, {Page}, {Sakamoto}, {Waxman}, {Zhang}, {Dai}, {Gehrels}, {Immler}, {Marshall}, {Mason}, {Moretti}, {O'Brien}, {Osborne}, {Page}, {Romano}, {Roming}, {Tagliaferri}, {Cominsky}, {Giommi}, {Godet}, {Kennea}, {Krimm}, {Angelini}, {Barthelmy}, {Boyd}, {Palmer}, {Wells}, \& {White}}]{Campana+06}
{Campana}, S., {Mangano}, V., {Blustin}, A.~J., {et~al.} 2006, \nat, 442, 1008, \dodoi{10.1038/nature04892}

\bibitem[{{Cano} {et~al.}(2011){Cano}, {Bersier}, {Guidorzi}, {Kobayashi}, {Levan}, {Tanvir}, {Wiersema}, {D'Avanzo}, {Fruchter}, {Garnavich}, {Gomboc}, {Gorosabel}, {Kasen}, {Kopa{\v{c}}}, {Margutti}, {Mazzali}, {Melandri}, {Mundell}, {Nugent}, {Pian}, {Smith}, {Steele}, {Wijers}, \& {Woosley}}]{Cano+11}
{Cano}, Z., {Bersier}, D., {Guidorzi}, C., {et~al.} 2011, \apj, 740, 41, \dodoi{10.1088/0004-637X/740/1/41}

\bibitem[{{Cano} {et~al.}(2017){Cano}, {Izzo}, {de Ugarte Postigo}, {Th{\"o}ne}, {Kr{\"u}hler}, {Heintz}, {Malesani}, {Geier}, {Fuentes}, {Chen}, {Covino}, {D'Elia}, {Fynbo}, {Goldoni}, {Gomboc}, {Hjorth}, {Jakobsson}, {Kann}, {Milvang-Jensen}, {Pugliese}, {S{\'a}nchez-Ram{\'\i}rez}, {Schulze}, {Sollerman}, {Tanvir}, \& {Wiersema}}]{Cano+17}
{Cano}, Z., {Izzo}, L., {de Ugarte Postigo}, A., {et~al.} 2017, \aap, 605, A107, \dodoi{10.1051/0004-6361/201731005}

\bibitem[{{Chornock} {et~al.}(2010){Chornock}, {Berger}, {Levesque}, {Soderberg}, {Foley}, {Fox}, {Frebel}, {Simon}, {Bochanski}, {Challis}, {Kirshner}, {Podsiadlowski}, {Roth}, {Rutledge}, {Schmidt}, {Sheppard}, \& {Simcoe}}]{Chornock+10}
{Chornock}, R., {Berger}, E., {Levesque}, E.~M., {et~al.} 2010, arXiv e-prints, arXiv:1004.2262, \dodoi{10.48550/arXiv.1004.2262}

\bibitem[{{Clocchiatti} {et~al.}(2011){Clocchiatti}, {Suntzeff}, {Covarrubias}, \& {Candia}}]{Clocchiatti+11}
{Clocchiatti}, A., {Suntzeff}, N.~B., {Covarrubias}, R., \& {Candia}, P. 2011, \aj, 141, 163, \dodoi{10.1088/0004-6256/141/5/163}

\bibitem[{{Cooke}(1976)}]{cooke76}
{Cooke}, B.~A. 1976, \nat, 261, 564, \dodoi{10.1038/261564a0}

\bibitem[{{Corsi} {et~al.}(2016){Corsi}, {Gal-Yam}, {Kulkarni}, {Frail}, {Mazzali}, {Cenko}, {Kasliwal}, {Cao}, {Horesh}, {Palliyaguru}, {Perley}, {Laher}, {Taddia}, {Leloudas}, {Maguire}, {Nugent}, {Sollerman}, \& {Sullivan}}]{Corsi+16}
{Corsi}, A., {Gal-Yam}, A., {Kulkarni}, S.~R., {et~al.} 2016, \apj, 830, 42, \dodoi{10.3847/0004-637X/830/1/42}

\bibitem[{{Dainotti} {et~al.}(2022){Dainotti}, {De Simone}, {Islam}, {Kawaguchi}, {Moriya}, {Takiwaki}, {Tominaga}, \& {Gangopadhyay}}]{Dainotti+22}
{Dainotti}, M.~G., {De Simone}, B., {Islam}, K.~M., {et~al.} 2022, \apj, 938, 41, \dodoi{10.3847/1538-4357/ac8b77}

\bibitem[{{Dessart} {et~al.}(2024){Dessart}, {Guti{\'e}rrez}, {Ercolino}, {Jin}, \& {Langer}}]{Dessart+24}
{Dessart}, L., {Guti{\'e}rrez}, C.~P., {Ercolino}, A., {Jin}, H., \& {Langer}, N. 2024, \aap, 685, A169, \dodoi{10.1051/0004-6361/202349066}

\bibitem[{{Dessart} {et~al.}(2020){Dessart}, {Yoon}, {Aguilera-Dena}, \& {Langer}}]{Dessart+20}
{Dessart}, L., {Yoon}, S.-C., {Aguilera-Dena}, D.~R., \& {Langer}, N. 2020, \aap, 642, A106, \dodoi{10.1051/0004-6361/202038763}

\bibitem[{{Dey} {et~al.}(2019){Dey}, {Schlegel}, {Lang}, {Blum}, {Burleigh}, {Fan}, {Findlay}, {Finkbeiner}, {Herrera}, {Juneau}, {Landriau}, {Levi}, {McGreer}, {Meisner}, {Myers}, {Moustakas}, {Nugent}, {Patej}, {Schlafly}, {Walker}, {Valdes}, {Weaver}, {Y{\`e}che}, {Zou}, {Zhou}, {Abareshi}, {Abbott}, {Abolfathi}, {Aguilera}, {Alam}, {Allen}, {Alvarez}, {Annis}, {Ansarinejad}, {Aubert}, {Beechert}, {Bell}, {BenZvi}, {Beutler}, {Bielby}, {Bolton}, {Brice{\~n}o}, {Buckley-Geer}, {Butler}, {Calamida}, {Carlberg}, {Carter}, {Casas}, {Castander}, {Choi}, {Comparat}, {Cukanovaite}, {Delubac}, {DeVries}, {Dey}, {Dhungana}, {Dickinson}, {Ding}, {Donaldson}, {Duan}, {Duckworth}, {Eftekharzadeh}, {Eisenstein}, {Etourneau}, {Fagrelius}, {Farihi}, {Fitzpatrick}, {Font-Ribera}, {Fulmer}, {G{\"a}nsicke}, {Gaztanaga}, {George}, {Gerdes}, {Gontcho}, {Gorgoni}, {Green}, {Guy}, {Harmer}, {Hernandez}, {Honscheid}, {Huang}, {James}, {Jannuzi}, {Jiang}, {Joyce}, {Karcher}, {Karkar}, {Kehoe}, {Kneib}, {Kueter-Young}, {Lan},
  {Lauer}, {Le Guillou}, {Le Van Suu}, {Lee}, {Lesser}, {Perreault Levasseur}, {Li}, {Mann}, {Marshall}, {Mart{\'\i}nez-V{\'a}zquez}, {Martini}, {du Mas des Bourboux}, {McManus}, {Meier}, {M{\'e}nard}, {Metcalfe}, {Mu{\~n}oz-Guti{\'e}rrez}, {Najita}, {Napier}, {Narayan}, {Newman}, {Nie}, {Nord}, {Norman}, {Olsen}, {Paat}, {Palanque-Delabrouille}, {Peng}, {Poppett}, {Poremba}, {Prakash}, {Rabinowitz}, {Raichoor}, {Rezaie}, {Robertson}, {Roe}, {Ross}, {Ross}, {Rudnick}, {Safonova}, {Saha}, {S{\'a}nchez}, {Savary}, {Schweiker}, {Scott}, {Seo}, {Shan}, {Silva}, {Slepian}, {Soto}, {Sprayberry}, {Staten}, {Stillman}, {Stupak}, {Summers}, {Sien Tie}, {Tirado}, {Vargas-Maga{\~n}a}, {Vivas}, {Wechsler}, {Williams}, {Yang}, {Yang}, {Yapici}, {Zaritsky}, {Zenteno}, {Zhang}, {Zhang}, {Zhou}, \& {Zhou}}]{Dey+19}
{Dey}, A., {Schlegel}, D.~J., {Lang}, D., {et~al.} 2019, \aj, 157, 168, \dodoi{10.3847/1538-3881/ab089d}

\bibitem[{{Drout} {et~al.}(2011){Drout}, {Soderberg}, {Gal-Yam}, {Cenko}, {Fox}, {Leonard}, {Sand}, {Moon}, {Arcavi}, \& {Green}}]{Drout+11}
{Drout}, M.~R., {Soderberg}, A.~M., {Gal-Yam}, A., {et~al.} 2011, \apj, 741, 97, \dodoi{10.1088/0004-637X/741/2/97}

\bibitem[{{Evans} {et~al.}(2007){Evans}, {Beardmore}, {Page}, {Tyler}, {Osborne}, {Goad}, {O'Brien}, {Vetere}, {Racusin}, {Morris}, {Burrows}, {Capalbi}, {Perri}, {Gehrels}, \& {Romano}}]{Evans+07}
{Evans}, P.~A., {Beardmore}, A.~P., {Page}, K.~L., {et~al.} 2007, \aap, 469, 379, \dodoi{10.1051/0004-6361:20077530}

\bibitem[{{Evans} {et~al.}(2009){Evans}, {Beardmore}, {Page}, {Osborne}, {O'Brien}, {Willingale}, {Starling}, {Burrows}, {Godet}, {Vetere}, {Racusin}, {Goad}, {Wiersema}, {Angelini}, {Capalbi}, {Chincarini}, {Gehrels}, {Kennea}, {Margutti}, {Morris}, {Mountford}, {Pagani}, {Perri}, {Romano}, \& {Tanvir}}]{Evans+09}
---. 2009, \mnras, 397, 1177, \dodoi{10.1111/j.1365-2966.2009.14913.x}

\bibitem[{{Eyles-Ferris} {et~al.}(2025){Eyles-Ferris}, {Jonker}, {Levan}, {Bj{\o}rn Malesani}, {Sarin}, {Fryer}, {Rastinejad}, {Burns}, {Tanvir}, {O'Brien}, {Fong}, {Mandel}, {Gompertz}, {Kilpatrick}, {Bloemen}, {Bright}, {Carotenuto}, {Corcoran}, {Cotter}, {Groot}, {Izzo}, {Laskar}, {Martin-Carrillo}, {Palmerio}, {Ravasio}, {van Roestel}, {Saccardi}, {Starling}, {Linesh Thakur}, {Vergani}, {Vreeswijk}, {Bauer}, {Campana}, {Chac{\'o}n}, {Chrimes}, {Covino}, {van Dalen}, {D'Elia}, {De Pasquale}, {Habeeb}, {Hartmann}, {van Hoof}, {Jakobsson}, {Julakanti}, {Leloudas}, {Mata S{\'a}nchez}, {Nixon}, {Pieterse}, {Pugliese}, {Quirola-V{\'a}squez}, {Rayson}, {Salvaterra}, {Schneider}, {Torres}, \& {Zafar}}]{EylesFerris25}
{Eyles-Ferris}, R. A.~J., {Jonker}, P.~G., {Levan}, A.~J., {et~al.} 2025, arXiv e-prints, arXiv:2504.08886, \dodoi{10.48550/arXiv.2504.08886}

\bibitem[{{Finneran} \& {Martin-Carrillo}(2024)}]{Finneran+24}
{Finneran}, G., \& {Martin-Carrillo}, A. 2024, arXiv e-prints, arXiv:2411.12574, \dodoi{10.48550/arXiv.2411.12574}

\bibitem[{{Fitzpatrick}(1999)}]{Fitzpatrick1999}
{Fitzpatrick}, E.~L. 1999, \pasp, 111, 63, \dodoi{10.1086/316293}

\bibitem[{{Flewelling} {et~al.}(2020){Flewelling}, {Magnier}, {Chambers}, {Heasley}, {Holmberg}, {Huber}, {Sweeney}, {Waters}, {Calamida}, {Casertano}, {Chen}, {Farrow}, {Hasinger}, {Henderson}, {Long}, {Metcalfe}, {Narayan}, {Nieto-Santisteban}, {Norberg}, {Rest}, {Saglia}, {Szalay}, {Thakar}, {Tonry}, {Valenti}, {Werner}, {White}, {Denneau}, {Draper}, {Hodapp}, {Jedicke}, {Kaiser}, {Kudritzki}, {Price}, {Wainscoat}, {Chastel}, {McLean}, {Postman}, \& {Shiao}}]{Flewelling+20}
{Flewelling}, H.~A., {Magnier}, E.~A., {Chambers}, K.~C., {et~al.} 2020, \apjs, 251, 7, \dodoi{10.3847/1538-4365/abb82d}

\bibitem[{{Foley} {et~al.}(2003){Foley}, {Papenkova}, {Swift}, {Filippenko}, {Li}, {Mazzali}, {Chornock}, {Leonard}, \& {Van Dyk}}]{Foley+02}
{Foley}, R.~J., {Papenkova}, M.~S., {Swift}, B.~J., {et~al.} 2003, \pasp, 115, 1220, \dodoi{10.1086/378242}

\bibitem[{{Fontana} {et~al.}(2014){Fontana}, {Dunlop}, {Paris}, {Targett}, {Boutsia}, {Castellano}, {Galametz}, {Grazian}, {McLure}, {Merlin}, {Pentericci}, {Wuyts}, {Almaini}, {Caputi}, {Chary}, {Cirasuolo}, {Conselice}, {Cooray}, {Daddi}, {Dickinson}, {Faber}, {Fazio}, {Ferguson}, {Giallongo}, {Giavalisco}, {Grogin}, {Hathi}, {Koekemoer}, {Koo}, {Lucas}, {Nonino}, {Rix}, {Renzini}, {Rosario}, {Santini}, {Scarlata}, {Sommariva}, {Stark}, {van der Wel}, {Vanzella}, {Wild}, {Yan}, \& {Zibetti}}]{Fontana2014a}
{Fontana}, A., {Dunlop}, J.~S., {Paris}, D., {et~al.} 2014, \aap, 570, A11, \dodoi{10.1051/0004-6361/201423543}

\bibitem[{{Frederiks} {et~al.}(2024){Frederiks}, {Lysenko}, {Ridnaia}, {Svinkin}, {Tsvetkova}, {Ulanov}, {Cline}, \& {Konus-Wind Team}}]{GCN.37071}
{Frederiks}, D., {Lysenko}, A., {Ridnaia}, A., {et~al.} 2024, GRB Coordinates Network, 37071, 1

\bibitem[{{Freudling} {et~al.}(2013){Freudling}, {Romaniello}, {Bramich}, {Ballester}, {Forchi}, {Garc{\'\i}a-Dabl{\'o}}, {Moehler}, \& {Neeser}}]{esoreflex}
{Freudling}, W., {Romaniello}, M., {Bramich}, D.~M., {et~al.} 2013, \aap, 559, A96, \dodoi{10.1051/0004-6361/201322494}

\bibitem[{{Frey} {et~al.}(2013){Frey}, {Fryer}, \& {Young}}]{2013ApJ...773L...7F}
{Frey}, L.~H., {Fryer}, C.~L., \& {Young}, P.~A. 2013, \apjl, 773, L7, \dodoi{10.1088/2041-8205/773/1/L7}

\bibitem[{{Fryer}(1999)}]{1999ApJ...522..413F}
{Fryer}, C.~L. 1999, \apj, 522, 413, \dodoi{10.1086/307647}

\bibitem[{{Fryer} {et~al.}(2024){Fryer}, {Burns}, {Ho}, {Corsi}, {Lien}, {Perley}, {Vail}, \& {Villar}}]{2024arXiv241010378F}
{Fryer}, C.~L., {Burns}, E., {Ho}, A. Y.~Q., {et~al.} 2024, arXiv e-prints, arXiv:2410.10378, \dodoi{10.48550/arXiv.2410.10378}

\bibitem[{{Fryer} {et~al.}(2020){Fryer}, {Fontes}, {Warsa}, {Roming}, {Coffing}, \& {Wood}}]{Fryer+20}
{Fryer}, C.~L., {Fontes}, C.~J., {Warsa}, J.~S., {et~al.} 2020, \apj, 898, 123, \dodoi{10.3847/1538-4357/ab99a7}

\bibitem[{{Fryer} \& {Woosley}(1998)}]{1998ApJ...502L...9F}
{Fryer}, C.~L., \& {Woosley}, S.~E. 1998, \apjl, 502, L9, \dodoi{10.1086/311493}

\bibitem[{{Fryer} {et~al.}(2023){Fryer}, {Keiter}, {Sharma}, {Leveillee}, {Meyerhofer}, {Barnak}, {Byvank}, {Elshafiey}, {Fontes}, {Johns}, {Kozlowski}, \& {Urbatsch}}]{Fryer+23}
{Fryer}, C.~L., {Keiter}, P.~A., {Sharma}, V., {et~al.} 2023, arXiv e-prints, arXiv:2312.16677, \dodoi{10.48550/arXiv.2312.16677}

\bibitem[{{Galama} {et~al.}(1998){Galama}, {Vreeswijk}, {van Paradijs}, {Kouveliotou}, {Augusteijn}, {B{\"o}hnhardt}, {Brewer}, {Doublier}, {Gonzalez}, {Leibundgut}, {Lidman}, {Hainaut}, {Patat}, {Heise}, {in't Zand}, {Hurley}, {Groot}, {Strom}, {Mazzali}, {Iwamoto}, {Nomoto}, {Umeda}, {Nakamura}, {Young}, {Suzuki}, {Shigeyama}, {Koshut}, {Kippen}, {Robinson}, {de Wildt}, {Wijers}, {Tanvir}, {Greiner}, {Pian}, {Palazzi}, {Frontera}, {Masetti}, {Nicastro}, {Feroci}, {Costa}, {Piro}, {Peterson}, {Tinney}, {Boyle}, {Cannon}, {Stathakis}, {Sadler}, {Begam}, \& {Ianna}}]{Galama+98}
{Galama}, T.~J., {Vreeswijk}, P.~M., {van Paradijs}, J., {et~al.} 1998, \nat, 395, 670, \dodoi{10.1038/27150}

\bibitem[{{Gargiulo} {et~al.}(2022){Gargiulo}, {Fumana}, {Bisogni}, {Franzetti}, {Cassar{\`a}}, {Garilli}, {Scodeggio}, \& {Vietri}}]{sipgi2022}
{Gargiulo}, A., {Fumana}, M., {Bisogni}, S., {et~al.} 2022, \mnras, 514, 2902, \dodoi{10.1093/mnras/stac1065}

\bibitem[{{Glennie} {et~al.}(2015){Glennie}, {Jonker}, {Fender}, {Nagayama}, \& {Pretorius}}]{Glennie2015}
{Glennie}, A., {Jonker}, P.~G., {Fender}, R.~P., {Nagayama}, T., \& {Pretorius}, M.~L. 2015, \mnras, 450, 3765, \dodoi{10.1093/mnras/stv801}

\bibitem[{{Greiner} {et~al.}(2015){Greiner}, {Mazzali}, {Kann}, {Kr{\"u}hler}, {Pian}, {Prentice}, {Olivares E.}, {Rossi}, {Klose}, {Taubenberger}, {Knust}, {Afonso}, {Ashall}, {Bolmer}, {Delvaux}, {Diehl}, {Elliott}, {Filgas}, {Fynbo}, {Graham}, {Guelbenzu}, {Kobayashi}, {Leloudas}, {Savaglio}, {Schady}, {Schmidl}, {Schweyer}, {Sudilovsky}, {Tanga}, {Updike}, {van Eerten}, \& {Varela}}]{Greiner+15}
{Greiner}, J., {Mazzali}, P.~A., {Kann}, D.~A., {et~al.} 2015, \nat, 523, 189, \dodoi{10.1038/nature14579}

\bibitem[{{Grichener}(2025)}]{2025Ap&SS.370...11G}
{Grichener}, A. 2025, \apss, 370, 11, \dodoi{10.1007/s10509-025-04402-1}

\bibitem[{{Groot} {et~al.}(2024){Groot}, {Bloemen}, {Vreeswijk}, {van Roestel}, {Jonker}, {Nelemans}, {Klein-Wolt}, {Lepoole}, {Pieterse}, {Rodenhuis}, {Boland}, {Haverkorn}, {Aerts}, {Bakker}, {Balster}, {Bekema}, {Dijkstra}, {Dolron}, {Elswijk}, {van Elteren}, {Engels}, {Fokker}, {de Haan}, {Hahn}, {ter Horst}, {Lesman}, {Kragt}, {Morren}, {Nillissen}, {Pessemier}, {Raskin}, {de Rijke}, {Scheers}, {Schuil}, {Timmer}, {Antunes Amaral}, {Arancibia-Rojas}, {Arcavi}, {Blagorodnova}, {Biswas}, {Breton}, {Dawson}, {Dayal}, {De Wet}, {Duffy}, {Faris}, {Fausnaugh}, {Gal-Yam}, {Geier}, {Horesh}, {Johnston}, {Katusiime}, {Kelley}, {Kosakowski}, {Kupfer}, {Leloudas}, {Levan}, {Modiano}, {Mogawana}, {Munday}, {Paice}, {Patat}, {Pelisoli}, {Ramsay}, {Ranaivomanana}, {Ruiz-Carmona}, {Schaffenroth}, {Scaringi}, {Stoppa}, {Street}, {Tranin}, {Uzundag}, {Valenti}, {Veresvarska}, {Vuc̆kovi{\'c}}, {Wichern}, {Wijers}, {Wijnands}, \& {Zimmerman}}]{BlackGEM24}
{Groot}, P.~J., {Bloemen}, S., {Vreeswijk}, P.~M., {et~al.} 2024, \pasp, 136, 115003, \dodoi{10.1088/1538-3873/ad8b6a}

\bibitem[{{Harutyunyan} {et~al.}(2008){Harutyunyan}, {Pfahler}, {Pastorello}, {Taubenberger}, {Turatto}, {Cappellaro}, {Benetti}, {Elias-Rosa}, {Navasardyan}, {Valenti}, {Stanishev}, {Patat}, {Riello}, {Pignata}, \& {Hillebrandt}}]{Harutyunyan+08}
{Harutyunyan}, A.~H., {Pfahler}, P., {Pastorello}, A., {et~al.} 2008, \aap, 488, 383, \dodoi{10.1051/0004-6361:20078859}

\bibitem[{{Heise} {et~al.}(2001){Heise}, {Zand}, {Kippen}, \& {Woods}}]{Heise+01}
{Heise}, J., {Zand}, J.~I., {Kippen}, R.~M., \& {Woods}, P.~M. 2001, in Gamma-ray Bursts in the Afterglow Era, ed. E.~{Costa}, F.~{Frontera}, \& J.~{Hjorth}, 16, \dodoi{10.1007/10853853_4}

\bibitem[{{Hirai}(2023)}]{2023MNRAS.523.6011H}
{Hirai}, R. 2023, \mnras, 523, 6011, \dodoi{10.1093/mnras/stad1856}

\bibitem[{{Hirai} {et~al.}(2018){Hirai}, {Podsiadlowski}, \& {Yamada}}]{2018ApJ...864..119H}
{Hirai}, R., {Podsiadlowski}, P., \& {Yamada}, S. 2018, \apj, 864, 119, \dodoi{10.3847/1538-4357/aad6a0}

\bibitem[{{Hirai} {et~al.}(2020){Hirai}, {Sato}, {Podsiadlowski}, {Vigna-G{\'o}mez}, \& {Mandel}}]{2020MNRAS.499.1154H}
{Hirai}, R., {Sato}, T., {Podsiadlowski}, P., {Vigna-G{\'o}mez}, A., \& {Mandel}, I. 2020, \mnras, 499, 1154, \dodoi{10.1093/mnras/staa2898}

\bibitem[{{Hiramatsu} {et~al.}(2020){Hiramatsu}, {Arcavi}, {Burke}, {Howell}, {McCully}, {Pellegrino}, \& {Valenti}}]{2020bvc_spec}
{Hiramatsu}, D., {Arcavi}, I., {Burke}, J., {et~al.} 2020, Transient Name Server Classification Report, 2020-403, 1

\bibitem[{{Ho} {et~al.}(2020){Ho}, {Kulkarni}, {Perley}, {Cenko}, {Corsi}, {Schulze}, {Lunnan}, {Sollerman}, {Gal-Yam}, {Anand}, {Barbarino}, {Bellm}, {Bruch}, {Burns}, {De}, {Dekany}, {Delacroix}, {Duev}, {Frederiks}, {Fremling}, {Goldstein}, {Golkhou}, {Graham}, {Hale}, {Kasliwal}, {Kupfer}, {Laher}, {Martikainen}, {Masci}, {Neill}, {Ridnaia}, {Rusholme}, {Savchenko}, {Shupe}, {Soumagnac}, {Strotjohann}, {Svinkin}, {Taggart}, {Tartaglia}, {Yan}, \& {Zolkower}}]{Ho+20}
{Ho}, A. Y.~Q., {Kulkarni}, S.~R., {Perley}, D.~A., {et~al.} 2020, \apj, 902, 86, \dodoi{10.3847/1538-4357/aba630}

\bibitem[{{Hu} {et~al.}(2021){Hu}, {Castro-Tirado}, {Kumar}, {Gupta}, {Valeev}, {Pandey}, {Kann}, {Castell{\'o}n}, {Agudo}, {Aryan}, {Caballero-Garc{\'\i}a}, {Guziy}, {Martin-Carrillo}, {Oates}, {Pian}, {S{\'a}nchez-Ram{\'\i}rez}, {Sokolov}, \& {Zhang}}]{Hu+21}
{Hu}, Y.~D., {Castro-Tirado}, A.~J., {Kumar}, A., {et~al.} 2021, \aap, 646, A50, \dodoi{10.1051/0004-6361/202039349}

\bibitem[{{Izzo} {et~al.}(2020){Izzo}, {Auchettl}, {Hjorth}, {De Colle}, {Gall}, {Angus}, {Raimundo}, \& {Ramirez-Ruiz}}]{Izzo+20}
{Izzo}, L., {Auchettl}, K., {Hjorth}, J., {et~al.} 2020, \aap, 639, L11, \dodoi{10.1051/0004-6361/202038152}

\bibitem[{{Izzo} {et~al.}(2019){Izzo}, {de Ugarte Postigo}, {Maeda}, {Th{\"o}ne}, {Kann}, {Della Valle}, {Sagues Carracedo}, {Micha{\l}owski}, {Schady}, {Schmidl}, {Selsing}, {Starling}, {Suzuki}, {Bensch}, {Bolmer}, {Campana}, {Cano}, {Covino}, {Fynbo}, {Hartmann}, {Heintz}, {Hjorth}, {Japelj}, {Kami{\'n}ski}, {Kaper}, {Kouveliotou}, {Kru{\.Z}y{\'n}ski}, {Kwiatkowski}, {Leloudas}, {Levan}, {Malesani}, {Micha{\l}owski}, {Piranomonte}, {Pugliese}, {Rossi}, {S{\'a}nchez-Ram{\'\i}rez}, {Schulze}, {Steeghs}, {Tanvir}, {Ulaczyk}, {Vergani}, \& {Wiersema}}]{Izzo+19}
{Izzo}, L., {de Ugarte Postigo}, A., {Maeda}, K., {et~al.} 2019, \nat, 565, 324, \dodoi{10.1038/s41586-018-0826-3}

\bibitem[{{Izzo} {et~al.}(2025){Izzo}, {Martin-Carrillo}, {Malesani}, {Levan}, {Jonker}, {Cotter}, {van Dalen}, {Corcoran}, {Wiersema}, \& {Bauer}}]{2025GCN.39851}
{Izzo}, L., {Martin-Carrillo}, A., {Malesani}, D.~B., {et~al.} 2025, GRB Coordinates Network, 39851, 1

\bibitem[{{Jonker} {et~al.}(2013){Jonker}, {Glennie}, {Heida}, {Maccarone}, {Hodgkin}, {Nelemans}, {Miller-Jones}, {Torres}, \& {Fender}}]{Jonker2013}
{Jonker}, P.~G., {Glennie}, A., {Heida}, M., {et~al.} 2013, \apj, 779, 14, \dodoi{10.1088/0004-637X/779/1/14}

\bibitem[{{Khatami} \& {Kasen}(2019)}]{Khatami2019}
{Khatami}, D.~K., \& {Kasen}, D.~N. 2019, \apj, 878, 56, \dodoi{10.3847/1538-4357/ab1f09}

\bibitem[{{Labrie} {et~al.}(2019){Labrie}, {Anderson}, {C{\'a}rdenes}, {Simpson}, \& {Turner}}]{DRAGONS19}
{Labrie}, K., {Anderson}, K., {C{\'a}rdenes}, R., {Simpson}, C., \& {Turner}, J. E.~H. 2019, in Astronomical Society of the Pacific Conference Series, Vol. 523, Astronomical Data Analysis Software and Systems XXVII, ed. P.~J. {Teuben}, M.~W. {Pound}, B.~A. {Thomas}, \& E.~M. {Warner}, 321

\bibitem[{{Levan} {et~al.}(2025{\natexlab{a}}){Levan}, {Malesani}, {Jonker}, {Quirola-V{\'a}squez}, {S{\'a}nchez-Sierras}, {Martin-Carrillo}, \& {Bauer}}]{EP250207b_z}
{Levan}, A.~J., {Malesani}, D.~B., {Jonker}, P.~G., {et~al.} 2025{\natexlab{a}}, GRB Coordinates Network, 39278, 1

\bibitem[{{Levan} {et~al.}(2024{\natexlab{a}}){Levan}, {Quirola-Vasquez}, {Rastinejad}, {Malesani}, {Jonker}, \& {Martin-Carrillo}}]{GCN38593}
{Levan}, A.~J., {Quirola-Vasquez}, J.~A., {Rastinejad}, J.~C., {et~al.} 2024{\natexlab{a}}, GRB Coordinates Network, 38593, 1

\bibitem[{{Levan} {et~al.}(2025{\natexlab{b}}){Levan}, {Rastinejad}, {Malesani}, {Fong}, {Tanvir}, {Jonker}, \& {Eyles-Ferris}}]{ep250108a_gmos_sn}
{Levan}, A.~J., {Rastinejad}, J.~C., {Malesani}, D.~B., {et~al.} 2025{\natexlab{b}}, GRB Coordinates Network, 38987, 1

\bibitem[{{Levan} {et~al.}(2014){Levan}, {Tanvir}, {Starling}, {Wiersema}, {Page}, {Perley}, {Schulze}, {Wynn}, {Chornock}, {Hjorth}, {Cenko}, {Fruchter}, {O'Brien}, {Brown}, {Tunnicliffe}, {Malesani}, {Jakobsson}, {Watson}, {Berger}, {Bersier}, {Cobb}, {Covino}, {Cucchiara}, {de Ugarte Postigo}, {Fox}, {Gal-Yam}, {Goldoni}, {Gorosabel}, {Kaper}, {Kr{\"u}hler}, {Karjalainen}, {Osborne}, {Pian}, {S{\'a}nchez-Ram{\'\i}rez}, {Schmidt}, {Skillen}, {Tagliaferri}, {Th{\"o}ne}, {Vaduvescu}, {Wijers}, \& {Zauderer}}]{Levan+14_ULGRBs}
{Levan}, A.~J., {Tanvir}, N.~R., {Starling}, R.~L.~C., {et~al.} 2014, \apj, 781, 13, \dodoi{10.1088/0004-637X/781/1/13}

\bibitem[{{Levan} {et~al.}(2024{\natexlab{b}}){Levan}, {Jonker}, {Saccardi}, {Bj{\o}rn Malesani}, {Tanvir}, {Izzo}, {Heintz}, {Mata S{\'a}nchez}, {Quirola-V{\'a}squez}, {Torres}, {Vergani}, {Schulze}, {Rossi}, {D'Avanzo}, {Gompertz}, {Martin-Carrillo}, {de Ugarte Postigo}, {Schneider}, {Yuan}, {Ling}, {Zhang}, {Mao}, {Liu}, {Sun}, {Xu}, {Zhu}, {Ag{\"u}{\'\i} Fern{\'a}ndez}, {Amati}, {Bauer}, {Campana}, {Carotenuto}, {Chrimes}, {van Dalen}, {D'Elia}, {Della Valle}, {De Pasquale}, {Dhillon}, {Galbany}, {Gaspari}, {Gianfagna}, {Gomboc}, {Habeeb}, {van Hoof}, {Hu}, {Jakobsson}, {Julakanti}, {Korth}, {Kouveliotou}, {Laskar}, {Littlefair}, {Maiorano}, {Mao}, {Melandri}, {Miller}, {Mukherjee}, {Oates}, {O'Brien}, {Palmerio}, {Parviainen}, {Pieterse}, {Piranomonte}, {Piro}, {Pugliese}, {Ravasio}, {Rayson}, {Salvaterra}, {S{\'a}nchez-Ram{\'\i}rez}, {Sarin}, {Shilling}, {Starling}, {Tagliaferri}, {Linesh Thakur}, {Th{\"o}ne}, {Wiersema}, {Worssam}, \& {Zafar}}]{Levan+24_ep}
{Levan}, A.~J., {Jonker}, P.~G., {Saccardi}, A., {et~al.} 2024{\natexlab{b}}, arXiv e-prints, arXiv:2404.16350, \dodoi{10.48550/arXiv.2404.16350}

\bibitem[{{Levan} {et~al.}(2024{\natexlab{c}}){Levan}, {Gompertz}, {Salafia}, {Bulla}, {Burns}, {Hotokezaka}, {Izzo}, {Lamb}, {Malesani}, {Oates}, {Ravasio}, {Rouco Escorial}, {Schneider}, {Sarin}, {Schulze}, {Tanvir}, {Ackley}, {Anderson}, {Brammer}, {Christensen}, {Dhillon}, {Evans}, {Fausnaugh}, {Fong}, {Fruchter}, {Fryer}, {Fynbo}, {Gaspari}, {Heintz}, {Hjorth}, {Kennea}, {Kennedy}, {Laskar}, {Leloudas}, {Mandel}, {Martin-Carrillo}, {Metzger}, {Nicholl}, {Nugent}, {Palmerio}, {Pugliese}, {Rastinejad}, {Rhodes}, {Rossi}, {Saccardi}, {Smartt}, {Stevance}, {Tohuvavohu}, {van der Horst}, {Vergani}, {Watson}, {Barclay}, {Bhirombhakdi}, {Breedt}, {Breeveld}, {Brown}, {Campana}, {Chrimes}, {D'Avanzo}, {D'Elia}, {De Pasquale}, {Dyer}, {Galloway}, {Garbutt}, {Green}, {Hartmann}, {Jakobsson}, {Kerry}, {Kouveliotou}, {Langeroodi}, {Le Floc'h}, {Leung}, {Littlefair}, {Munday}, {O'Brien}, {Parsons}, {Pelisoli}, {Sahman}, {Salvaterra}, {Sbarufatti}, {Steeghs}, {Tagliaferri}, {Th{\"o}ne}, {de Ugarte Postigo}, \&
  {Kann}}]{Levan+24}
{Levan}, A.~J., {Gompertz}, B.~P., {Salafia}, O.~S., {et~al.} 2024{\natexlab{c}}, \nat, 626, 737, \dodoi{10.1038/s41586-023-06759-1}

\bibitem[{{Li} {et~al.}(2025{\natexlab{a}}){Li}, {Chen}, {Chatterjee}, {Hua}, {Liu}, \& {Einstein Probe Team}}]{GCN_EP250108a_disc}
{Li}, R.~Z., {Chen}, X.~L., {Chatterjee}, K., {et~al.} 2025{\natexlab{a}}, GRB Coordinates Network, 38861, 1

\bibitem[{{Li} {et~al.}(2025{\natexlab{b}}){Li}, {Zhu}, {Zou}, {Geng}, {Liu}, {Wang}, {Li}, {Xu}, {Sun}, {Wang}, {Yu}, {Zhang}, {Wu}, {Yang}, {Filippenko}, {Liu}, {Yuan}, {Aguado}, {An}, {An}, {Buckley}, {Castro-Tirado}, {Fu}, {Fynbo}, {Howell}, {Hu}, {Jiang}, {Kumar}, {Mao}, {Maund}, {Liu}, {Mockler}, {Moskvitin}, {Andrews}, {Bom}, {Brink}, {Chatterjee}, {Chen}, {Cheng}, {Cooke}, {Dai}, {Du}, {Erasmus}, {Fang}, {Farah}, {Goranskij}, {Gritsevich}, {Gu}, {Guo}, {Hsiao}, {Hu}, {Hua}, {Jacobson-Gal{\'a}n}, {Jia}, {Jin}, {Kasliwal}, {Kilpatrick}, {Kumar}, {Lei}, {Li}, {Li}, {Li}, {Ling}, {Liu}, {Liu}, {Liu}, {L{\'o}pez-Oramas}, {Maslennikova}, {McCully}, {Monageng}, {Newsone}, {Padilla Gonzalez}, {Pan}, {Peng}, {Pignata}, {Poidevin}, {Potter}, {P{\'e}rez-Fournon}, {Santana-Silva}, {Santos}, {Song}, {Song}, {Spiridonova}, {Sun}, {Sun}, {Terreran}, {Wang}, {Wang}, {Wang}, {Wang}, {Wu}, {Xiang}, {Xiao}, {Xu}, {Xue}, {Yan}, {Yang}, {Yu}, {Zhang}, {Zhang}, {Zhang}, {Zhang}, {Zhang}, {Zheng}, \&
  {Zou}}]{Li+25_ep250108a}
{Li}, W.~X., {Zhu}, Z.~P., {Zou}, X.~Z., {et~al.} 2025{\natexlab{b}}, arXiv e-prints, arXiv:2504.17034, \dodoi{10.48550/arXiv.2504.17034}

\bibitem[{{Liljegren} {et~al.}(2023){Liljegren}, {Jerkstrand}, {Barklem}, {Nyman}, {Brady}, \& {Yurchenko}}]{Liljegren23}
{Liljegren}, S., {Jerkstrand}, A., {Barklem}, P.~S., {et~al.} 2023, \aap, 674, A184, \dodoi{10.1051/0004-6361/202243491}

\bibitem[{{Lin} {et~al.}(2022){Lin}, {Irwin}, {Berger}, \& {Nguyen}}]{2022ApJ...927..211L}
{Lin}, D., {Irwin}, J.~A., {Berger}, E., \& {Nguyen}, R. 2022, \apj, 927, 211, \dodoi{10.3847/1538-4357/ac4fc6}

\bibitem[{{Liu} {et~al.}(2023){Liu}, {Miller}, {Polin}, {Nugent}, {De}, {Nugent}, {Schulze}, {Gal-Yam}, {Fremling}, {Anand}, {Andreoni}, {Blanchard}, {Brink}, {Dhawan}, {Filippenko}, {Maguire}, {Schweyer}, {Sears}, {Sharma}, {Graham}, {Groom}, {Hale}, {Kasliwal}, {Masci}, {Purdum}, {Racine}, {Sollerman}, \& {Kulkarni}}]{Liu+23}
{Liu}, C., {Miller}, A.~A., {Polin}, A., {et~al.} 2023, \apj, 946, 83, \dodoi{10.3847/1538-4357/acbb5e}

\bibitem[{{Liu} {et~al.}(2025){Liu}, {Sun}, {Xu}, {Svinkin}, {Delaunay}, {Tanvir}, {Gao}, {Zhang}, {Chen}, {Wu}, {Zhang}, {Yuan}, {An}, {Bruni}, {Frederiks}, {Ghirlanda}, {Hu}, {Li}, {Li}, {Li}, {Malesani}, {Piro}, {Raman}, {Ricci}, {Troja}, {Vergani}, {Wu}, {Yang}, {Zhang}, {Zhu}, {de Ugarte Postigo}, {Demin}, {Dobie}, {Fan}, {Fu}, {Fynbo}, {Geng}, {Gianfagna}, {Hu}, {Huang}, {Jiang}, {Jonker}, {Julakanti}, {Kennea}, {Kokomov}, {Kuulkers}, {Lei}, {Leung}, {Levan}, {Li}, {Li}, {Littlefair}, {Liu}, {Lysenko}, {Ma}, {Martin-Carrillo}, {O'Brien}, {Parsotan}, {Quirola-V{\'a}squez}, {Ridnaia}, {Ronchini}, {Rossi}, {Mata-S{\'a}nchez}, {Schneider}, {Shen}, {Thakur}, {Tohuvavohu}, {Torres}, {Tsvetkova}, {Ulanov}, {Wei}, {Xiao}, {Yin}, {Bai}, {Burwitz}, {Cai}, {Chen}, {Chen}, {Chen}, {Chen}, {Chen}, {Chen}, {Cheng}, {Cordier}, {Cui}, {Cui}, {Dai}, {Dai}, {Eder}, {Eyles-Ferris}, {Fan}, {Feldman}, {Feng}, {Feng}, {Friedrich}, {Gao}, {Gonzalez}, {Guan}, {Han}, {Han}, {Hou}, {Hu}, {Hu}, {Huang}, {Huo}, {Hutchinson}, {Ji},
  {Jia}, {Jia}, {Jiang}, {Jin}, {Jin}, {Jin}, {Keereman}, {Lerman}, {Li}, {Li}, {Li}, {Li}, {Li}, {Lian}, {Liang}, {Ling}, {Liu}, {Liu}, {Liu}, {Liu}, {Liu}, {Lu}, {L{\"u}}, {Luo}, {Ma}, {Ma}, {Mao}, {Mao}, {McHugh}, {Meidinger}, {Nandra}, {Osborne}, {Pan}, {Pan}, {Ravasio}, {Rau}, {Rea}, {Rehman}, {Sanders}, {Santovincenzo}, {Song}, {Su}, {Sun}, {Sun}, {Sun}, {Tan}, {Tang}, {Tao}, {Tong}, {Wang}, {Wang}, {Wang}, {Wang}, {Wang}, {Wang}, {Wang}, {Wang}, {Wang}, {Wei}, {Willingale}, {Xiong}, {Xu}, {Xu}, {Xu}, {Xu}, {Xu}, {Xue}, {Xue}, {Yan}, {Yang}, {Yang}, {Yang}, {Yang}, {Yu}, {Zhang}, {Zhang}, {Zhang}, {Zhang}, {Zhang}, {Zhang}, {Zhang}, {Zhang}, {Zhang}, {Zhao}, {Zhao}, {Zhao}, {Zhao}, {Zhou}, {Zhou}, {Zhu}, {Zhu}, \& {Zuo}}]{Liu+25}
{Liu}, Y., {Sun}, H., {Xu}, D., {et~al.} 2025, Nature Astronomy, \dodoi{10.1038/s41550-024-02449-8}

\bibitem[{{Lucy}(1991)}]{Lucy91}
{Lucy}, L.~B. 1991, \apj, 383, 308, \dodoi{10.1086/170787}

\bibitem[{{Magnier} \& {Cuillandre}(2004)}]{elixir+04}
{Magnier}, E.~A., \& {Cuillandre}, J.~C. 2004, \pasp, 116, 449, \dodoi{10.1086/420756}

\bibitem[{{Maguire} {et~al.}(2016){Maguire}, {Taubenberger}, {Sullivan}, \& {Mazzali}}]{Maguire+16}
{Maguire}, K., {Taubenberger}, S., {Sullivan}, M., \& {Mazzali}, P.~A. 2016, \mnras, 457, 3254, \dodoi{10.1093/mnras/stv2991}

\bibitem[{{Margutti} {et~al.}(2019){Margutti}, {Metzger}, {Chornock}, {Vurm}, {Roth}, {Grefenstette}, {Savchenko}, {Cartier}, {Steiner}, {Terreran}, {Margalit}, {Migliori}, {Milisavljevic}, {Alexander}, {Bietenholz}, {Blanchard}, {Bozzo}, {Brethauer}, {Chilingarian}, {Coppejans}, {Ducci}, {Ferrigno}, {Fong}, {G{\"o}tz}, {Guidorzi}, {Hajela}, {Hurley}, {Kuulkers}, {Laurent}, {Mereghetti}, {Nicholl}, {Patnaude}, {Ubertini}, {Banovetz}, {Bartel}, {Berger}, {Coughlin}, {Eftekhari}, {Frederiks}, {Kozlova}, {Laskar}, {Svinkin}, {Drout}, {MacFadyen}, \& {Paterson}}]{Margutti+19}
{Margutti}, R., {Metzger}, B.~D., {Chornock}, R., {et~al.} 2019, \apj, 872, 18, \dodoi{10.3847/1538-4357/aafa01}

\bibitem[{{Matheson} {et~al.}(2003){Matheson}, {Garnavich}, {Stanek}, {Bersier}, {Holland}, {Krisciunas}, {Caldwell}, {Berlind}, {Bloom}, {Bolte}, {Bonanos}, {Brown}, {Brown}, {Calkins}, {Challis}, {Chornock}, {Echevarria}, {Eisenstein}, {Everett}, {Filippenko}, {Flint}, {Foley}, {Freedman}, {Hamuy}, {Harding}, {Hathi}, {Hicken}, {Hoopes}, {Impey}, {Jannuzi}, {Jansen}, {Jha}, {Kaluzny}, {Kannappan}, {Kirshner}, {Latham}, {Lee}, {Leonard}, {Li}, {Luhman}, {Martini}, {Mathis}, {Maza}, {Megeath}, {Miller}, {Minniti}, {Olszewski}, {Papenkova}, {Phillips}, {Pindor}, {Sasselov}, {Schild}, {Schweiker}, {Spahr}, {Thomas-Osip}, {Thompson}, {Weisz}, {Windhorst}, \& {Zaritsky}}]{Matheson+03}
{Matheson}, T., {Garnavich}, P.~M., {Stanek}, K.~Z., {et~al.} 2003, \apj, 599, 394, \dodoi{10.1086/379228}

\bibitem[{{Matzner} \& {McKee}(1999)}]{matznermckee99}
{Matzner}, C.~D., \& {McKee}, C.~F. 1999, \apjl, 526, L109, \dodoi{10.1086/312376}

\bibitem[{{Meikle} {et~al.}(1989){Meikle}, {Allen}, {Spyromilio}, \& {Varani}}]{Meikle89}
{Meikle}, W.~P.~S., {Allen}, D.~A., {Spyromilio}, J., \& {Varani}, G.~F. 1989, \mnras, 238, 193, \dodoi{10.1093/mnras/238.1.193}

\bibitem[{{Metzger}(2019)}]{metzger19}
{Metzger}, B.~D. 2019, Living Reviews in Relativity, 23, 1, \dodoi{10.1007/s41114-019-0024-0}

\bibitem[{{Mirabal} {et~al.}(2006){Mirabal}, {Halpern}, {An}, {Thorstensen}, \& {Terndrup}}]{Mirabal+06}
{Mirabal}, N., {Halpern}, J.~P., {An}, D., {Thorstensen}, J.~R., \& {Terndrup}, D.~M. 2006, \apjl, 643, L99, \dodoi{10.1086/505177}

\bibitem[{{Modjaz} {et~al.}(2016){Modjaz}, {Liu}, {Bianco}, \& {Graur}}]{Modjaz+16}
{Modjaz}, M., {Liu}, Y.~Q., {Bianco}, F.~B., \& {Graur}, O. 2016, \apj, 832, 108, \dodoi{10.3847/0004-637X/832/2/108}

\bibitem[{{Modjaz} {et~al.}(2006){Modjaz}, {Stanek}, {Garnavich}, {Berlind}, {Blondin}, {Brown}, {Calkins}, {Challis}, {Diamond-Stanic}, {Hao}, {Hicken}, {Kirshner}, \& {Prieto}}]{Modjaz+06}
{Modjaz}, M., {Stanek}, K.~Z., {Garnavich}, P.~M., {et~al.} 2006, \apjl, 645, L21, \dodoi{10.1086/505906}

\bibitem[{{Modjaz} {et~al.}(2009){Modjaz}, {Li}, {Butler}, {Chornock}, {Perley}, {Blondin}, {Bloom}, {Filippenko}, {Kirshner}, {Kocevski}, {Poznanski}, {Hicken}, {Foley}, {Stringfellow}, {Berlind}, {Barrado y Navascues}, {Blake}, {Bouy}, {Brown}, {Challis}, {Chen}, {de Vries}, {Dufour}, {Falco}, {Friedman}, {Ganeshalingam}, {Garnavich}, {Holden}, {Illingworth}, {Lee}, {Liebert}, {Marion}, {Olivier}, {Prochaska}, {Silverman}, {Smith}, {Starr}, {Steele}, {Stockton}, {Williams}, \& {Wood-Vasey}}]{Modjaz+09}
{Modjaz}, M., {Li}, W., {Butler}, N., {et~al.} 2009, \apj, 702, 226, \dodoi{10.1088/0004-637X/702/1/226}

\bibitem[{{Modjaz} {et~al.}(2014){Modjaz}, {Blondin}, {Kirshner}, {Matheson}, {Berlind}, {Bianco}, {Calkins}, {Challis}, {Garnavich}, {Hicken}, {Jha}, {Liu}, \& {Marion}}]{Modjaz+14}
{Modjaz}, M., {Blondin}, S., {Kirshner}, R.~P., {et~al.} 2014, \aj, 147, 99, \dodoi{10.1088/0004-6256/147/5/99}

\bibitem[{{Nagy}(2018)}]{Nagy2018}
{Nagy}, A.~P. 2018, \apj, 862, 143, \dodoi{10.3847/1538-4357/aace56}

\bibitem[{{Niblett} {et~al.}(2025){Niblett}, {Fryer}, \& {Fryer}}]{Niblett2025}
{Niblett}, A.~E., {Fryer}, D.~A., \& {Fryer}, C.~L. 2025, arXiv e-prints, arXiv:2501.15702, \dodoi{10.48550/arXiv.2501.15702}

\bibitem[{{Nicholl} {et~al.}(2023){Nicholl}, {Srivastav}, {Fulton}, {Gomez}, {Huber}, {Oates}, {Ramsden}, {Rhodes}, {Smartt}, {Smith}, {Aamer}, {Anderson}, {Bauer}, {Berger}, {de Boer}, {Chambers}, {Charalampopoulos}, {Chen}, {Fender}, {Fraser}, {Gao}, {Green}, {Galbany}, {Gompertz}, {Gromadzki}, {Guti{\'e}rrez}, {Howell}, {Inserra}, {Jonker}, {Kopsacheili}, {Lowe}, {Magnier}, {McCully}, {McGee}, {Moore}, {M{\"u}ller-Bravo}, {Newsome}, {Gonzalez}, {Pellegrino}, {Pessi}, {Pursiainen}, {Rest}, {Ridley}, {Shappee}, {Sheng}, {Smith}, {Terreran}, {Tucker}, {Vink{\'o}}, {Wainscoat}, {Wiseman}, \& {Young}}]{Nicholl23}
{Nicholl}, M., {Srivastav}, S., {Fulton}, M.~D., {et~al.} 2023, \apjl, 954, L28, \dodoi{10.3847/2041-8213/acf0ba}

\bibitem[{{Novara} {et~al.}(2020){Novara}, {Esposito}, {Tiengo}, {Vianello}, {Salvaterra}, {Belfiore}, {De Luca}, {D'Avanzo}, {Greiner}, {Scodeggio}, {Rosen}, {Delvaux}, {Pian}, {Campana}, {Lisini}, {Mereghetti}, \& {Israel}}]{Novara+20}
{Novara}, G., {Esposito}, P., {Tiengo}, A., {et~al.} 2020, \apj, 898, 37, \dodoi{10.3847/1538-4357/ab98f8}

\bibitem[{{O'Connor} {et~al.}(2025){O'Connor}, {Pasham}, {Andreoni}, {Hare}, {Beniamini}, {Troja}, {Ricci}, {Dobie}, {Chakraborty}, {Ng}, {Klingler}, {Karambelkar}, {Rose}, {Schulze}, {Ryan}, {Dichiara}, {Monageng}, {Buckley}, {Hu}, {Srinivasaragavan}, {Bruni}, {Cabrera}, {Cenko}, {van Eerten}, {Freeburn}, {Hammerstein}, {Kasliwal}, {Kouveliotou}, {Kunnumkai}, {Leung}, {Lien}, {Palmese}, \& {Sakamoto}}]{O'Connor+25}
{O'Connor}, B., {Pasham}, D., {Andreoni}, I., {et~al.} 2025, \apjl, 979, L30, \dodoi{10.3847/2041-8213/ada7f5}

\bibitem[{{Olivares E.} {et~al.}(2012){Olivares E.}, {Greiner}, {Schady}, {Rau}, {Klose}, {Kr{\"u}hler}, {Afonso}, {Updike}, {Nardini}, {Filgas}, {Nicuesa Guelbenzu}, {Clemens}, {Elliott}, {Kann}, {Rossi}, \& {Sudilovsky}}]{Olivares+12}
{Olivares E.}, F., {Greiner}, J., {Schady}, P., {et~al.} 2012, \aap, 539, A76, \dodoi{10.1051/0004-6361/201117929}

\bibitem[{{Owocki} \& {Rybicki}(1984)}]{Owocki+84}
{Owocki}, S.~P., \& {Rybicki}, G.~B. 1984, \apj, 284, 337, \dodoi{10.1086/162412}

\bibitem[{{Patat} {et~al.}(2001){Patat}, {Cappellaro}, {Danziger}, {Mazzali}, {Sollerman}, {Augusteijn}, {Brewer}, {Doublier}, {Gonzalez}, {Hainaut}, {Lidman}, {Leibundgut}, {Nomoto}, {Nakamura}, {Spyromilio}, {Rizzi}, {Turatto}, {Walsh}, {Galama}, {van Paradijs}, {Kouveliotou}, {Vreeswijk}, {Frontera}, {Masetti}, {Palazzi}, \& {Pian}}]{Patat+01}
{Patat}, F., {Cappellaro}, E., {Danziger}, J., {et~al.} 2001, \apj, 555, 900, \dodoi{10.1086/321526}

\bibitem[{{Perley} {et~al.}(2014){Perley}, {Cenko}, {Corsi}, {Tanvir}, {Levan}, {Kann}, {Sonbas}, {Wiersema}, {Zheng}, {Zhao}, {Bai}, {Bremer}, {Castro-Tirado}, {Chang}, {Clubb}, {Frail}, {Fruchter}, {G{\"o}{\u{g}}{\"u}{\c{s}}}, {Greiner}, {G{\"u}ver}, {Horesh}, {Filippenko}, {Klose}, {Mao}, {Morgan}, {Pozanenko}, {Schmidl}, {Stecklum}, {Tanga}, {Volnova}, {Volvach}, {Wang}, {Winters}, \& {Xin}}]{Perley+14}
{Perley}, D.~A., {Cenko}, S.~B., {Corsi}, A., {et~al.} 2014, \apj, 781, 37, \dodoi{10.1088/0004-637X/781/1/37}

\bibitem[{{Perley} {et~al.}(2019){Perley}, {Mazzali}, {Yan}, {Cenko}, {Gezari}, {Taggart}, {Blagorodnova}, {Fremling}, {Mockler}, {Singh}, {Tominaga}, {Tanaka}, {Watson}, {Ahumada}, {Anupama}, {Ashall}, {Becerra}, {Bersier}, {Bhalerao}, {Bloom}, {Butler}, {Copperwheat}, {Coughlin}, {De}, {Drake}, {Duev}, {Frederick}, {Gonz{\'a}lez}, {Goobar}, {Heida}, {Ho}, {Horst}, {Hung}, {Itoh}, {Jencson}, {Kasliwal}, {Kawai}, {Khanam}, {Kulkarni}, {Kumar}, {Kumar}, {Kutyrev}, {Lee}, {Maeda}, {Mahabal}, {Murata}, {Neill}, {Ngeow}, {Penprase}, {Pian}, {Quimby}, {Ramirez-Ruiz}, {Richer}, {Rom{\'a}n-Z{\'u}{\~n}iga}, {Sahu}, {Srivastav}, {Socia}, {Sollerman}, {Tachibana}, {Taddia}, {Tinyanont}, {Troja}, {Ward}, {Wee}, \& {Yu}}]{Perley+19}
{Perley}, D.~A., {Mazzali}, P.~A., {Yan}, L., {et~al.} 2019, \mnras, 484, 1031, \dodoi{10.1093/mnras/sty3420}

\bibitem[{{Pian} {et~al.}(2000){Pian}, {Amati}, {Antonelli}, {Butler}, {Costa}, {Cusumano}, {Danziger}, {Feroci}, {Fiore}, {Frontera}, {Giommi}, {Masetti}, {Muller}, {Nicastro}, {Oosterbroek}, {Orlandini}, {Owens}, {Palazzi}, {Parmar}, {Piro}, {in't Zand}, {Castro-Tirado}, {Coletta}, {Dal Fiume}, {Del Sordo}, {Heise}, {Soffitta}, \& {Torroni}}]{Pian+00}
{Pian}, E., {Amati}, L., {Antonelli}, L.~A., {et~al.} 2000, \apj, 536, 778, \dodoi{10.1086/308978}

\bibitem[{{Pian} {et~al.}(2006){Pian}, {Mazzali}, {Masetti}, {Ferrero}, {Klose}, {Palazzi}, {Ramirez-Ruiz}, {Woosley}, {Kouveliotou}, {Deng}, {Filippenko}, {Foley}, {Fynbo}, {Kann}, {Li}, {Hjorth}, {Nomoto}, {Patat}, {Sauer}, {Sollerman}, {Vreeswijk}, {Guenther}, {Levan}, {O'Brien}, {Tanvir}, {Wijers}, {Dumas}, {Hainaut}, {Wong}, {Baade}, {Wang}, {Amati}, {Cappellaro}, {Castro-Tirado}, {Ellison}, {Frontera}, {Fruchter}, {Greiner}, {Kawabata}, {Ledoux}, {Maeda}, {M{\o}ller}, {Nicastro}, {Rol}, \& {Starling}}]{Pian+06}
{Pian}, E., {Mazzali}, P.~A., {Masetti}, N., {et~al.} 2006, \nat, 442, 1011, \dodoi{10.1038/nature05082}

\bibitem[{{Planck Collaboration} {et~al.}(2020){Planck Collaboration}, {Aghanim}, {Akrami}, {Ashdown}, {Aumont}, {Baccigalupi}, {Ballardini}, {Banday}, {Barreiro}, {Bartolo}, {Basak}, {Battye}, {Benabed}, {Bernard}, {Bersanelli}, {Bielewicz}, {Bock}, {Bond}, {Borrill}, {Bouchet}, {Boulanger}, {Bucher}, {Burigana}, {Butler}, {Calabrese}, {Cardoso}, {Carron}, {Challinor}, {Chiang}, {Chluba}, {Colombo}, {Combet}, {Contreras}, {Crill}, {Cuttaia}, {de Bernardis}, {de Zotti}, {Delabrouille}, {Delouis}, {Di Valentino}, {Diego}, {Dor{\'e}}, {Douspis}, {Ducout}, {Dupac}, {Dusini}, {Efstathiou}, {Elsner}, {En{\ss}lin}, {Eriksen}, {Fantaye}, {Farhang}, {Fergusson}, {Fernandez-Cobos}, {Finelli}, {Forastieri}, {Frailis}, {Fraisse}, {Franceschi}, {Frolov}, {Galeotta}, {Galli}, {Ganga}, {G{\'e}nova-Santos}, {Gerbino}, {Ghosh}, {Gonz{\'a}lez-Nuevo}, {G{\'o}rski}, {Gratton}, {Gruppuso}, {Gudmundsson}, {Hamann}, {Handley}, {Hansen}, {Herranz}, {Hildebrandt}, {Hivon}, {Huang}, {Jaffe}, {Jones}, {Karakci}, {Keih{\"a}nen},
  {Keskitalo}, {Kiiveri}, {Kim}, {Kisner}, {Knox}, {Krachmalnicoff}, {Kunz}, {Kurki-Suonio}, {Lagache}, {Lamarre}, {Lasenby}, {Lattanzi}, {Lawrence}, {Le Jeune}, {Lemos}, {Lesgourgues}, {Levrier}, {Lewis}, {Liguori}, {Lilje}, {Lilley}, {Lindholm}, {L{\'o}pez-Caniego}, {Lubin}, {Ma}, {Mac{\'\i}as-P{\'e}rez}, {Maggio}, {Maino}, {Mandolesi}, {Mangilli}, {Marcos-Caballero}, {Maris}, {Martin}, {Martinelli}, {Mart{\'\i}nez-Gonz{\'a}lez}, {Matarrese}, {Mauri}, {McEwen}, {Meinhold}, {Melchiorri}, {Mennella}, {Migliaccio}, {Millea}, {Mitra}, {Miville-Desch{\^e}nes}, {Molinari}, {Montier}, {Morgante}, {Moss}, {Natoli}, {N{\o}rgaard-Nielsen}, {Pagano}, {Paoletti}, {Partridge}, {Patanchon}, {Peiris}, {Perrotta}, {Pettorino}, {Piacentini}, {Polastri}, {Polenta}, {Puget}, {Rachen}, {Reinecke}, {Remazeilles}, {Renzi}, {Rocha}, {Rosset}, {Roudier}, {Rubi{\~n}o-Mart{\'\i}n}, {Ruiz-Granados}, {Salvati}, {Sandri}, {Savelainen}, {Scott}, {Shellard}, {Sirignano}, {Sirri}, {Spencer}, {Sunyaev}, {Suur-Uski}, {Tauber}, {Tavagnacco},
  {Tenti}, {Toffolatti}, {Tomasi}, {Trombetti}, {Valenziano}, {Valiviita}, {Van Tent}, {Vibert}, {Vielva}, {Villa}, {Vittorio}, {Wandelt}, {Wehus}, {White}, {White}, {Zacchei}, \& {Zonca}}]{Planck20}
{Planck Collaboration}, {Aghanim}, N., {Akrami}, Y., {et~al.} 2020, \aap, 641, A6, \dodoi{10.1051/0004-6361/201833910}

\bibitem[{{Prentice} {et~al.}(2018){Prentice}, {Maguire}, {Smartt}, {Magee}, {Schady}, {Sim}, {Chen}, {Clark}, {Colin}, {Fulton}, {McBrien}, {O'Neill}, {Smith}, {Ashall}, {Chambers}, {Denneau}, {Flewelling}, {Heinze}, {Holoien}, {Huber}, {Kochanek}, {Mazzali}, {Prieto}, {Rest}, {Shappee}, {Stalder}, {Stanek}, {Stritzinger}, {Thompson}, \& {Tonry}}]{Prentice+18}
{Prentice}, S.~J., {Maguire}, K., {Smartt}, S.~J., {et~al.} 2018, \apjl, 865, L3, \dodoi{10.3847/2041-8213/aadd90}

\bibitem[{{Prochaska} {et~al.}(2020){Prochaska}, {Hennawi}, {Westfall}, {Cooke}, {Wang}, {Hsyu}, {Davies}, {Farina}, \& {Pelliccia}}]{PypeIt}
{Prochaska}, J., {Hennawi}, J., {Westfall}, K., {et~al.} 2020, The Journal of Open Source Software, 5, 2308, \dodoi{10.21105/joss.02308}

\bibitem[{{Puls} {et~al.}(2008){Puls}, {Vink}, \& {Najarro}}]{Puls+08}
{Puls}, J., {Vink}, J.~S., \& {Najarro}, F. 2008, \aapr, 16, 209, \dodoi{10.1007/s00159-008-0015-8}

\bibitem[{{Quataert} {et~al.}(2016){Quataert}, {Fern{\'a}ndez}, {Kasen}, {Klion}, \& {Paxton}}]{Quataert+16}
{Quataert}, E., {Fern{\'a}ndez}, R., {Kasen}, D., {Klion}, H., \& {Paxton}, B. 2016, \mnras, 458, 1214, \dodoi{10.1093/mnras/stw365}

\bibitem[{{Quirola-V{\'a}squez} {et~al.}(2022){Quirola-V{\'a}squez}, {Bauer}, {Jonker}, {Brandt}, {Yang}, {Levan}, {Xue}, {Eappachen}, {Zheng}, \& {Luo}}]{QuirolaVazquez+22}
{Quirola-V{\'a}squez}, J., {Bauer}, F.~E., {Jonker}, P.~G., {et~al.} 2022, \aap, 663, A168, \dodoi{10.1051/0004-6361/202243047}

\bibitem[{{Quirola-V{\'a}squez} {et~al.}(2023){Quirola-V{\'a}squez}, {Bauer}, {Jonker}, {Brandt}, {Yang}, {Levan}, {Xue}, {Eappachen}, {Camacho}, {Ravasio}, {Zheng}, \& {Luo}}]{QuirolaVasquez+23}
---. 2023, \aap, 675, A44, \dodoi{10.1051/0004-6361/202345912}

\bibitem[{{Rastinejad} {et~al.}(2022){Rastinejad}, {Gompertz}, {Levan}, {Fong}, {Nicholl}, {Lamb}, {Malesani}, {Nugent}, {Oates}, {Tanvir}, {de Ugarte Postigo}, {Kilpatrick}, {Moore}, {Metzger}, {Ravasio}, {Rossi}, {Schroeder}, {Jencson}, {Sand}, {Smith}, {Ag{\"u}{\'\i} Fern{\'a}ndez}, {Berger}, {Blanchard}, {Chornock}, {Cobb}, {De Pasquale}, {Fynbo}, {Izzo}, {Kann}, {Laskar}, {Marini}, {Paterson}, {Escorial}, {Sears}, \& {Th{\"o}ne}}]{Rastinejad+22}
{Rastinejad}, J.~C., {Gompertz}, B.~P., {Levan}, A.~J., {et~al.} 2022, \nat, 612, 223, \dodoi{10.1038/s41586-022-05390-w}

\bibitem[{{Rastinejad} {et~al.}(2024){Rastinejad}, {Fong}, {Levan}, {Tanvir}, {Kilpatrick}, {Fruchter}, {Anand}, {Bhirombhakdi}, {Covino}, {Fynbo}, {Halevi}, {Hartmann}, {Heintz}, {Izzo}, {Jakobsson}, {Kangas}, {Lamb}, {Malesani}, {Melandri}, {Metzger}, {Milvang-Jensen}, {Pian}, {Pugliese}, {Rossi}, {Siegel}, {Singh}, \& {Stratta}}]{Rastinejad+24}
{Rastinejad}, J.~C., {Fong}, W., {Levan}, A.~J., {et~al.} 2024, \apj, 968, 14, \dodoi{10.3847/1538-4357/ad409c}

\bibitem[{{Ravasio} {et~al.}(2025){Ravasio}, {Burns}, {Wilson-Hodge}, {Jonker}, \& {Fermi-GBM Team}}]{EP250108a_fermi}
{Ravasio}, M.~E., {Burns}, E., {Wilson-Hodge}, C., {Jonker}, P.~G., \& {Fermi-GBM Team}. 2025, GRB Coordinates Network, 39146, 1

\bibitem[{{Reichert} {et~al.}(2023){Reichert}, {Obergaulinger}, {Aloy}, {Gabler}, \& et~al.}]{Reichert2023}
{Reichert}, M., {Obergaulinger}, M., {Aloy}, M.~{\'A}., {Gabler}, M., \& et~al. 2023, \mnras, 518, 1557, \dodoi{10.1093/mnras/stac3185}

\bibitem[{{Rest} {et~al.}(2005){Rest}, {Stubbs}, {Becker}, {Miknaitis}, {Miceli}, {Covarrubias}, {Hawley}, {Smith}, {Suntzeff}, {Olsen}, {Prieto}, {Hiriart}, {Welch}, {Cook}, {Nikolaev}, {Huber}, {Prochtor}, {Clocchiatti}, {Minniti}, {Garg}, {Challis}, {Keller}, \& {Schmidt}}]{Rest+05}
{Rest}, A., {Stubbs}, C., {Becker}, A.~C., {et~al.} 2005, \apj, 634, 1103, \dodoi{10.1086/497060}

\bibitem[{{Rodr{\'\i}guez} {et~al.}(2023){Rodr{\'\i}guez}, {Maoz}, \& {Nakar}}]{Rodriguez+23}
{Rodr{\'\i}guez}, {\'O}., {Maoz}, D., \& {Nakar}, E. 2023, \apj, 955, 71, \dodoi{10.3847/1538-4357/ace2bd}

\bibitem[{{Ror} {et~al.}(2025){Ror}, {Gupta}, {Kiran}, {Pandey}, \& {Mishra}}]{ep250108a_dfot}
{Ror}, A.~K., {Gupta}, A., {Kiran}, {Pandey}, S.~B., \& {Mishra}, K. 2025, GRB Coordinates Network, 39002, 1

\bibitem[{{Sakamoto} {et~al.}(2005){Sakamoto}, {Lamb}, {Kawai}, {Yoshida}, {Graziani}, {Fenimore}, {Donaghy}, {Matsuoka}, {Suzuki}, {Ricker}, {Atteia}, {Shirasaki}, {Tamagawa}, {Torii}, {Galassi}, {Doty}, {Vanderspek}, {Crew}, {Villasenor}, {Butler}, {Prigozhin}, {Jernigan}, {Barraud}, {Boer}, {Dezalay}, {Olive}, {Hurley}, {Levine}, {Monnelly}, {Martel}, {Morgan}, {Woosley}, {Cline}, {Braga}, {Manchanda}, {Pizzichini}, {Takagishi}, \& {Yamauchi}}]{Sakamoto+05}
{Sakamoto}, T., {Lamb}, D.~Q., {Kawai}, N., {et~al.} 2005, \apj, 629, 311, \dodoi{10.1086/431235}

\bibitem[{{Santos} {et~al.}(2024){Santos}, {Kilpatrick}, {Bom}, {Darc}, {Herpich}, {Lacerda}, {Sartori}, {Alvarez-Candal}, {Mendes de Oliveira}, {Kanaan}, {Ribeiro}, \& {Schoenell}}]{Santos+24}
{Santos}, A., {Kilpatrick}, C.~D., {Bom}, C.~R., {et~al.} 2024, \mnras, 529, 59, \dodoi{10.1093/mnras/stae466}

\bibitem[{{Sarin} {et~al.}(2024){Sarin}, {H{\"u}bner}, {Omand}, {Setzer}, {Schulze}, {Adhikari}, {Sagu{\'e}s-Carracedo}, {Galaudage}, {Wallace}, {Lamb}, \& {Lin}}]{Sarin+24}
{Sarin}, N., {H{\"u}bner}, M., {Omand}, C. M.~B., {et~al.} 2024, \mnras, 531, 1203, \dodoi{10.1093/mnras/stae1238}

\bibitem[{{Schlafly} \& {Finkbeiner}(2011)}]{SchlaflyFinkbeiner11}
{Schlafly}, E.~F., \& {Finkbeiner}, D.~P. 2011, \apj, 737, 103, \dodoi{10.1088/0004-637X/737/2/103}

\bibitem[{{Science Software Branch at STScI}(2012)}]{Pyraf2012}
{Science Software Branch at STScI}. 2012, {PyRAF: Python alternative for IRAF}, Astrophysics Source Code Library, record ascl:1207.011

\bibitem[{{Shahbandeh} {et~al.}(2022){Shahbandeh}, {Hsiao}, {Ashall}, {Teffs}, {Hoeflich}, {Morrell}, {Phillips}, {Anderson}, {Baron}, {Burns}, {Contreras}, {Davis}, {Diamond}, {Folatelli}, {Galbany}, {Gall}, {Hachinger}, {Holmbo}, {Karamehmetoglu}, {Kasliwal}, {Kirshner}, {Krisciunas}, {Kumar}, {Lu}, {Marion}, {Mazzali}, {Piro}, {Sand}, {Stritzinger}, {Suntzeff}, {Taddia}, \& {Uddin}}]{Shahbandeh+22}
{Shahbandeh}, M., {Hsiao}, E.~Y., {Ashall}, C., {et~al.} 2022, \apj, 925, 175, \dodoi{10.3847/1538-4357/ac4030}

\bibitem[{{Shahbandeh} {et~al.}(2024){Shahbandeh}, {Ashall}, {Hoeflich}, {Baron}, {Fox}, {Mera}, {DerKacy}, {Stritzinger}, {Shappee}, {Law}, {Morrison}, {Pauly}, {Pierel}, {Medler}, {Andrews}, {Baade}, {Bostroem}, {Brown}, {Burns}, {Burrow}, {Cikota}, {Cross}, {Davis}, {de Jaeger}, {Do}, {Dong}, {Hsiao}, {Dominguez}, {Galbany}, {Janzen}, {Jencson}, {Hoang}, {Karamehmetoglu}, {Khaghani}, {Krisciunas}, {Kumar}, {Lu}, {Mazzali}, {Morrell}, {Patat}, {Pearson}, {Pfeffer}, {Wang}, {Yang}, {Cai}, {Camacho-Neves}, {Elias-Rosa}, {Lundquist}, {Maund}, {Phillips}, {Rest}, {Retamal}, {Stangl}, {Shrestha}, {Stevens}, {Suntzeff}, {Telesco}, {Tucker}, {Foley}, {Jha}, {Kwok}, {Larison}, {LeBaron}, {Moran}, {Rho}, {Salmaso}, {Schmidt}, \& {Tinyanont}}]{Shahbandeh+24}
{Shahbandeh}, M., {Ashall}, C., {Hoeflich}, P., {et~al.} 2024, arXiv e-prints, arXiv:2401.14474, \dodoi{10.48550/arXiv.2401.14474}

\bibitem[{{Siegel} {et~al.}(2019){Siegel}, {Barnes}, \& {Metzger}}]{SiegelBarnesMetzger2019}
{Siegel}, D.~M., {Barnes}, J., \& {Metzger}, B.~D. 2019, \nat, 569, 241, \dodoi{10.1038/s41586-019-1136-0}

\bibitem[{{Siegel} \& {Metzger}(2017)}]{2017PhRvL.119w1102S}
{Siegel}, D.~M., \& {Metzger}, B.~D. 2017, \prl, 119, 231102, \dodoi{10.1103/PhysRevLett.119.231102}

\bibitem[{{Skrutskie} {et~al.}(2006){Skrutskie}, {Cutri}, {Stiening}, {Weinberg}, {Schneider}, {Carpenter}, {Beichman}, {Capps}, {Chester}, {Elias}, {Huchra}, {Liebert}, {Lonsdale}, {Monet}, {Price}, {Seitzer}, {Jarrett}, {Kirkpatrick}, {Gizis}, {Howard}, {Evans}, {Fowler}, {Fullmer}, {Hurt}, {Light}, {Kopan}, {Marsh}, {McCallon}, {Tam}, {Van Dyk}, \& {Wheelock}}]{2MASS}
{Skrutskie}, M.~F., {Cutri}, R.~M., {Stiening}, R., {et~al.} 2006, \aj, 131, 1163, \dodoi{10.1086/498708}

\bibitem[{{Soderberg} {et~al.}(2006){Soderberg}, {Berger}, {Kasliwal}, {Frail}, {Price}, {Schmidt}, {Kulkarni}, {Fox}, {Cenko}, {Gal-Yam}, {Nakar}, \& {Roth}}]{Soderberg+06}
{Soderberg}, A.~M., {Berger}, E., {Kasliwal}, M., {et~al.} 2006, \apj, 650, 261, \dodoi{10.1086/506429}

\bibitem[{{Soderberg} {et~al.}(2008){Soderberg}, {Berger}, {Page}, {Schady}, {Parrent}, {Pooley}, {Wang}, {Ofek}, {Cucchiara}, {Rau}, {Waxman}, {Simon}, {Bock}, {Milne}, {Page}, {Barentine}, {Barthelmy}, {Beardmore}, {Bietenholz}, {Brown}, {Burrows}, {Burrows}, {Byrngelson}, {Cenko}, {Chandra}, {Cummings}, {Fox}, {Gal-Yam}, {Gehrels}, {Immler}, {Kasliwal}, {Kong}, {Krimm}, {Kulkarni}, {Maccarone}, {M{\'e}sz{\'a}ros}, {Nakar}, {O'Brien}, {Overzier}, {de Pasquale}, {Racusin}, {Rea}, \& {York}}]{Soderberg+08}
{Soderberg}, A.~M., {Berger}, E., {Page}, K.~L., {et~al.} 2008, \nat, 453, 469, \dodoi{10.1038/nature06997}

\bibitem[{{Sollerman} {et~al.}(1998){Sollerman}, {Cumming}, \& {Lundqvist}}]{Sollerman98}
{Sollerman}, J., {Cumming}, R.~J., \& {Lundqvist}, P. 1998, \apj, 493, 933, \dodoi{10.1086/305163}

\bibitem[{{Song} {et~al.}(2025){Song}, {Li}, {Wang}, {Mao}, {Lu}, \& {Bai}}]{ep250108a_gmg}
{Song}, F.~F., {Li}, R.~Z., {Wang}, B.~T., {et~al.} 2025, GRB Coordinates Network, 38972, 1

\bibitem[{{Srinivasaragavan} {et~al.}(2024){Srinivasaragavan}, {Yang}, {Anand}, {Sollerman}, {Ho}, {Corsi}, {Cenko}, {Perley}, {Schulze}, {Sanchez-Fleming}, {Pope}, {Sarin}, {Omand}, {Das}, {Fremling}, {Andreoni}, {Bruch}, {Burdge}, {De}, {Gal-Yam}, {Gangopadhyay}, {Graham}, {Jencson}, {Karambelkar}, {Kasliwal}, {Kulkarni}, {Martikainen}, {Sharma}, {Tzanidakis}, {Yan}, {Yao}, {Bellm}, {Groom}, {Masci}, {Nir}, {Purdum}, {Smith}, \& {Sravan}}]{Srinivasaragavan+24}
{Srinivasaragavan}, G.~P., {Yang}, S., {Anand}, S., {et~al.} 2024, \apj, 976, 71, \dodoi{10.3847/1538-4357/ad7fde}

\bibitem[{{Srinivasaragavan} {et~al.}(2025){Srinivasaragavan}, {Hamidani}, {Schroeder}, {Sarin}, {Ho}, {Piro}, {Cenko}, {Anand}, {Sollerman}, {Perley}, {Maeda}, {O'Connor}, {Kuncarayakti}, {Miller}, {Ahumada}, {Vail}, {Duffell}, {Ghosh Dastidar}, {Andreoni}, {Bochenek}, {Brennan}, {Carney}, {Chen}, {Freeburn}, {Gal-Yam}, {Jacobson-Gal{\'a}n}, {Kasliwal}, {Li}, {Li}, {Sravan}, \& {Warshofsky}}]{Srinivasaragavan+25}
{Srinivasaragavan}, G.~P., {Hamidani}, H., {Schroeder}, G., {et~al.} 2025, arXiv e-prints, arXiv:2504.17516, \dodoi{10.48550/arXiv.2504.17516}

\bibitem[{{Srivastav} {et~al.}(2025){Srivastav}, {Chen}, {Gillanders}, {Rhodes}, {Smartt}, {Huber}, {Aryan}, {Yang}, {Beri}, {Cooper}, {Nicholl}, {Smith}, {Stevance}, {Carotenuto}, {Chambers}, {Aamer}, {Angus}, {Fulton}, {Moore}, {Smith}, {Young}, {de Boer}, {Gao}, {Lin}, {Lowe}, {Magnier}, {Minguez}, {Pan}, \& {Wainscoat}}]{SCG+25}
{Srivastav}, S., {Chen}, T.~W., {Gillanders}, J.~H., {et~al.} 2025, \apjl, 978, L21, \dodoi{10.3847/2041-8213/ad9c75}

\bibitem[{{Sun} {et~al.}(2015){Sun}, {Zhang}, \& {Li}}]{Sun15}
{Sun}, H., {Zhang}, B., \& {Li}, Z. 2015, \apj, 812, 33, \dodoi{10.1088/0004-637X/812/1/33}

\bibitem[{{Sun} {et~al.}(2024){Sun}, {Li}, {Liu}, {Gao}, {Wang}, {Yuan}, {Zhang}, {Filippenko}, {Xu}, {An}, {Ai}, {Brink}, {Liu}, {Liu}, {Wang}, {Wu}, {Wu}, {Yang}, {Zhang}, {Zheng}, {Ahumada}, {Dai}, {Delaunay}, {Elias-Rosa}, {Benetti}, {Fu}, {Howell}, {Huang}, {Kasliwal}, {Karambelkar}, {Stein}, {Lei}, {Lian}, {Peng}, {Ridnaia}, {Svinkin}, {Wang}, {Wang}, {Wei}, {An}, {Andrews}, {Bai}, {Dai}, {Ehgamberdiev}, {Fan}, {Farah}, {Feng}, {Fynbo}, {Guo}, {Guo}, {Hu}, {Hu}, {Jiang}, {Jin}, {Li}, {Li}, {Li}, {Liang}, {Ling}, {Liu}, {Mao}, {McCully}, {Mirzaqulov}, {Newsome}, {Padilla Gonzalez}, {Pan}, {Terreran}, {Tinyanont}, {Wang}, {Wang}, {Wen}, {Xiang}, {Xue}, {Yang}, {Zhu}, {Cai}, {Castro-Tirado}, {Chen}, {Chen}, {Chen}, {Chen}, {Chen}, {Chen}, {Chen}, {Cheng}, {Cordier}, {Cui}, {Cui}, {Dai}, {Fan}, {Feng}, {Guan}, {Han}, {Hou}, {Hu}, {Huang}, {Huo}, {Jia}, {Jia}, {Jiang}, {Jin}, {Jin}, {Kuulkers}, {Li}, {Li}, {Li}, {Li}, {Li}, {Li}, {Li}, {Liu}, {Liu}, {Liu}, {Liu}, {Lu}, {Luo}, {Ma}, {Mao}, {Nandra},
  {O'Brien}, {Pan}, {Rau}, {Rea}, {Sanders}, {Song}, {Sun}, {Sun}, {Tan}, {Tang}, {Tao}, {Wang}, {Wang}, {Wang}, {Wang}, {Wang}, {Wang}, {Xiong}, {Xu}, {Xu}, {Xu}, {Xu}, {Xu}, {Xue}, {Xue}, {Yan}, {Yang}, {Yang}, {Yang}, {Zhang}, {Zhang}, {Zhang}, {Zhang}, {Zhang}, {Zhang}, {Zhang}, {Zhang}, {Zhang}, {Zhang}, {Zhao}, {Zhao}, {Zhao}, {Zhao}, {Zhou}, {Zhu}, \& {Zhu}}]{Sun+24}
{Sun}, H., {Li}, W.~X., {Liu}, L.~D., {et~al.} 2024, arXiv e-prints, arXiv:2410.02315, \dodoi{10.48550/arXiv.2410.02315}

\bibitem[{{Surman} {et~al.}(2006){Surman}, {McLaughlin}, \& {Hix}}]{2006ApJ...643.1057S}
{Surman}, R., {McLaughlin}, G.~C., \& {Hix}, W.~R. 2006, \apj, 643, 1057, \dodoi{10.1086/501116}

\bibitem[{{Taddia} {et~al.}(2019){Taddia}, {Sollerman}, {Fremling}, {Barbarino}, \& et~al.}]{Taddia2019}
{Taddia}, F., {Sollerman}, J., {Fremling}, C., {Barbarino}, C., \& et~al. 2019, \aap, 621, A71, \dodoi{10.1051/0004-6361/201834429}

\bibitem[{{Tinyanont} {et~al.}(2024){Tinyanont}, {Foley}, {Taggart}, {Davis}, {LeBaron}, {Andrews}, {Bustamante-Rosell}, {Camacho-Neves}, {Chornock}, {Coulter}, {Galbany}, {Jha}, {Kilpatrick}, {Kwok}, {Larison}, {Pierel}, {Siebert}, {Aldering}, {Auchettl}, {Bloom}, {Dhawan}, {Filippenko}, {French}, {Gagliano}, {Grayling}, {Howell}, {Jacobson-Gal{\'a}n}, {Jones}, {Le Saux}, {Macias}, {Mandel}, {McCully}, {Padilla Gonzalez}, {Rest}, {Rho}, {Rojas-Bravo}, {Skrutskie}, {Thorp}, {Wang}, \& {Ward}}]{Tinyanont+24}
{Tinyanont}, S., {Foley}, R.~J., {Taggart}, K., {et~al.} 2024, \pasp, 136, 014201, \dodoi{10.1088/1538-3873/ad1b39}

\bibitem[{{Tonry} {et~al.}(2018){Tonry}, {Denneau}, {Heinze}, {Stalder}, {Smith}, {Smartt}, {Stubbs}, {Weiland }, \& {Rest}}]{Tonry+18}
{Tonry}, J.~L., {Denneau}, L., {Heinze}, A.~N., {et~al.} 2018, \pasp, 130, 064505, \dodoi{10.1088/1538-3873/aabadf}

\bibitem[{{Troja} {et~al.}(2022){Troja}, {Fryer}, {O'Connor}, {Ryan}, {Dichiara}, {Kumar}, {Ito}, {Gupta}, {Wollaeger}, {Norris}, {Kawai}, {Butler}, {Aryan}, {Misra}, {Hosokawa}, {Murata}, {Niwano}, {Pandey}, {Kutyrev}, {van Eerten}, {Chase}, {Hu}, {Caballero-Garcia}, \& {Castro-Tirado}}]{Troja+22}
{Troja}, E., {Fryer}, C.~L., {O'Connor}, B., {et~al.} 2022, \nat, 612, 228, \dodoi{10.1038/s41586-022-05327-3}

\bibitem[{{Tucker} {et~al.}(2020){Tucker}, {Shappee}, {Vallely}, {Stanek}, {Prieto}, {Botyanszki}, {Kochanek}, {Anderson}, {Brown}, {Galbany}, {Holoien}, {Hsiao}, {Kumar}, {Kuncarayakti}, {Morrell}, {Phillips}, {Stritzinger}, \& {Thompson}}]{Tucker+20}
{Tucker}, M.~A., {Shappee}, B.~J., {Vallely}, P.~J., {et~al.} 2020, \mnras, 493, 1044, \dodoi{10.1093/mnras/stz3390}

\bibitem[{{van Dalen} {et~al.}(2025){van Dalen}, {Levan}, {Jonker}, {Malesani}, {Izzo}, {Sarin}, {Quirola-V{\'a}squez}, {S{\'a}nchez}, {de Ugarte Postigo}, {van Hoof}, {Torres}, {Schulze}, {Littlefair}, {Chrimes}, {Ravasio}, {Bauer}, {Martin-Carrillo}, {Fraser}, {van der Horst}, {Jakobsson}, {O'Brien}, {De Pasquale}, {Pugliese}, {Sollerman}, {Tanvir}, {Zafar}, {Anderson}, {Galbany}, {Gal-Yam}, {Gromadzki}, {M{\"u}ller-Bravo}, {Ragosta}, \& {Terwel}}]{vanDalen+24}
{van Dalen}, J. N.~D., {Levan}, A.~J., {Jonker}, P.~G., {et~al.} 2025, \apjl, 982, L47, \dodoi{10.3847/2041-8213/adbc7e}

\bibitem[{{van Dokkum}(2001)}]{van_Dokkum2001}
{van Dokkum}, P.~G. 2001, \pasp, 113, 1420, \dodoi{10.1086/323894}

\bibitem[{{Weaver} {et~al.}(1977){Weaver}, {McCray}, {Castor}, {Shapiro}, \& {Moore}}]{1977ApJ...218..377W}
{Weaver}, R., {McCray}, R., {Castor}, J., {Shapiro}, P., \& {Moore}, R. 1977, \apj, 218, 377, \dodoi{10.1086/155692}

\bibitem[{{Woosley} {et~al.}(2002){Woosley}, {Heger}, \& {Weaver}}]{2002RvMP...74.1015W}
{Woosley}, S.~E., {Heger}, A., \& {Weaver}, T.~A. 2002, Reviews of Modern Physics, 74, 1015, \dodoi{10.1103/RevModPhys.74.1015}

\bibitem[{{Xiang} {et~al.}(2021){Xiang}, {Wang}, {Lin}, {Mo}, {Lin}, {Burke}, {Hiramatsu}, {Hosseinzadeh}, {Howell}, {McCully}, {Valenti}, {Vink{\'o}}, {Wheeler}, {Ehgamberdiev}, {Mirzaqulov}, {B{\'o}di}, {Bogn{\'a}r}, {Cseh}, {Hanyecz}, {Ign{\'a}cz}, {Kalup}, {K{\"o}nyves-T{\'o}th}, {Kriskovics}, {Ordasi}, {P{\'a}l}, {S{\'a}rneczky}, {Seli}, {Szak{\'a}ts}, {Arranz-Heras}, {Benavides-Palencia}, {Cejudo-Mart{\'\i}nez}, {De la Fuente-Fern{\'a}ndez}, {Escart{\'\i}n-P{\'e}rez}, {Garc{\'\i}a-De la Cuesta}, {Gonz{\'a}lez-Carballo}, {Gonz{\'a}lez-Farf{\'a}n}, {Lim{\'o}n-Mart{\'\i}nez}, {Mantero}, {Naves-Nogu{\'e}s}, {Morales-Aimar}, {Ru{\'\i}z-Ru{\'\i}z}, {Sold{\'a}n-Alfaro}, {Valero-P{\'e}rez}, {Violat-Bordonau}, {Zhang}, {Zhang}, {Li}, {Chen}, {Sai}, \& {Li}}]{Xiang+21}
{Xiang}, D., {Wang}, X., {Lin}, W., {et~al.} 2021, \apj, 910, 42, \dodoi{10.3847/1538-4357/abdeba}

\bibitem[{{Xu} {et~al.}(2025){Xu}, {Zhu}, {Liu}, {Fynbo}, {Zou}, {Kumar}, {Liu}, {Song}, {Li}, {Mao}, {Liu}, {An}, {Li}, {Wang}, {Geng}, {Wu}, {Sun}, {Yuan}, \& {Zhang}}]{ep250108a_not_sn}
{Xu}, D., {Zhu}, Z.~P., {Liu}, X., {et~al.} 2025, GRB Coordinates Network, 38984, 1

\bibitem[{{Yan} {et~al.}(2017){Yan}, {Lunnan}, {Perley}, {Gal-Yam}, {Yaron}, {Roy}, {Quimby}, {Sollerman}, {Fremling}, {Leloudas}, {Cenko}, {Vreeswijk}, {Graham}, {Howell}, {De Cia}, {Ofek}, {Nugent}, {Kulkarni}, {Hosseinzadeh}, {Masci}, {McCully}, {Rebbapragada}, \& {Wo{\'z}niak}}]{Yan+17}
{Yan}, L., {Lunnan}, R., {Perley}, D.~A., {et~al.} 2017, \apj, 848, 6, \dodoi{10.3847/1538-4357/aa8993}

\bibitem[{{Yang} {et~al.}(2022){Yang}, {Ai}, {Zhang}, {Zhang}, {Liu}, {Wang}, {Yang}, {Yin}, {Li}, \& {L{\"u}}}]{Yang+22}
{Yang}, J., {Ai}, S., {Zhang}, B.-B., {et~al.} 2022, \nat, 612, 232, \dodoi{10.1038/s41586-022-05403-8}

\bibitem[{{Yang} {et~al.}(2025){Yang}, {Becerra}, {Watson}, {Troja}, {Dichiara}, \& {Hu}}]{ep250207b_gtc}
{Yang}, Y.-H., {Becerra}, R.~L., {Watson}, A.~M., {et~al.} 2025, GRB Coordinates Network, 39305, 1

\bibitem[{{Yaron} \& {Gal-Yam}(2012)}]{Yaron+12}
{Yaron}, O., \& {Gal-Yam}, A. 2012, \pasp, 124, 668, \dodoi{10.1086/666656}

\bibitem[{{Yin} {et~al.}(2024){Yin}, {Zhang}, {Yang}, {Sun}, {Zhang}, {Shao}, {Hu}, {Zhu}, {Xu}, {An}, {Gao}, {Wu}, {Zhang}, {Castro-Tirado}, {Pandey}, {Rau}, {Lei}, {Xie}, {Ghirlanda}, {Piro}, {O'Brien}, {Troja}, {Jonker}, {Yu}, {An}, {Chen}, {Chen}, {Dong}, {Eyles-Ferris}, {Fan}, {Fu}, {Fynbo}, {Gao}, {Huang}, {Jiang}, {Jiang}, {Julakanti}, {Kuulkers}, {Lao}, {Li}, {Ling}, {Liu}, {Liu}, {Mou}, {Pan}, {Wei}, {Wu}, {Yadav}, {Yang}, {Yuan}, \& {Zhang}}]{Yin+24}
{Yin}, Y.-H.~I., {Zhang}, B.-B., {Yang}, J., {et~al.} 2024, \apjl, 975, L27, \dodoi{10.3847/2041-8213/ad8652}

\bibitem[{{Yuan} {et~al.}(2022){Yuan}, {Zhang}, {Chen}, \& {Ling}}]{Yuan+22}
{Yuan}, W., {Zhang}, C., {Chen}, Y., \& {Ling}, Z. 2022, in Handbook of X-ray and Gamma-ray Astrophysics, ed. C.~{Bambi} \& A.~{Sangangelo}, 86, \dodoi{10.1007/978-981-16-4544-0_151-1}

\bibitem[{{Yuan} {et~al.}(2015){Yuan}, {Zhang}, {Feng}, {Zhang}, {Ling}, {Zhao}, {Deng}, {Qiu}, {Osborne}, {O'Brien}, {Willingale}, {Lapington}, {Fraser}, \& {the Einstein Probe team}}]{Yuan+15}
{Yuan}, W., {Zhang}, C., {Feng}, H., {et~al.} 2015, arXiv e-prints, arXiv:1506.07735, \dodoi{10.48550/arXiv.1506.07735}

\bibitem[{{Zenati} {et~al.}(2020){Zenati}, {Siegel}, {Metzger}, \& {Perets}}]{Zenati+20}
{Zenati}, Y., {Siegel}, D.~M., {Metzger}, B.~D., \& {Perets}, H.~B. 2020, \mnras, 499, 4097, \dodoi{10.1093/mnras/staa3002}

\bibitem[{{Zhang} {et~al.}(2024){Zhang}, {Mao}, {Zhang}, {Liu}, {Liu}, {Zhang}, {Ling}, {Jin}, {Cheng}, {Chen}, {Cui}, {Fan}, {Hu}, {Hu}, {Huang}, {Li}, {Lian}, {Liu}, {Lv}, {Pan}, {Pan}, {Sun}, {Wang}, {Wang}, {Wu}, {Xu}, {Xu}, {Yang}, {Zhang}, {Zhang}, {Zhang}, {Zhao}, {Kuulkers}, {O'Brien}, {Yuan}, \& {Einstein Probe team}}]{240315a_disc}
{Zhang}, W.~J., {Mao}, X., {Zhang}, W.~D., {et~al.} 2024, GRB Coordinates Network, 35931, 1

\bibitem[{{Zhu} {et~al.}(2024){Zhu}, {Liu}, {Yu}, {Mandel}, {Hirai}, {Zhang}, \& {Chen}}]{Zhu+24}
{Zhu}, J.-P., {Liu}, L.-D., {Yu}, Y.-W., {et~al.} 2024, \apjl, 970, L42, \dodoi{10.3847/2041-8213/ad63a8}

\bibitem[{{Zou} {et~al.}(2025){Zou}, {Liu}, {Kumar}, {Chen}, {Du}, {Wang}, {Lagioia}, {Fang}, {Pan}, {Han}, {Zhang}, {Xin}, {Wu}, {Liu}, {Liu}, \& {Mephisto Team}}]{ep250108a_mephisto}
{Zou}, X., {Liu}, C., {Kumar}, B., {et~al.} 2025, GRB Coordinates Network, 38914, 1

\end{thebibliography}

\newpage
\section*{Appendix}

\subsection{Nickel Mixing Model}
\label{subsec:mixing}

To better ascertain the quantity ${}^{56}$Ni and to also explore the impact of mixing on our light curve fits and estimated parameters, we also fit a ${}^{56}$Ni-mixing model. We outline the details of this model below, but detailed derivations and comparison to the one-zone model will be presented in a forthcoming publication (Sarin et al., in prep). This derivation is largely inspired by the kilonova model outlined in~\citet{metzger19}. We begin by assuming a power-law distribution of masses under the assumption of homologous expansion

\begin{equation}
M_{i} = M_{*}(v_{i}/v_0)^{-\beta}, 
\end{equation}

\noindent where $M_{*}$ is an arbitrary coefficient such that the sum of all layers is the total ejecta mass, $M$, $v_0$ is the minimum velocity of the outflow, and $M_{i}$ describes the mass in each `layer'. 
The number of layers is arbitrary, provided that they are sufficiently thin to accurately capture the dynamical evolution of the ejecta. A better description of SN ejecta is likely a broken power law~\citep[e.g.,][]{matznermckee99}. However, the deviations in light curve for different density profiles are smaller than the impact of mixing and other uncertain physics such as $\gamma$-ray leakage \citep[e.g.,][]{Sollerman98}.
The thermal energy of each layer evolves following the first law of thermodynamics, 

\begin{equation}
\frac{dE_{i}}{dt} = \dot{Q}_{i} -\frac{E_i}{R_i}\frac{dR_{i}}{dt} - L_{i}.
\end{equation}

\noindent where $\dot{Q}_{i}$ describes the energy input into each layer from radioactive decay, the second term on the right-hand-side describes losses due to PdV expansion and $L_{i}$ describes radiative losses from each layer, which needs to account for energy losses due to the light crossing times~\citep[e.g.,][]{metzger19}. These equations can be numerically solved while distributing the total ${}^{56}$Ni into layers (following the same distribution as the original mass distribution) up to a certain mass layer set by an additional parameter $f_{\rm mix}$ to capture nickel mixing. The photospheric radius is set as the radius of the mass shell at which the $\tau = 1$, which starts to recede back into the ejecta when the temperature drops below $T_{\rm floor}$. 
We further improve the physics of the model by incorporating a temperature-dependent opacity following,

\begin{equation}
\kappa_{\rm eff} = \kappa_{\rm min} + 0.5 (\kappa_{\rm max} - \kappa_{\rm min}) \times 
\left(1 + \tanh\left(\frac{T - T_{\rm floor}}{\Delta T}\right)\right), 
\end{equation}

\noindent where $\kappa_{\rm eff}$ is the effective opacity, which smoothly transitions from the maximum opacity, $\kappa_{\rm max} = 0.7~\rm{cm}^2~g^{-1}$ to the minimum opacity, $\kappa_{\rm min} = 0.07~\rm{cm}^2~g^{-1}$ as the temperature starts to approach $T_{\rm floor}$. This functional form and $\Delta T$ are chosen to replicate the opacity evolution in detailed numerical simulations~\citep[e.g.,][]{Nagy2018}. We note that this function is merely meant to include temperature dependence and ignores the more complex dependence on density, i.e., the opacity does not vary between layers. 

We fit this model to the observed light curve as we did in Section~\ref{subsec:arnett}. Compared to the one-zone model, our estimated parameters better match intuition and expectations. In particular, we infer a total ejecta mass of $0.8^{+0.5}_{-0.2}~M_{\odot}$ with a ${}^{56}$Ni mass of $0.21 \pm 0.04~M_{\odot}$. We also find evidence for significant mixing, with $f_{\rm mix} = 62\pm 20\%$, that is, ${}^{56}$Ni is distributed out into $\sim 60\%$ of the mass layers. We find the cumulative mass out to the layers with ${}^{56}$Ni is $0.6^{+0.5}_{-0.2} M_\odot$. The lower ${}^{56}$Ni estimate compared to our one-zone model is driven by the high mixing, which reduces the need for a larger central ${}^{56}$Ni as the mixed material has to overcome less adiabatic losses and diffuse through less ejecta. Comparing the Bayesian evidences, we find that the mixing model is a better fit to the data, with a $\ln \rm{BF} = 1.4$ in favour of the mixing model.


\subsection{Priors for Modeling Near-IR Absorption Features}
\label{sec:priors_appendix}

We give the priors used in our fitting of the 1 and 2 $\mu$m features in Section~\ref{subsec:helium} in Table~\ref{tab:priors}.

\begin{deluxetable}{cc}
\tablecaption{Priors Used for Each Line Fit in Section~\ref{subsec:helium} \label{tab:priors}}
\tablehead{
    \multicolumn{2}{c}{\textbf{ He I / C I / Mg II Priors}} \\
    \hline
    \colhead{Parameter} & 
    \colhead{Prior}
}
\startdata
$\mu_v\,\mathrm{[km\,s^{-1}]}$ & $\mathcal{N}(15000,\,1000)$ \\
$\log(\sigma_v\,\mathrm{[km\,s^{-1}]})$ & $\mathcal{N}(3.5,\,0.5)$ \\
$A$ & $\mathcal{U}(0,\,5)$ \\
$\log\left(\frac{A(2\,\micron)}{A(1\,\micron)}\right)$ & $\mathcal{U}(-2,\,0)$ 
\enddata
\end{deluxetable}

\subsection{Tables of Observations}
\label{sec:obstables}

Here we provide a log of the photometric programs used in this work, a lot of our spectroscopic observations, and a table of our photometric observations of SN\,2025kg.

\begin{deluxetable}{ccCcc}[H]
\tabletypesize{\small}
\centering
\tablecolumns{10}
\tabcolsep0.1in
\tablecaption{Photometric Programs Employed in this Work
\label{tab:photprog}}
\tablehead {
\colhead {Telescope}		&
\colhead {Instruments} &
\colhead {Filters} &
\colhead {Program ID(s)} &
\colhead {P.I.(s)} }
\startdata 
BlackGEM & - & q & Local Transient Survey & - \\
CFHT & MegaCam & gri & K1-03-00209 & A. Aryan \\
Gemini-North & GMOS-N & griz & GN-2024B-Q-107 & J. Rastinejad \\
Gemini-South & GMOS-S, FLAMINGOS2 & grizJHK & GS-2024B-Q-105, GS-2025A-Q-107 & J. Rastinejad \\
LBT & MODS & g'r'i' &  IT-2024B-023 & E. Maiorano \\
LT & IO:O & griz & PL24B06, PL25A25 & R. Eyles-Ferris \\
SLT & - & gri & -- & T. Chen \\
NOT & ALFOSC & griz & 70-301 & P. Jonker \\
MMT & MMIRS & K & UAO-G206-24B & J. Rastinejad  \\
Pan-STARRS & - & griz & - & S. Smartt \\
LOT & - & gri & R09 & A. Aryan \\
SOAR & Goodman & riz & SOAR2024B-016 & F. Bauer \\
T80S & T80S-Cam & griz & T80S-09 & C. Bom, C. Kilpatrick \\
\enddata
\end{deluxetable}

\startlongtable
\begin{deluxetable*}{crccccc}
\tabletypesize{\footnotesize}
\centering
\tablecolumns{7}
\tabcolsep0.04in
\tablecaption{Photometric Observations of SN\,2025kg
\label{tab:phot}}
\tablehead {
\colhead {Date}		&
\colhead {$\delta t$}		&
\colhead {Tel./Instum.}		&
\colhead {Band}		& 
\colhead {Exp. Time} &
\colhead {Magnitude}	&
\colhead {Ref.}  \\
\colhead {} &
\colhead {(days)} &
\colhead {} &
\colhead {(s)} &
\colhead {(AB mag)}	&
\colhead {} 
}
\startdata
2025 Jan 14.93401 & 6.41285 & LT/IO:O & \textit{g} & $6\times200$ & $21.11 \pm 0.13$ & This work. \\
2025 Jan 15.21383 & 6.69268 & Gemini-South/GMOS-S & \textit{g} & 60 & $20.94 \pm 0.05$ & This work. \\
2025 Jan 15.89495 & 7.37379 & LT/IO:O & \textit{g} & $6\times150$ & $20.93 \pm 0.08$ & This work. \\
2025 Jan 16.47753 & 7.956 & LOT, Lulin/Driver for Teledyne Princeton Instruments cameras & \textit{g} & $3\times300$ & $20.83 \pm 0.11$ & This work. \\
2025 Jan 16.52491 & 8.00375 & GMG-2.4m & \textit{g} & --- & $20.78 \pm 0.09$ & 1 \\
2025 Jan 16.85265 & 8.33149 & LT/IO:O & \textit{g} & $4\times150$ & $20.84 \pm 0.04$ & This work. \\
2025 Jan 17.87915 & 9.35799 & LT/IO:O & \textit{g} & $3\times150$ & $20.81 \pm 0.04$ & This work. \\
2025 Jan 17.90028 & 9.37911 & NOT/ALFOSC & \textit{g} & $4\times300$ & $20.84 \pm 0.03$ & This work. \\
2025 Jan 18.23900 & 9.71784 & Pan-STARRS & \textit{g} & 300 & $20.83 \pm 0.17$ & This work. \\
2025 Jan 18.83988 & 10.31872 & LT/IO:O & \textit{g} & $6\times120$ & $20.67 \pm 0.04$ & This work. \\
2025 Jan 19.52045 & 10.999 & SLT, Lulin/Andor CCD/EMCCD (SDK2) & \textit{g} & $18\times300$ & $20.92 \pm 0.24$ & This work. \\
2025 Jan 19.82822 & 11.30706 & LT/IO:O & \textit{g} & $6\times100$ & $20.68 \pm 0.06$ & This work. \\
2025 Jan 19.91938 & 11.39821 & NOT/ALFOSC & \textit{g} & $2\times300$ & $20.67 \pm 0.03$ & This work. \\
2025 Jan 20.24500 & 11.72384 & Pan-STARRS & \textit{g} & 400 & $20.65 \pm 0.10$ & This work. \\
2025 Jan 22.09702 & 13.57586 & LCO/Sinistro & \textit{g} & $3\times300$ & $20.47 \pm 0.05$ & This work. \\
2025 Jan 22.19166 & 13.671 & LBT/MODS & \textit{g} & $3\times120$ & $20.39 \pm 0.02$ & This work. \\
2025 Jan 22.23900 & 13.71784 & Pan-STARRS & \textit{g} & 300 & $20.56 \pm 0.09$ & This work. \\
2025 Jan 23.093 & 14.57184 & T80S/T80S-Cam & \textit{g} & 180 & $20.73 \pm 0.19$ & This work. \\
2025 Jan 23.09692 & 14.57576 & T80S/T80S-Cam & \textit{g} & 180 & $20.77 \pm 0.28$ & This work. \\
2025 Jan 23.10191 & 14.58075 & LCO/Sinistro & \textit{g} & $3\times300$ & $20.60 \pm 0.04$ & This work. \\
2025 Jan 24.17622 & 15.65506 & LCO/Sinistro & \textit{g} & $3\times300$ & $20.51 \pm 0.04$ & This work. \\
2025 Jan 24.25 & 15.72884 & Pan-STARRS & \textit{g} & 300 & $20.69 \pm 0.15$ & This work. \\
2025 Jan 25.83566 & 17.3145 & LT/IO:O & \textit{g} & $2\times60$ & $20.72 \pm 0.12$ & This work. \\
2025 Jan 26.23599 & 17.71484 & Pan-STARRS & \textit{g} & 300 & $20.61 \pm 0.13$ & This work. \\
2025 Jan 26.83535 & 18.31419 & LT/IO:O & \textit{g} & $6\times100$ & $20.62 \pm 0.04$ & This work. \\
2025 Jan 27.06368 & 18.54252 & T80S/T80S-Cam & \textit{g} & 180 & $20.83 \pm 0.15$ & This work. \\
2025 Jan 27.06768 & 18.54652 & T80S/T80S-Cam & \textit{g} & 180 & $20.94 \pm 0.16$ & This work. \\
2025 Jan 27.8232 & 19.30204 & LT/IO:O & \textit{g} & $6\times100$ & $20.90 \pm 0.11$ & This work. \\
2025 Jan 28.83482 & 20.31366 & LT/IO:O & \textit{g} & $6\times100$ & $20.82 \pm 0.05$ & This work. \\
2025 Jan 29.07637 & 20.55521 & T80S/T80S-Cam & \textit{g} & 180 & $21.11 \pm 0.22$ & This work. \\
2025 Jan 29.0803 & 20.55914 & T80S/T80S-Cam & \textit{g} & 180 & $21.06 \pm 0.21$ & This work. \\
2025 Jan 29.29000 & 20.76884 & Pan-STARRS & \textit{g} & 300 & $20.88 \pm 0.07$ & This work. \\
2025 Jan 29.83971 & 21.31854 & NOT/ALFOSC & \textit{g} & $2\times300$ & $20.83 \pm 0.04$ & This work. \\
2025 Jan 30.05121 & 21.53005 & T80S/T80S-Cam & \textit{g} & 180 & $21.33 \pm 0.27$ & This work. \\
2025 Jan 30.05526 & 21.5341 & T80S/T80S-Cam & \textit{g} & 180 & $21.27 \pm 0.23$ & This work. \\
2025 Feb 3.87949 & 26.35832 & NOT/ALFOSC & \textit{g} & $3\times300$ & $21.27 \pm 0.04$ & This work. \\
2025 Feb 05.23099 & 27.70984 & Pan-STARRS & \textit{g} & 300 & $>21.18$ & This work. \\
2025 Feb 8.86684 & 31.34567 & NOT/ALFOSC & \textit{g} & $4\times300$ & $21.72 \pm 0.11$ & This work. \\
2025 Feb 16.15225 & 36.6311 & ZTF & \textit{g} & 30 & $>21.77$ & This work. \\
2025 Feb 21.23332 & 43.712 & CFHT/MegaCam & \textit{g} & $1\times60$ & $22.34 \pm 0.13$ & This work. \\
2025 Mar 16.00208 & 66.48091 & Gemini-South/GMOS-S & \textit{g} & $7\times120$ & $22.75 \pm 0.06$ & This work. \\
\tableline
2025 Jan 19.08561 & 10.56445 & BlackGEM & \textit{q} & 60 & $20.51 \pm 0.12$ & This work. \\
2025 Jan 22.08465 & 13.5635 & BlackGEM & \textit{q} & 60 & $20.20 \pm 0.11$ & This work. \\
2025 Jan 23.08454 & 14.56339 & BlackGEM & \textit{q} & 60 & $19.93 \pm 0.09$ & This work. \\
2025 Jan 24.10005 & 15.57889 & BlackGEM & \textit{q} & 60 & $20.35 \pm 0.11$ & This work. \\
2025 Jan 25.08014 & 16.55898 & BlackGEM & \textit{q} & 60 & $20.12 \pm 0.10$ & This work. \\
2025 Jan 26.07680 & 17.55565 & BlackGEM & \textit{q} & 60 & $20.38 \pm 0.12$ & This work. \\
2025 Jan 27.07300 & 18.55185 & BlackGEM & \textit{q} & 60 & $20.31 \pm 0.11$ & This work. \\
\tableline
2025 Jan 15.17612 & 6.65497 & Gemini-South/GMOS-S & \textit{r} & 100 & $20.68 \pm 0.12$ & This work. \\
2025 Jan 15.60866 & 7.0875 & GMG-2.4m & \textit{r} & --- & $20.75 \pm 0.16$ & 1 \\
2025 Jan 15.90684 & 7.38568 & LT/IO:O & \textit{r} & $6\times150$ & $20.53 \pm 0.05$ & This work. \\
2025 Jan 16.26299 & 7.74184 & Pan-STARRS & \textit{r} & 300 & $20.41 \pm 0.17$ & This work. \\
2025 Jan 16.46645 & 7.945 & LOT, Lulin/Driver for Teledyne Princeton Instruments cameras & \textit{r} & $3\times300$ & $20.55 \pm 0.07$ & This work. \\
2025 Jan 16.52824 & 8.00708 & GMG-2.4m & \textit{r} & --- & $20.28 \pm 0.07$ & 1 \\
2025 Jan 16.86631 & 8.34515 & LT/IO:O & \textit{r} & $5\times150$ & $20.45 \pm 0.06$ & This work. \\
2025 Jan 17.89496 & 9.3738 & LT/IO:O & \textit{r} & $7\times150$ & $20.39 \pm 0.03$ & This work. \\
2025 Jan 17.91044 & 9.38927 & NOT/ALFOSC & \textit{r} & $2\times180$ & $20.30 \pm 0.03$ & This work. \\
2025 Jan 18.24299 & 9.72184 & Pan-STARRS & \textit{r} & 300 & $20.32 \pm 0.15$ & This work. \\
2025 Jan 18.84968 & 10.32852 & LT/IO:O & \textit{r} & $6\times120$ & $20.30 \pm 0.06$ & This work. \\
2025 Jan 19.05309 & 10.53194 & Gemini-South/GMOS-S & \textit{r} & 60 & $20.31 \pm 0.05$ & This work. \\
2025 Jan 19.1257 & 10.60454 & T80S/T80S-Cam & \textit{r} & 180 & $20.31 \pm 0.15$ & This work. \\
2025 Jan 19.12961 & 10.60845 & T80S/T80S-Cam & \textit{r} & 180 & $20.45 \pm 0.19$ & This work. \\
2025 Jan 19.48267 & 10.962 & LOT, Lulin/Driver for Teledyne Princeton Instruments cameras & \textit{r} & $6\times300$ & $20.24 \pm 0.06$ & This work. \\
2025 Jan 19.83732 & 11.31616 & LT/IO:O & \textit{r} & $5\times100$ & $20.20 \pm 0.04$ & This work. \\
2025 Jan 19.92580 & 11.40463 & NOT/ALFOSC & \textit{r} & $2\times180$ & $20.25 \pm 0.03$ & This work. \\
2025 Jan 20.24900 & 11.72784 & Pan-STARRS & \textit{r} & 300 & $20.21 \pm 0.06$ & This work. \\
2025 Jan 22.03038 & 13.50923 & Gemini-South/GMOS-S & \textit{r} & 60 & $20.15 \pm 0.12$ & This work. \\
2025 Jan 22.09848 & 13.57732 & T80S/T80S-Cam & \textit{r} & 180 & $20.21 \pm 0.16$ & This work. \\
2025 Jan 22.10085 & 13.57969 & LCO/Sinistro & \textit{r} & $3\times300$ & $20.12 \pm 0.04$ & This work. \\
2025 Jan 22.10241 & 13.58125 & T80S/T80S-Cam & \textit{r} & 180 & $20.23 \pm 0.16$ & This work. \\
2025 Jan 22.19305 & 13.672 & LBT/MODS & \textit{r} & $6\times120$ & $20.12 \pm 0.01$ & This work. \\
2025 Jan 22.24299 & 13.72184 & Pan-STARRS & \textit{r} & 300 & $20.09 \pm 0.08$ & This work. \\
2025 Jan 23.07442 & 14.55327 & Gemini-South/GMOS-S & \textit{r} & 60 & $20.15 \pm 0.10$ & This work. \\
2025 Jan 23.101 & 14.57984 & T80S/T80S-Cam & \textit{r} & 180 & $20.26 \pm 0.15$ & This work. \\
2025 Jan 23.10493 & 14.58377 & T80S/T80S-Cam & \textit{r} & 180 & $20.21 \pm 0.13$ & This work. \\
2025 Jan 23.10575 & 14.58459 & LCO/Sinistro & \textit{r} & $3\times300$ & $20.09 \pm 0.04$ & This work. \\
2025 Jan 24.18007 & 15.65891 & LCO/Sinistro & \textit{r} & $3\times300$ & $20.06 \pm 0.05$ & This work. \\
2025 Jan 24.25400 & 15.73284 & Pan-STARRS & \textit{r} & 300 & $19.99 \pm 0.10$ & This work. \\
2025 Jan 25.05144 & 16.53028 & Gemini-South/GMOS-S & \textit{r} & 60 & $20.08 \pm 0.14$ & This work. \\
2025 Jan 25.83161 & 17.31045 & LT/IO:O & \textit{r} & $12\times60$ & $20.21 \pm 0.09$ & This work. \\
2025 Jan 26.06544 & 17.54428 & T80S/T80S-Cam & \textit{r} & 180 & $20.20 \pm 0.10$ & This work. \\
2025 Jan 26.23999 & 17.71884 & Pan-STARRS & \textit{r} & 300 & $20.06 \pm 0.10$ & This work. \\
2025 Jan 26.84378 & 18.32262 & LT/IO:O & \textit{r} & $6\times100$ & $20.07 \pm 0.05$ & This work. \\
2025 Jan 27.07169 & 18.55053 & T80S/T80S-Cam & \textit{r} & 180 & $20.22 \pm 0.10$ & This work. \\
2025 Jan 27.07566 & 18.5545 & T80S/T80S-Cam & \textit{r} & 180 & $20.18 \pm 0.09$ & This work. \\
2025 Jan 27.08282 & 18.56598 & SOAR/Goodman & \textit{r} & $5\times140$ & $20.11 \pm 0.02$ & This work. \\
2025 Jan 27.83161 & 19.31045 & LT/IO:O & \textit{r} & $6\times100$ & $20.12 \pm 0.04$ & This work. \\
2025 Jan 28.14083 & 19.61968 & Gemini-South/GMOS-S & \textit{r} & 60 & $20.16 \pm 0.12$ & This work. \\
2025 Jan 28.84324 & 20.32208 & LT/IO:O & \textit{r} & $6\times100$ & $20.17 \pm 0.03$ & This work. \\
2025 Jan 29.08431 & 20.56315 & T80S/T80S-Cam & \textit{r} & 180 & $20.33 \pm 0.10$ & This work. \\
2025 Jan 29.08825 & 20.56709 & T80S/T80S-Cam & \textit{r} & 180 & $20.29 \pm 0.10$ & This work. \\
2025 Jan 29.24197 & 20.72082 & Gemini-North/GMOS-N & \textit{r} & 50 & $20.17 \pm 0.07$ & This work. \\
2025 Jan 29.29400 & 20.77284 & Pan-STARRS & \textit{r} & 300 & $20.06 \pm 0.05$ & This work. \\
2025 Jan 29.84613 & 21.32497 & NOT/ALFOSC & \textit{r} & $2\times180$ & $20.20 \pm 0.02$ & This work. \\
2025 Jan 30.0593 & 21.53814 & T80S/T80S-Cam & \textit{r} & 180 & $20.24 \pm 0.10$ & This work. \\
2025 Jan 30.0633 & 21.54214 & T80S/T80S-Cam & \textit{r} & 180 & $20.31 \pm 0.11$ & This work. \\
2025 Feb 3.89605 & 26.37488 & NOT/ALFOSC & \textit{r} & $2\times180$ & $20.44 \pm 0.03$ & This work. \\
2025 Feb 05.23400 & 27.71284 & Pan-STARRS & \textit{r} & 300 & $20.47 \pm 0.24$ & This work. \\
2025 Feb 8.87705 & 31.35588 & NOT/ALFOSC & \textit{r} & $2\times180$ & $20.77 \pm 0.06$ & This work. \\
2025 Feb 19.11887 & 40.59772 & ZTF & \textit{r} & 30 & $>21.09$ & This work. \\
2025 Feb 18.44954 & 40.928 & LOT, Lulin/Driver for Teledyne Princeton Instruments cameras & \textit{r} & $6\times300$ & $21.43 \pm 0.10$ & This work. \\
2025 Feb 20.01365 & 42.49249 & Gemini-South/GMOS-S & \textit{r} & 60 & $21.59 \pm 0.12$ & This work. \\
2025 Feb 21.22865 & 43.708 & CFHT/MegaCam & \textit{r} & $3\times60$ & $21.66 \pm 0.08$ & This work. \\
2025 Feb27.83787 & 50.31671 & NOT/ALFOSC & \textit{r} & $4\times180$ & $21.90 \pm 0.07$ & This work. \\
2025 Mar 16.01389 & 66.49272 & Gemini-South/GMOS-S & \textit{r} & $7\times120$ & $22.19 \pm 0.07$ & This work. \\
\tableline
2025 Jan 19.67752 & 11.15636 & DFOT & \textit{R} & $12\times300$ & $19.89 \pm 0.04$ & This work. \\
\tableline
2025 Jan 15.44627 & 6.92512 & ATLAS & \textit{o} & $7\times30$ & $>19.70$ & This work. \\
2025 Jan 16.85751 & 8.33636 & ATLAS & \textit{o} & $4\times30$ & $20.00 \pm 0.21$ & This work. \\
2025 Jan 18.09971 & 9.57855 & ATLAS & \textit{o} & $4\times30$ & $>19.90$ & This work. \\
2025 Jan 19.86149 & 11.34034 & ATLAS & \textit{o} & $4\times30$ & $>20.27$ & This work. \\
2025 Jan 20.87214 & 12.35098 & ATLAS & \textit{o} & $4\times30$ & $>19.84$ & This work. \\
2025 Jan 23.86630 & 15.34515 & ATLAS & \textit{o} & $3\times30$ & $19.99 \pm 0.20$ & This work. \\
2025 Jan 24.83415 & 16.313 & ATLAS & \textit{o} & $4\times30$ & $20.09 \pm 0.17$ & This work. \\
2025 Jan 28.83719 & 20.31603 & ATLAS & \textit{o} & $4\times30$ & $>20.05$ & This work. \\
2025 Feb 01.80872 & 24.28757 & ATLAS & \textit{o} & $4\times30$ & $20.17 \pm 0.21$ & This work. \\
2025 Feb 04.46553 & 26.94438 & ATLAS & \textit{o} & $8\times30$ & $20.23 \pm 0.19$ & This work. \\
\tableline
2025 Jan 15.22278 & 6.70162 & Gemini-South/GMOS-S & \textit{i} & 60 & $20.95 \pm 0.07$ & This work. \\
2025 Jan 15.9187 & 7.39754 & LT/IO:O & \textit{i} & $6\times150$ & $20.81 \pm 0.04$ & This work. \\
2025 Jan 16.26700 & 7.74584 & Pan-STARRS & \textit{i} & 300 & $20.86 \pm 0.23$ & This work. \\
2025 Jan 16.45980 & 7.939 & SLT, Lulin/Andor CCD/EMCCD (SDK2) & \textit{i} & $24\times300$ & $>19.80$ & This work. \\
2025 Jan 16.87837 & 8.35721 & LT/IO:O & \textit{i} & $6\times150$ & $20.63 \pm 0.04$ & This work. \\
2025 Jan 17.51716 & 8.996 & SLT, Lulin/Andor CCD/EMCCD (SDK2) & \textit{i} & $35\times300$ & $20.55 \pm 0.21$ & This work. \\
2025 Jan 17.90765 & 9.38649 & LT/IO:O & \textit{i} & $5\times150$ & $20.38 \pm 0.06$ & This work. \\
2025 Jan 17.91660 & 9.39543 & NOT/ALFOSC & \textit{i} & $3\times180$ & $20.50 \pm 0.04$ & This work. \\
2025 Jan 18.24700 & 9.72584 & Pan-STARRS & \textit{i} & 300 & $20.48 \pm 0.15$ & This work. \\
2025 Jan 18.85947 & 10.33831 & LT/IO:O & \textit{i} & $6\times120$ & $20.40 \pm 0.04$ & This work. \\
2025 Jan 19.13363 & 10.61247 & T80S/T80S-Cam & \textit{i} & 180 & $20.60 \pm 0.27$ & This work. \\
2025 Jan 19.13759 & 10.61643 & T80S/T80S-Cam & \textit{i} & 180 & $20.66 \pm 0.27$ & This work. \\
2025 Jan 19.84502 & 11.32386 & LT/IO:O & \textit{i} & $6\times100$ & $20.30 \pm 0.03$ & This work. \\
2025 Jan 19.93194 & 11.41078 & NOT/ALFOSC & \textit{i} & $3\times180$ & $20.38 \pm 0.03$ & This work. \\
2025 Jan 20.25299 & 11.73184 & Pan-STARRS & \textit{i} & 300 & $20.42 \pm 0.13$ & This work. \\
2025 Jan 22.10646 & 13.5853 & T80S/T80S-Cam & \textit{i} & 180 & $20.43 \pm 0.19$ & This work. \\
2025 Jan 22.11039 & 13.58923 & T80S/T80S-Cam & \textit{i} & 180 & $20.45 \pm 0.17$ & This work. \\
2025 Jan 22.19444 & 13.673 & LBT/MODS & \textit{i} & $6\times120$ & $20.28 \pm 0.02$ & This work. \\
2025 Jan 22.24599 & 13.72484 & Pan-STARRS & \textit{i} & 300 & $20.29 \pm 0.10$ & This work. \\
2025 Jan 23.10904 & 14.58788 & T80S/T80S-Cam & \textit{i} & 180 & $20.32 \pm 0.15$ & This work. \\
2025 Jan 23.10961 & 14.58845 & LCO/Sinistro & \textit{i} & $3\times300$ & $20.19 \pm 0.06$ & This work. \\
2025 Jan 23.11296 & 14.5918 & T80S/T80S-Cam & \textit{i} & 180 & $20.48 \pm 0.20$ & This work. \\
2025 Jan 23.50364 & 14.982 & SLT, Lulin/Andor CCD/EMCCD (SDK2) & \textit{i} & $36\times300$ & $20.36 \pm 0.14$ & This work. \\
2025 Jan 24.10162 & 15.59028 & SOAR/Goodman & \textit{i} & $12\times120$ & $20.17 \pm 0.01$ & This work. \\
2025 Jan 24.18393 & 15.66277 & LCO/Sinistro & \textit{i} & $3\times300$ & $20.13 \pm 0.07$ & This work. \\
2025 Jan 24.25799 & 15.73684 & Pan-STARRS & \textit{i} & 300 & $20.09 \pm 0.16$ & This work. \\
2025 Jan 25.83837 & 17.31721 & LT/IO:O & \textit{i} & $12\times100$ & $20.14 \pm 0.05$ & This work. \\
2025 Jan 26.07339 & 17.55223 & T80S/T80S-Cam & \textit{i} & 180 & $20.85 \pm 0.19$ & This work. \\
2025 Jan 26.24299 & 17.72184 & Pan-STARRS & \textit{i} & 300 & $20.23 \pm 0.14$ & This work. \\
2025 Jan 26.85217 & 18.33101 & LT/IO:O & \textit{i} & $6\times100$ & $20.15 \pm 0.04$ & This work. \\
2025 Jan 27.07968 & 18.55852 & T80S/T80S-Cam & \textit{i} & 180 & $20.27 \pm 0.11$ & This work. \\
2025 Jan 27.08373 & 18.56257 & T80S/T80S-Cam & \textit{i} & 180 & $20.33 \pm 0.11$ & This work. \\
2025 Jan 27.84 & 19.31884 & LT/IO:O & \textit{i} & $6\times100$ & $20.12 \pm 0.03$ & This work. \\
2025 Jan 28.85165 & 20.33049 & LT/IO:O & \textit{i} & $6\times100$ & $20.20 \pm 0.03$ & This work. \\
2025 Jan 29.09227 & 20.57111 & T80S/T80S-Cam & \textit{i} & 180 & $20.56 \pm 0.16$ & This work. \\
2025 Jan 29.0962 & 20.57504 & T80S/T80S-Cam & \textit{i} & 180 & $20.37 \pm 0.13$ & This work. \\
2025 Jan 29.29799 & 20.77684 & Pan-STARRS & \textit{i} & 300 & $20.26 \pm 0.07$ & This work. \\
2025 Jan 29.85230 & 21.33113 & NOT/ALFOSC & \textit{i} & $3\times180$ & $20.28 \pm 0.03$ & This work. \\
2025 Jan 30.06734 & 21.54618 & T80S/T80S-Cam & \textit{i} & 180 & $20.62 \pm 0.18$ & This work. \\
2025 Jan 30.07129 & 21.55013 & T80S/T80S-Cam & \textit{i} & 180 & $20.61 \pm 0.18$ & This work. \\
2025 Feb 3.90219 & 26.38102 & NOT/ALFOSC & \textit{i} & $3\times180$ & $20.38 \pm 0.04$ & This work. \\
2025 Feb 05.23799 & 27.71684 & Pan-STARRS & \textit{i} & 300 & $>20.67$ & This work. \\
2025 Feb 8.88318 & 31.36201 & NOT/ALFOSC & \textit{i} & $3\times180$ & $20.59 \pm 0.07$ & This work. \\
2025 Feb 21.23631 & 43.715 & CFHT/MegaCam & \textit{i} & $5\times60$ & $21.35 \pm 0.08$ & This work. \\
2026 Feb27.84760 & 50.32644 & NOT/ALFOSC & \textit{i} & $3\times360$ & $21.54 \pm 0.06$ & This work. \\
2025 Mar 16.02813 & 66.50699 & Gemini-South/GMOS-S & \textit{i} & $7\times100$ & $22.04 \pm 0.09$ & This work. \\
\tableline
2025 Jan 14.96824 & 6.44708 & LT/IO:O & \textit{z} & $6\times200$ & $21.06 \pm 0.18$ & This work. \\
2025 Jan 15.22722 & 6.70606 & Gemini-South/GMOS-S & \textit{z} & 60 & $21.25 \pm 0.14$ & This work. \\
2025 Jan 15.93347 & 7.41231 & LT/IO:O & \textit{z} & $6\times250$ & $20.82 \pm 0.09$ & This work. \\
2025 Jan 16.27099 & 7.74984 & Pan-STARRS & \textit{z} & 300 & $>20.82$ & This work. \\
2025 Jan 16.89159 & 8.37043 & LT/IO:O & \textit{z} & $3\times250$ & $21.07 \pm 0.12$ & This work. \\
2025 Jan 17.92261 & 9.40145 & LT/IO:O & \textit{z} & $6\times250$ & $20.89 \pm 0.10$ & This work. \\
2025 Jan 17.92701 & 9.40584 & NOT/ALFOSC & \textit{z} & $5\times200$ & $21.03 \pm 0.10$ & This work. \\
2025 Jan 18.25 & 9.72884 & Pan-STARRS & \textit{z} & 400 & $>21.01$ & This work. \\
2025 Jan 18.8703 & 10.34914 & LT/IO:O & \textit{z} & $5\times200$ & $20.74 \pm 0.12$ & This work. \\
2025 Jan 19.85485 & 11.33369 & LT/IO:O & \textit{z} & $6\times150$ & $20.80 \pm 0.07$ & This work. \\
2025 Jan 19.94758 & 11.42641 & NOT/ALFOSC & \textit{z} & $9\times200$ & $20.85 \pm 0.07$ & This work. \\
2025 Jan 20.25700 & 11.73584 & Pan-STARRS & \textit{z} & 400 & $20.69 \pm 0.23$ & This work. \\
2025 Jan 22.25 & 13.72884 & Pan-STARRS & \textit{z} & 400 & $20.66 \pm 0.19$ & This work. \\
2025 Jan 23.11703 & 14.59587 & T80S/T80S-Cam & \textit{z} & 180 & $>21.14$ & This work. \\
2025 Jan 23.12095 & 14.59979 & T80S/T80S-Cam & \textit{z} & 180 & $20.57 \pm 0.24$ & This work. \\
2025 Jan 24.05224 & 15.54025 & SOAR/Goodman & \textit{z} & $12\times120$ & $20.43 \pm 0.03$ & This work. \\
2025 Jan 24.26099 & 15.73984 & Pan-STARRS & \textit{z} & 400 & $>20.55$ & This work. \\
2025 Jan 25.85688 & 17.33572 & LT/IO:O & \textit{z} & $6\times100$ & $20.68 \pm 0.13$ & This work. \\
2025 Jan 26.07752 & 17.55636 & T80S/T80S-Cam & \textit{z} & 180 & $20.19 \pm 0.18$ & This work. \\
2025 Jan 26.08148 & 17.56032 & T80S/T80S-Cam & \textit{z} & 180 & $20.64 \pm 0.29$ & This work. \\
2025 Jan 26.24700 & 17.72584 & Pan-STARRS & \textit{z} & 400 & $20.60 \pm 0.28$ & This work. \\
2025 Jan 26.86055 & 18.33939 & LT/IO:O & \textit{z} & $6\times100$ & $20.39 \pm 0.09$ & This work. \\
2025 Jan 27.08774 & 18.56658 & T80S/T80S-Cam & \textit{z} & 180 & $20.46 \pm 0.21$ & This work. \\
2025 Jan 27.84839 & 19.32723 & LT/IO:O & \textit{z} & $6\times100$ & $20.54 \pm 0.10$ & This work. \\
2025 Jan 28.86005 & 20.33889 & LT/IO:O & \textit{z} & $6\times100$ & $20.56 \pm 0.08$ & This work. \\
2025 Jan 29.10028 & 20.57912 & T80S/T80S-Cam & \textit{z} & 180 & $20.45 \pm 0.25$ & This work. \\
2025 Jan 29.30099 & 20.77984 & Pan-STARRS & \textit{z} & 400 & $20.36 \pm 0.21$ & This work. \\
2025 Jan 29.86795 & 21.34678 & NOT/ALFOSC & \textit{z} & $9\times200$ & $20.56 \pm 0.09$ & This work. \\
2025 Jan 30.07532 & 21.55416 & T80S/T80S-Cam & \textit{z} & 180 & $20.29 \pm 0.24$ & This work. \\
2025 Feb 3.91785 & 26.39668 & NOT/ALFOSC & \textit{z} & $9\times200$ & $20.66 \pm 0.08$ & This work. \\
2025 Feb 05.24200 & 27.72084 & Pan-STARRS & \textit{z} & 400 & $>20.35$ & This work. \\
2025 Feb 8.89881 & 31.37764 & NOT/ALFOSC & \textit{z} & $9\times200$ & $20.94 \pm 0.09$ & This work. \\
2025 Feb 10.27033 & 32.74917333 & Pan-STARRS & \textit{z} & 1800 & $20.85 \pm 0.22$ & This work. \\
2025 Feb 14.24841 & 36.72725667 & Pan-STARRS & \textit{z} & 1800 & $21.00 \pm 0.06$ & This work. \\
2025 Feb 21.23366 & 43.71250667 & Pan-STARRS & \textit{z} & 1800 & $21.55 \pm 0.23$ & This work. \\
2027 Feb27.86133 & 50.34017 & NOT/ALFOSC & \textit{z} & $5\times300$ & $21.56 \pm 0.21$ & This work. \\
2025 Mar 01.24099 & 51.71984 & Pan-STARRS & \textit{z} & 1800 & $21.74 \pm 0.33$ & This work. \\
2025 Mar 16.23999 & 66.71884 & Pan-STARRS & \textit{z} & 1800 & $>21.41$ & This work. \\
\tableline
2025 Jan 20.03253 & 11.51137 & Gemini-South/F2 & \textit{J} & $26\times30$ & $20.76 \pm 0.03$ & This work. \\
2025 Jan 29.10275 & 20.58159 & Gemini-South/F2 & \textit{J} & $25\times30$ & $20.61 \pm 0.06$ & This work. \\
\tableline
2025 Jan 20.05285 & 11.53169 & Gemini-South/F2 & \textit{H} & $38\times10$ & $21.25 \pm 0.11$ & This work. \\
2025 Jan 23.08603 & 14.56486 & Gemini-South/F2 & \textit{H} & $10\times10$ & $21.22 \pm 0.12$ & This work. \\
2025 Jan 25.08371 & 16.56255 & Gemini-South/F2 & \textit{H} & $32\times10$ & $21.09 \pm 0.12$ & This work. \\
2025 Jan 28.08175 & 19.56059 & Gemini-South/F2 & \textit{H} & $22\times10$ & $21.21 \pm 0.11$ & This work. \\
2025 Feb 09.12099 & 31.59983 & Gemini-South/F2 & \textit{H} & $66\times10$ & $21.20 \pm 0.13$ & This work. \\
2025 Feb 22.01750 & 44.49633 & Gemini-South/F2 & \textit{H} & $38\times10$ & $21.76 \pm 0.12$ & This work. \\
2025 Mar 08.00869 & 58.48752 & Gemini-South/F2 & \textit{H} & $41\times10$ & $22.25 \pm 0.19$ & This work. \\
\tableline
2025 Jan 20.14749 & 11.62633 & MMT/MMIRS & \textit{K} & $93\times20$ & $21.45 \pm 0.15$ & This work. \\
2025 Jan 29.11899 & 20.59782 & Gemini-South/F2 & \textit{K} & $38\times10$ & $21.02 \pm 0.07$ & This work.
\enddata
\tablecomments{
Observations are not corrected for Galactic nor local extinction. \\ Times are presented in the observer frame. \\ References: (1) \citet{ep250108a_gmg} }
\end{deluxetable*}

\begin{deluxetable*}{clccCC}
\tabletypesize{\small}
\centering
\tablecolumns{10}
\tabcolsep0.12in
\tablecaption{Log of SN\,2025kg Spectroscopic Observations
\label{tab:spec}}
\tablehead {
\colhead {Date \& Mid-Time}		&
\colhead {$\delta t$}		& 
\colhead {Tel.} &
\colhead {Instrum.} &
\colhead {Exp. Time} &
\colhead {Wavelength Range} 
\\
\colhead {(UTC)}		&
\colhead {(days)}		&
\colhead {} &
\colhead {} &
\colhead {(s)} &
\colhead {(\AA)} 	
		}
\startdata
2025-01-19 01:41:47 & 10.5 & Gemini-South & GMOS & 4 \times 400 & 4300 - 8400 \\
2025-01-19 07:42:41 & 10.8 & Gemini-North & GMOS & 4 \times 1200 & 4800 - 9000 \\
2025-01-19 21:17:24 & 11.4 & GTC & OSIRIS & 3 \times 1200, 3 \times 900 &  3600 - 9500 \\
2025-01-22 01:12:58 & 13.5 & Gemini-South & GMOS & 4 \times 250 & 4300 - 9000 \\
2025-01-22 04:10:00 & 13.7 & LBT & MODS & 2 \times 900 & 3200-10000 \\
2025-01-22 05:57:25 & 13.7 & Gemini-North & GNIRS & 11 \times 300 & 7000 - 21400 \\
2025-01-24 02:25:46 & 15.6 & VLT & MUSE & 4 \times 608 & 4700 - 9300 \\
2025-01-25 01:35:30 & 16.5 & Gemini-South & GMOS & 4 \times 400 & 4300 - 8400 \\
2025-01-25 16:43:34 & 17.2 & JWST & NIRSpec & 6302.4 &  6000 - 53000 \\
2025-01-26 05:45:45 & 17.7 & Keck I & LRIS &  3 \times 900 &  3000 - 10200 \\
2025-01-29 06:25:50 & 20.7 & Gemini-South & GMOS & 4 \times 800 & 4900 - 9400 \\
2025-01-29 05:25:51 & 20.7 & Keck I & LRIS & 2 \times 900 & 3000 - 10200 \\
2025-02-05 21:02:32 & 28.4 & GTC & OSIRIS & 3 \times 900 & 3200 - 9400 \\
2025-02-20 01:14:56 & 42.5 & Gemini-South & GMOS & 4 \times 800 & 5500 - 9400 \\
2025-02-26 05:57:25 & 47.4 & Keck I & LRIS & 3 \times 1200 & 3000 - 8900 \\
\enddata
\end{deluxetable*}

\end{document}